\documentclass[12pt]{article}
\usepackage{epsfig}

\title{Wick-Cutkosky model: an introduction}
\author{Z.K.~Silagadze \vspace*{3mm} \\
\small \em
Budker Institute of Nuclear Physics \\
\small \em 630 090, Novosibirsk, Russia  \vspace*{3mm} \\}
\date{}
\newif\iftranslate
\translatetrue
\begin{document}
\large
\iftranslate
\maketitle

\begin{abstract}
A general pedagogical survey of the basics of the Bethe-Salpeter equation
and in particular Wick-Cutkosky model is given in great technical details.
\end{abstract}


\section*{Bethe-Salpeter amplitude}
The Bethe-Salpeter approach to the relativistic two-body bound state problem
assumes that all needed information about a bound state $|B>$ is contained in
the B-S amplitude
\begin{eqnarray}
\Phi (x_1,x_2;P_B)=<0|T\phi_1(x_1)\phi_2(x_2)|B>
\label{eq1} \end{eqnarray}
\noindent
and in its conjugate
\begin{eqnarray}
\bar \Phi (x_1,x_2;P_B)=<B|T\phi_1^+(x_1)\phi_2^+(x_2)|0> \; \; .
\label{eq2} \end{eqnarray}
Here $P_B$ is the bound state 4-momentum and field operators are in the
Heisenberg representation.

Note, that
$$ <B|T\phi_1^+(x_1)\phi_2^+(x_2)|0>= <0|\bar T\phi_1(x_1)\phi_2(x_2)|B>^*$$
\noindent
$\bar T$ being antichronological operator. So we can say that $\bar \Phi$
is obtained from $\Phi$ through time reversal.

It is clear that instead of individual $x_1,x_2$ coordinates more useful
are variables which characterize two particle relative motion and the
motion of the system as a whole.

The coordinate corresponding to the relative motion is of course
$x=x_1-x_2$. Then any linear independent combination $X=\eta_1x_1+\eta_2x_2$
can serve for a description of the system as a whole. Linear independence
means $\eta_1+\eta_2 \not= 0$. In order to have a correspondence to the
nonrelativistic center of mass definition, we take
\begin{eqnarray}
\eta_1=\frac{m_1}{m_1+m_2} \; \; , \; \; \eta_2=\frac{m_2}{m_1+m_2} \; \; .
\label{eq3} \end{eqnarray}
$x_1=X+\eta_2x \; , \; x_2=X-\eta_1x$ equalities and the fact that $\hat P$
4-momentum operator is the space-time translation operator enables us to
write
$$\phi_1(x_1)=e^{i\hat P X}\phi_1(\eta_2x)e^{-i\hat P X} \; \; , \; \;
\phi_2(x_2)=e^{i\hat P X}\phi_2(-\eta_1x)e^{-i\hat P X} \; .$$
\noindent
Inserting this expression into (1) we get
$$<0|T\phi_1(x_1)\phi_2(x_2)|B>=\Theta(x_0)<0|\phi_1(x_1)\phi_2(x_2)|B>+$$
$$ \Theta(-x_0)<0|\phi_2(x_2)\phi_1(x_1)|B>=$$
$$
\Theta(x_0)<0|e^{i\hat P X}\phi_1(\eta_2x)\phi_2(-\eta_1x)e^{-i\hat P X}|B>+
$$ $$
\Theta(-x_0)<0|e^{i\hat P X}\phi_2(-\eta_1x)\phi_1(\eta_2x)e^{-i\hat P X}|B>
= $$
$$e^{-iP_BX} \{ \Theta(x_0)<0|\phi_1(\eta_2x)\phi_2(-\eta_1x)|B>+ $$
$$ \Theta(-x_0)<0|\phi_2(-\eta_1x)\phi_1(\eta_2x)|B> \}= $$
$$e^{-iP_BX} <0|T\phi_1(\eta_2x)\phi_2(-\eta_1x)|B> \; . $$
\noindent
So, if the reduced B-S amplitude is introduced
\begin{eqnarray}
\Phi(x;P_B)=(2\pi)^{3/2} <0|T\phi_1(\eta_2x)\phi_2(-\eta_1x)|B>
\label{eq4} \end{eqnarray} \noindent
when
\begin{eqnarray}
\Phi(x_1,x_2;P_B)=(2\pi)^{-3/2} e^{-iP_BX} \Phi(x;P_B) \; .
\label{eq5} \end{eqnarray} \noindent
Analogously
\begin{eqnarray}
\bar \Phi(x_1,x_2;P_B)=(2\pi)^{-3/2} e^{iP_BX} \bar \Phi(x;P_B) \; ,
\label{eq6} \end{eqnarray} \noindent
where
$$\bar \Phi(x;P_B)=(2\pi)^{3/2} <B|T\phi_1^+(\eta_2x)\phi_2^+(-\eta_1x)|0>
\; . $$
An equation for the B-S amplitude can be obtained with the help of 4-point
Green's function.

\section*{Equation for Green's function}
Let us consider 4-point Green's function
$$G(x_1,x_2;y_1,y_2)=<0|T\phi_1(x_1)\phi_2(x_2)
\phi_1^+(y_1)\phi_2^+(y_2)|0> \; . $$
Field operators here are in the Heisenberg representation, so we have
the full Green's function, which can be expanded into a infinite
perturbation series. Let us rearrange the terms of this series: first of
all we sum up self-energy insertions in propagators of $\phi_1$ and $\phi_2$
particles, this will give us the full propagators for them, then from the
remaining diagrams we separate $(\phi_1+\phi_2)$-two-particle irreducible
ones, the sum of which will play the role of interaction operator
(a diagram is $(\phi_1+\phi_2)$-two-particle reducible, if by cutting one
$\phi_1$ and one $\phi_2$ of inner lines it can be divided into two
disconnected parts). More vividly this is illustrated graphically in Fig.1.
\begin{figure}[htb]
\begin{center}
\epsfig{figure=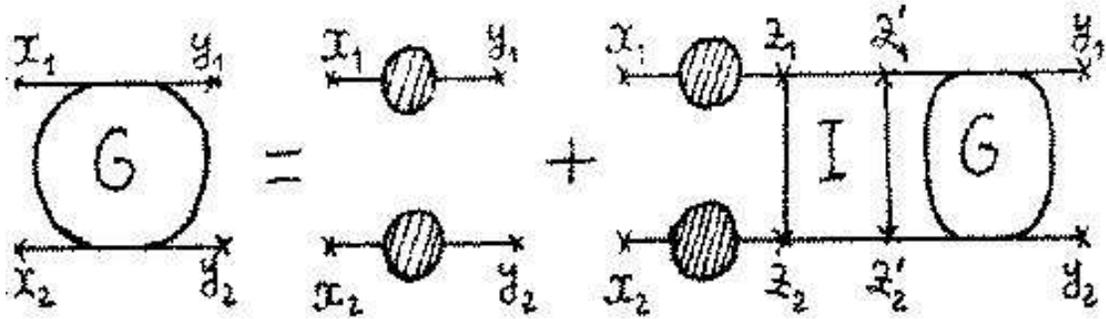,width=15cm}
\caption{Integral equation for the full Green's function.}
\end{center}
\label{Fig1}
\end{figure}

\noindent
Here $I$ is the sum of $(\phi_1+\phi_2)$-two-particle irreducible diagrams
without external propagators, as is shown in Fig.2.
\begin{figure}[htb]
\begin{center}
\epsfig{figure=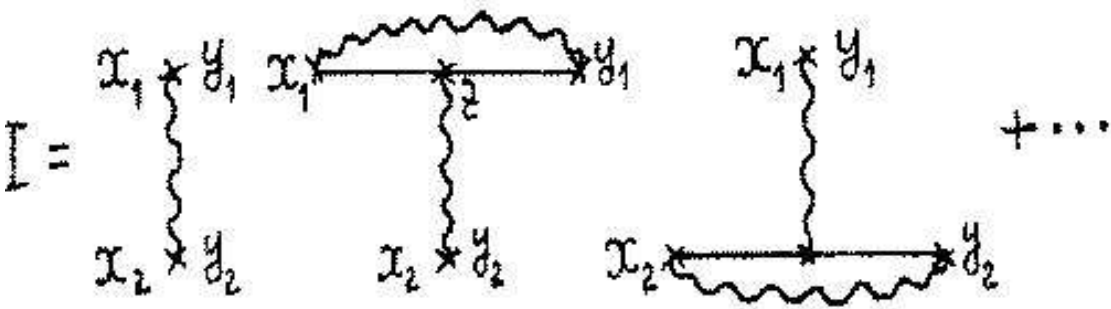,width=15cm}
\caption{Integral equation kernel.}
\end{center}
\label{Fig2}
\end{figure}

The analytical expression of the above integral equation (Fig.1) looks like
\begin{eqnarray}
G(x_1,x_2;y_1,y_2)=\Delta_1(x_1-y_1) \Delta_2(x_2-y_2)+  \nonumber \\
\int \, dz_1 \, dz_2 \, dz_1^\prime \, dz_2^\prime
\Delta_1(x_1-z_1)\Delta_2(x_2-z_2)
I(z_1,z_2;z_1^\prime,z_2^\prime)G(z_1^\prime,z_2^\prime;y_1,y_2) \; .
\label{eq7} \end{eqnarray}
We get the so called ladder approximation if the full propagator $\Delta
(x-y)$ is replaced by the free propagator and in the interaction function $I$
only the first single-particle-exchange term is left: 
\begin{figure}[htb]
\begin{center}
\epsfig{figure=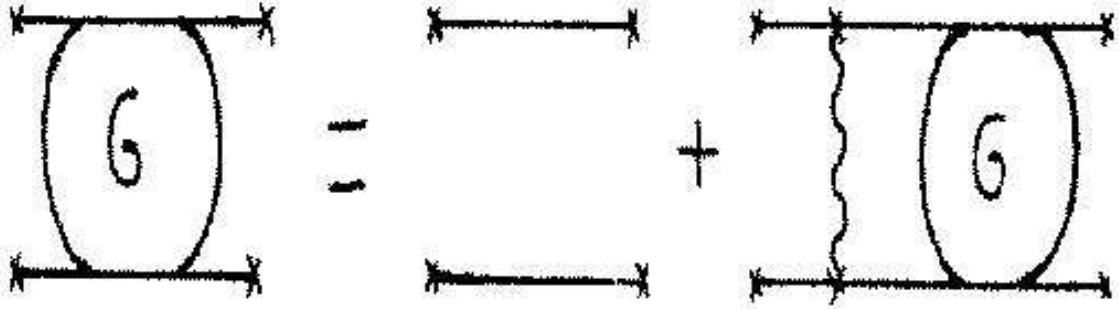,width=15cm}
\caption{Ladder approximation.}
\end{center}
\label{Fig3}
\end{figure}

An appearance of the iterative solution of this equation explains the
origin of the approximation name: 
\begin{figure}[htb]
\begin{center}
\epsfig{figure=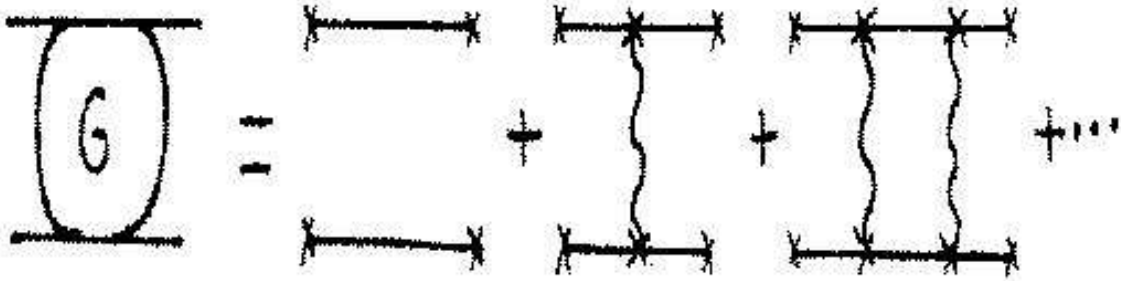,width=15cm}
\caption{Iterative solution in the ladder approximation.}
\end{center}
\label{Fig4}
\end{figure}

\section*{Momentum representation}
Eq.7 is written in the coordinate space. as is well known, a relativistic
particle can't be localized in the space-time with arbitrary precision.
Therefore in relativistic theory momentum space is often more useful
than coordinate space.

In order to rewrite Eq.7  in the momentum space, let us perform the Fourier
transformation
\begin{eqnarray}
G(x_1,x_2;y_1,y_2)=(2\pi)^{-8}\int dp \, dq \, dP \, e^{-ipx} e^{iqy}
e^{-iP(X-Y)}G(p,q;P) \; .
\label{eq8} \end{eqnarray}
\noindent
Because of the translational invariance, $G$ depends only on coordinate
differences. In the role of independent differences drawn up from the
$x_1,x_2,y_1,y_2$ coordinates, we have taken $x=x_1-x_2, \, y=y_1-y_2$
and $X-Y=(\eta_1x_1+\eta_2x_2)-(\eta_1y_1+\eta_2y_2).$

let us clear up some technical details of the transition to the momentum
space. First of all $\Delta_1(x_1-y_1) \Delta_2(x_2-y_2)$ product should
be rewritten in the form of Eq.8. Momentum space propagator is defined
through the Fourier transform
$$\Delta(x)=(2\pi)^{-4}\int dp \, e^{-ipx}\Delta (p) \; . $$
\noindent Then
$$\Delta_1(x_1-y_1) \Delta_2(x_2-y_2)=$$
$$(2\pi)^{-8}\int dp_1 \, dp_2 e^{-ip_1(x_1-y_1)}e^{-ip_2(x_2-y_2)}
\Delta_1(p_1) \Delta_2(p_2) \;. $$   \noindent
Let us make $p_1=\eta_1P+p, \, p_2=\eta_2P-p$ variable change in the
integral with $\frac{D(p_1,p_2)}{D(P,p)}=\eta_1+\eta_2=1$ unit Jacobian
$$\Delta_1(x_1-y_1) \Delta_2(x_2-y_2)=$$
$$(2\pi)^{-8}\int dP \, dp \, e^{-iP(X-Y)} e^{-ipx} e^{ipy}
\Delta_1(\eta_1P+p) \Delta_2(\eta_2P-p) \; , $$ \noindent
which, it is clear, can be rewritten also as
$$\Delta_1(x_1-y_1) \Delta_2(x_2-y_2)=$$
$$(2\pi)^{-8}\int dP \, dp \, dq \, e^{-iP(X-Y)} e^{-ipx} e^{iqy}
\delta (p-q) \Delta_1(\eta_1P+p) \Delta_2(\eta_2P-p) \; . $$
Now let us transform $\Delta_1\Delta_2IG$ integral
$$\Delta_1\Delta_2IG=$$
$$\int dz_1dz_2dz_1^\prime dz_2^\prime \Delta_1(x_1-z_1)\Delta_2(x_2-z_2)
I(z_1,z_2;z_1^\prime,z_2^\prime) G(z_1^\prime,z_2^\prime;y_1,y_2)=$$
$$(2\pi)^{-24}\int dz_1dz_2dz_1^\prime dz_2^\prime dP \, dp \, dq
\, dP^\prime \, dp^\prime \, dq^\prime
\, dP^{\prime \prime} \, dp^{\prime \prime} \, dq^{\prime \prime} \times$$
$$e^{-iP(X-Z)}e^{-ipx}e^{iq^\prime z} \delta (p-q^\prime)
\Delta_1(\eta_1P+p) \Delta_2(\eta_2P-p)  \times $$
$$e^{-iP^\prime (Z-Z^\prime)} e^{-ip^\prime z} e^{iq^{\prime \prime}z^\prime}
I(p^\prime,q^{\prime \prime};P^\prime)e^{-iP^{\prime \prime}(Z^\prime -Y)}
e^{-ip^{\prime \prime}z^\prime}e^{iqy}G(p^{\prime \prime},q;
P^{\prime \prime})= $$
$$(2\pi)^{-24} \int dz \, dz^\prime \, dZ \, dZ^\prime  \,
dP \, dp \, dq \, dP^\prime \, dp^\prime
\, dP^{\prime \prime} \, dp^{\prime \prime} \, dq^{\prime \prime} \times $$
$$ \Delta_1(\eta_1P+p) \Delta_2(\eta_2P-p)
I(p^\prime,q^{\prime \prime};P^\prime) \times $$
$$G(p^{\prime \prime},q;P^{\prime \prime})e^{-iz(p^\prime-p)}
e^{-iz^\prime(p^{\prime \prime}-q^{\prime \prime})}
e^{-iZ(P^\prime-P)} e^{-iZ^\prime (P^{\prime \prime}-P^\prime)} \; .$$
\noindent Coordinate integrations will produce $\delta$-functions and some
4-momentum integrals become trivial. At the end we get
$$\Delta_1\Delta_2IG=$$
$$(2\pi)^{-8}\int dP \, dp \, dq \, dq^\prime
\Delta_1(\eta_1P+p) \Delta_2(\eta_2P-p) I(p,q^\prime;P)G(q^\prime,q;P) \;.$$
After all of these, Eq.7 easily is rewritten in the momentum space
$$G(p,q;P)=\delta (p-q) \Delta_1(\eta_1P+p) \Delta_2(\eta_2P-p) +$$
$$\Delta_1(\eta_1P+p) \Delta_2(\eta_2P-p) \int dq^\prime  \,
I(p,q^\prime;P)G(q^\prime,q;P)  \; , $$ \noindent
or
\begin{eqnarray}  &&
[\Delta_1(\eta_1P+p) \Delta_2(\eta_2P-p)]^{-1}G(p,q;P)= \nonumber \\ &&
\delta (p-q) + \int dq^\prime  \,
I(p,q^\prime;P)G(q^\prime,q;P)  \; . \label{eq9} \end{eqnarray} \noindent
If the following definitions are introduced
$$(A\cdot B)(p,q;P)=\int dq^\prime \, A(p,q^\prime;P)B(q^\prime,q;P) \; ,$$
$$K(p,q;P)= [\Delta_1(\eta_1P+p) \Delta_2(\eta_2P-p)]^{-1} \delta (p-q) \; ,
$$ \noindent
this equation takes the form
\begin{eqnarray}
K \cdot G =1+I \cdot G \; .
\label{eq10} \end{eqnarray}

\section*{Two-particle bound states contribution 
\\ into Green's function}
Inserting $\sum |n><n|=1$ full set of states, the Green's function can be
rewritten as
$$G(x_1,x_2;y_1,y_2) \equiv <0|T\phi_1(x_1)\phi_2(x_2)
\phi_1^+(y_1)\phi_2^+(y_2)|0>=$$
$$\Theta ({\rm min}[(x_1)_0,(x_2)_0]-{\rm max}[(y_1)_0,(y_2)_0]) \times $$
$$<0|T\phi_1(x_1)\phi_2(x_2)|n><n|T \phi_1^+(y_1)\phi_2^+(y_2)|0> +$$
$$\Theta ({\rm min}[(x_1)_0,(y_1)_0]-{\rm max}[(x_2)_0,(y_2)_0]) \times $$
$$<0|T\phi_1(x_1)\phi_1^+(y_1)|n><n|T \phi_2(x_2)\phi_2^+(y_2)|0> +$$
$$\Theta ({\rm min}[(x_1)_0,(y_2)_0]-{\rm max}[(x_2)_0,(y_1)_0]) \times $$
$$<0|T\phi_1(x_1)\phi_2^+(y_2)|n><n|T \phi_2(x_2)\phi_1^+(y_1)|0> +$$
$$\Theta ({\rm min}[(x_2)_0,(y_1)_0]-{\rm max}[(x_1)_0,(y_2)_0]) \times $$
$$<0|T\phi_2(x_2)\phi_1^+(y_1)|n><n|T \phi_1(x_1)\phi_2^+(y_2)|0> +$$
$$\Theta ({\rm min}[(x_2)_0,(y_2)_0]-{\rm max}[(x_1)_0,(y_1)_0]) \times $$
$$<0|T\phi_2(x_2)\phi_2^+(y_2)|n><n|T \phi_1(x_1)\phi_1^+(y_1)|0> +$$
$$\Theta ({\rm min}[(y_1)_0,(y_2)_0]-{\rm max}[(x_1)_0,(x_2)_0]) \times $$
\begin{eqnarray}
<0|T\phi_1^+(y_1)\phi_2^+(y_2)|n><n|T \phi_1(x_1)\phi_2(x_2)|0>
\label{eq11} \end{eqnarray} \noindent
To separate the contribution from two-particle bound states $|B>$, let us
substitute
$$\sum |n><n| \longrightarrow \int dP \, \delta(P^2-m_B^2)\Theta (P_0)
|B><B| \; . $$ \noindent
($<\vec{P}|\vec{Q}>=2P_0\delta (\vec{P}-\vec{Q})$ normalization is assumed).

An action of two annihilation operators over $|B>$ is needed to obtain
the vacuum quantum numbers. Only in this case we get a nonzero matrix
element in (11). Therefore two-particle bound state contribution into
the Green's function is
$$B=\int dP \, <0|T\phi_1(x_1)\phi_2(x_2)|B><B|T \phi_1^+(y_1)
\phi_2^+(y_2)|0> \times $$
$$\delta(P^2-m_B^2)\Theta (P_0)
\Theta ({\rm min}[(x_1)_0,(x_2)_0]-{\rm max}[(y_1)_0,(y_2)_0]) \; . $$ \noindent
$dP_0$ integration can be performed by using
$$\delta(P^2-m_B^2)\Theta (P_0)=$$
$$\frac{\Theta (P_0)}{2P_0} \left \{ \delta(P_0-\sqrt{\vec{P}^2+m_B^2})+
\delta(P_0+\sqrt{\vec{P}^2+m_B^2}) \right \} = $$
$$\frac{\Theta (P_0)}{2P_0} \delta(P_0-\sqrt{\vec{P}^2+m_B^2}) $$ \noindent
and remembering (5) and (6) :
$$B=(2\pi)^{-3} \int \frac{d \vec{P}}{2\omega_B}\Phi (x;P_B)\bar \Phi (y,P_B)
e^{-i\omega_B (X_0-Y_0)}e^{-i\vec{P}(\vec{X}-\vec{Y})} \times $$
$$\Theta ({\rm min}[(x_1)_0,(x_2)_0]-{\rm max}[(y_1)_0,(y_2)_0]) \; . $$ \noindent
here $P_B=(\omega_B,\vec{P})$ and $\omega_B=\sqrt{\vec{P}^2+m_B^2}$;
 $m_B$-being the bound state mass.

The argument of $\Theta$ function can be transformed as
$${\rm min}[(x_1)_0,(x_2)_0]-{\rm max}[(y_1)_0,(y_2)_0]=$$
$${\rm min}[X_0+\eta_2x_0,X_0-\eta_1x_0]-{\rm max}[Y_0+\eta_2y_0,Y_0-\eta_1y_0]=$$
$$ \left \{  \matrix {
X_0-Y_0-\eta_1x_0-\eta_2y_0 \; , \; {\rm if} \; x_0>0, \, y_0>0 \cr
X_0-Y_0-\eta_1x_0+\eta_1y_0 \; , \; {\rm if} \; x_0>0, \, y_0<0 \cr
X_0-Y_0+\eta_2x_0-\eta_2y_0 \; , \; {\rm if} \; x_0<0, \, y_0>0 \cr
X_0-Y_0+\eta_2x_0+\eta_1y_0 \; , \; {\rm if} \; x_0<0, \, y_0<0  }
\right \} = $$
$$X_0-Y_0-\frac{1}{2}|x_0|-\frac{1}{2}|y_0|+\frac{1}{2}(\eta_2-\eta_1)
(x_0-y_0) \; .$$ \noindent
Besides, let us use its integral representation
$$\Theta (z)=\frac{i}{2\pi}\int\limits_{-\infty}^\infty dk \,
\frac{e^{-ikz}}{k+i\epsilon} \; , $$ \noindent
in which the variable change $k \to P_0-\omega_B$ is made. Then
$$\Theta(X_0-Y_0-\frac{1}{2}|x_0|-\frac{1}{2}|y_0|+\frac{1}{2}(\eta_2-\eta_1)
(x_0-y_0))=\frac{i}{2\pi} \int\limits_{-\infty}^\infty
\frac{dP_0}{P_0-\omega_B+i\epsilon} \times$$
$$\exp{ \{ -i(P_0-\omega_B)[X_0-Y_0-\frac{1}{2}|x_0|-\frac{1}{2}|y_0|+
\frac{1}{2}(\eta_2-\eta_1)(x_0-y_0)] \} } \; .$$ \noindent
inserting this into B, we get
$$B(x,y;X-Y)=i(2\pi)^{-4}\int dP \, \Phi (x;P_B) \bar \Phi (y;P_B)
\frac{e^{-iP(X-Y)}}{2\omega_B (P_0-\omega_B+i\epsilon)} \times $$
$$\exp{ \{ -i(P_0-\omega_B)[-\frac{1}{2}|x_0|-\frac{1}{2}|y_0|+
\frac{1}{2}(\eta_2-\eta_1)(x_0-y_0)] \} }= $$
$$i(2\pi)^{-12}\int dP \, dp \, dq \, e^{-iP(X-Y)}e^{-ipx}e^{iqy}
\frac{\Phi (p;P_B) \bar \Phi (q;P_B)}{2\omega_B (P_0-\omega_B+i\epsilon)}
\times $$
$$\exp{ \{ -i(P_0-\omega_B)[-\frac{1}{2}|x_0|-\frac{1}{2}|y_0|+
\frac{1}{2}(\eta_2-\eta_1)(x_0-y_0)] \} }  \; , $$ \noindent
where in the reduced B-S amplitudes the transition to the momentum space
was done:
$$\Phi (x,P_B)=(2\pi)^{-4} \int dp \, e^{-ipx} \Phi (p,P_B)$$
\noindent and
$$\bar \Phi (y,P_B)=(2\pi)^{-4} \int dq \, e^{iqy} \bar \Phi (q,P_B) \, .$$
\noindent If we designate for brevity
$$A(p,q;P)=\frac{\Phi (p;P_B) \bar \Phi (q;P_B)}{2\omega_B (P_0-
\omega_B+i\epsilon)} \, ,$$
$$f(x_0,y_0)=\frac{1}{2}(\eta_2-\eta_1)(x_0-y_0)
-\frac{1}{2}|x_0|-\frac{1}{2}|y_0| \, $$ \noindent
then the momentum transform of B will be
$$B(p,q,P)=(2\pi)^{-4} \int e^{ipx} e^{-iqy} e^{iPX} B(x,y;X) dx \, dy \, dX
=$$ $$i(2\pi)^{-16}\int dx \, dy \, dX \, dP^\prime \, dp^\prime \,
dq^\prime e^{i(p-p^\prime)x} e^{-i(q-q^\prime)y} \times $$
$$e^{i(P-P^\prime)X} A(p^\prime,q^\prime;P^\prime)
\exp { \{ -i(P^\prime_0-\omega^\prime_B)f(x_0,y_0) \} } =$$
$$i(2\pi)^{-12} \int dx \, dy \, dp^\prime \, dq^\prime
e^{i(p-p^\prime)x} e^{-i(q-q^\prime)y} A(p^\prime,q^\prime;P^\prime) \times
$$ $$\exp { \{ -i(P_0-\omega_B)f(x_0,y_0) \} }  \simeq
i(2\pi)^{-4} A(p,q;P) \; {\rm when} \, P_0 \to \omega_B \; .$$
\noindent Therefore, near the point $P_0=\omega_B$
$$B(p,q;P)=\frac{i}{(2\pi)^4} \frac{\Phi (p;P_B) \bar \Phi (q;P_B)}
{2\omega_B (P_0-\omega_B+i\epsilon)} \; , $$ \noindent
that is two-particle bound state contribution in $G(p,q;P)$ has a pole
at $P_0=\omega_B$. Other $|n>$ intermediate states will give a regular
contribution at this point, if their masses differ from $m_B$ (and we
assume that this is the case). So
\begin{eqnarray}
G(p,q;P) \longrightarrow \frac{i}{(2\pi)^4} \frac{\Phi (p;P_B)
\bar \Phi (q;P_B)} {2\omega_B (P_0-\omega_B+i\epsilon)} \; ,
\; {\rm when} \, P_0 \to \omega_B \; .
\label{eq12} \end{eqnarray}

\section*{Bethe-Salpeter equation}
Let us take $p \neq q ,\; P \to P_B$ in the equation (9) for Green's
function. Taking into account (12), we get
$$[\Delta_1(\eta_1P+p) \Delta_2(\eta_2P-p)]^{-1}
\frac{i}{(2\pi)^4} \frac{\Phi (p;P_B)
\bar \Phi (q;P_B)} {2\omega_B (P_0-\omega_B+i\epsilon)}=$$
$$\int dq^\prime \, I(p,q^\prime;P_B) \frac{i}{(2\pi)^4}
\frac{\Phi (q^\prime;P_B)\bar \Phi (q;P_B)}
{2\omega_B (P_0-\omega_B+i\epsilon)} \; .$$ \noindent
This implies the Bethe-Salpeter equation for the B-S amplitude
\begin{eqnarray}
[\Delta_1(\eta_1P_B+p) \Delta_2(\eta_2P_B-p)]^{-1}\Phi (p;P_B)=
\int dq^\prime \, I(p,q^\prime;P_B) \Phi (q^\prime;P_B) \; .
\label{eq13} \end{eqnarray}

\section*{Normalization condition}
The normalization of the B-S amplitude can not be determined from the
homogeneous B-S equation. It can be obtained in such a way. From (10)
$G \cdot (K-I)=1$. Differentiating with respect to $P_0$, we find
$$\frac {\partial G}{\partial P_0} (K-I)+G \left ( \frac {\partial K}
{\partial P_0} -\frac {\partial I}{\partial P_0} \right ) = 0 \; , $$
\noindent that is
$$\frac{\partial G}{\partial P_0}=- G~ \left ( \frac {\partial K}
{\partial P_0} -\frac {\partial I}{\partial P_0} \right )~ G \; . $$
Let us consider this equation near the point $P_0=\omega_B$, where
Eq.12 for Green's function is valid:
$$-\frac{i}{(2\pi)^4} \frac{\Phi (p;P_B)
\bar \Phi (q;P_B)} {2\omega_B (P_0-\omega_B+i\epsilon)^2} = -
\int dq^\prime \; dq^{\prime \prime} \frac{i}{(2\pi)^4} \frac{\Phi (p;P_B)
\bar \Phi (q^\prime;P_B)} {2\omega_B (P_0-\omega_B+i\epsilon)} \times $$
$$\left [ \frac{\partial}{\partial P_0} (K-I) \right ] (q^\prime,q^
{\prime \prime}) \frac{i}{(2\pi)^4} \frac{\Phi (q^{\prime \prime};P_B)
\bar \Phi (q;P_B)} {2\omega_B (P_0-\omega_B+i\epsilon)} \; ,$$ \noindent
which gives the following normalization condition
$$\frac{i}{(2\pi)^4}\int dp \; dq \bar \Phi (p;P_B)
\left [ \frac{\partial}{\partial P_0} (K-I) \right ] _{P_0=\omega_B}(p,q)
\Phi (q;P_B)=2\omega_B \equiv 2(P_B)_0  $$ \noindent
or symbolically
\begin{eqnarray}
\frac{i}{(2\pi)^4} \bar \Phi
\left ( \frac{\partial K}{\partial P_0} -\frac{\partial I}{\partial P_0}
\right )_{P_0=\omega_B} \Phi =2\omega_B \; .
\label{eq14} \end{eqnarray}

\section*{B-S equation in the ladder approximation}
Let us consider the B-S equation in the ladder approximation for two
scalar particles interacting via scalar quantum exchange. The interaction
kernel in this case can be calculated according to standard Feynman
rules: every vertex contributes $ig$ and every scalar propagator
$$\Delta(X)=(2\pi)^{-4}\int \frac{e^{-ipx}~dp}{i[\mu^2-p^2-i\epsilon]}
\; .$$ \noindent
Thus
[A$$I(x,y;X-Y)=-g_1g_2~\Delta(x)~\delta(x_1-x_2)~\delta(x_2-y_2)=$$
$$-g_1g_2~\Delta(x)~\delta(X-Y+\eta_2(x-y))~\delta(X-Y-\eta_1(x-y)) \;
. $$ \noindent
Let us note that
$$\delta(X-Y+\eta_2(x-y))~\delta(X-Y-\eta_1(x-y))=$$
$$\delta[\eta_1(x-y)+\eta_2(x-y)]~\delta[X-Y-\eta_1(x-y)]=
\delta(x-y)~\delta(X-Y) \; , $$ \noindent
that is
$$I(x,y;X-Y)=-g_1g_2\Delta(x)~\delta(x-y)~\delta(X-Y) \; ,$$ \noindent
its momentum space image being
$$I(p,q;P)=(2\pi)^{-4}\int e^{ipx} e^{-iqy} e^{-iPX}I(x,y;X)~dx~dy~dX=$$
$$-g_1g_2(2\pi)^{-4}\int dx~e^{i(p-q)x} \Delta (x) \; ,$$ \noindent
or
\begin{eqnarray}
I(p,q;P)=-g_1g_2(2\pi)^{-4}\Delta(p-q)=
\frac{ig_1g_2}{(2\pi)^4}\frac{1}{\mu^2-(p-q)^2-i\epsilon} \; .
\label{eq15} \end{eqnarray}
Let us substitute this into (13) and also let us take $\vec{P_B}=0$.
As a result, we get the rest frame B-S equation in the ladder approximation
\begin{eqnarray} &&
[m_1^2+{\vec{p}}\,^2-(\eta_1P_0+p_0)^2]
[m_2^2+{\vec{p}}\,^2-(\eta_2P_0-p_0)^2] \Phi (p;P_0)= \nonumber \\ &&
\frac{\lambda}{i\pi^2} \int dq \frac{\Phi (q;P_0)}
{\mu^2-(p-q)^2-i\epsilon} \; , \; \; \lambda=\frac{g_1g_2}{16\pi^2}
\; . \label{eq16} \end{eqnarray}
This integral equation has a singular kernel and so standard mathematical
tools are inapplicable to it. But the singularity in
$$\frac{1}{\mu^2-(p_0-q_0)^2+(\vec{p}-\vec{q})^2-i\epsilon}$$ \noindent
disappears if we suppose that $p_0$ and $q_0$ are pure imaginary. The
procedure of transition to such $p_0,q_0$ is called "Wick rotation" and
it assumes two things: the analytic continuation of $\Phi(p;P_B)$ in
the complex $p_0$-plane and a rotation of the $dq_0$ integration contour from
real to imaginary axis. During this  rotation the integration contour,
of course, should not hook on singularities of the integrand. So it is
needed to study the analytical properties of the $\Phi(p;P_B)$ amplitude.

\section*{Analytical properties of the B-S amplitude}
Let us consider the reduced B-S amplitude
\begin{eqnarray} &&
(2\pi)^{-3/2}\Phi(x;P)=<0|T\phi_1(\eta_2x)\phi_2(-\eta_1x)|B>= \nonumber 
\\ && \Theta(x_0)f(x;P)+\Theta(-x_0)g(x;P)  ,
\nonumber \end{eqnarray} \noindent
where
$$f(x;P)=<0|T\phi_1(\eta_2x)\phi_2(-\eta_1x)|B>$$
\noindent and
$$g(x;P)=<0|T\phi_2(-\eta_1x)\phi_1(\eta_2x)|B>  .$$
Each of them can be transformed by using a complete set of states
$$1=\sum |n><n| \equiv \sum_\alpha \int dp |p,\alpha><p,\alpha| $$
\noindent ($\alpha$ represents discrete quantum numbers of the state 
$|n>$ and $p$ is its ~4-momentum). Taking into account identities
$$\phi_1(\eta_2x)=e^{i\eta_2x\hat P }\phi_1(0)e^{-i\eta_2x\hat P } \; 
{\rm and}  \; 
\phi_2(-\eta_1x)=e^{-i\eta_1x\hat P }\phi_2(0)e^{i\eta_1x\hat P } \; ,$$
\noindent we get 
\begin{eqnarray} &&
f(x;P)=\sum_\alpha \int dp<0|\phi_1(\eta_2x)|p,\alpha><|p,\alpha|
\phi_2(-\eta_1x)|B>= \nonumber \\ &&
\sum_\alpha \int dp e^{-i(p-\eta_1 P)x}
<0|\phi_1(0)|p,\alpha><|p,\alpha|\phi_2(0)|B> \; , \nonumber \\ &&
g(x;P)=\sum_\alpha \int dp<0|\phi_2(-\eta_1x)|p,\alpha><|p,\alpha|
\phi_1(\eta_2x)|B>= \nonumber \\ &&
\sum_\alpha \int dp e^{-i(\eta_2 P-p)x}
<0|\phi_2(0)|p,\alpha><|p,\alpha|\phi_1(0)|B> \; , 
\nonumber \end{eqnarray} 
\noindent or
\begin{eqnarray}
f(x;P)= \int dp e^{-i(p-\eta_1 P)x} f(p;P) \; , \;               
g(x;P)= \int dp e^{-i(\eta_2 P-p)x} g(p;P) ,              
\nonumber \end{eqnarray}     
\noindent where we have designated
\begin{eqnarray} &&
f(p;P)= \sum_\alpha                
<0|\phi_1(0)|p,\alpha><|p,\alpha|\phi_2(0)|B> \; , \nonumber \\ &&
g(p;P)=
\sum_\alpha                
<0|\phi_2(0)|p,\alpha><|p,\alpha|\phi_1(0)|B> \; .    
\nonumber \end{eqnarray}

The matrix element $<0|\phi_1(0)|p,\alpha>$ is different from zero
only then $|p,\alpha>$ has the same quantum numbers as the $\phi_1$
field quantum (otherwise $\phi_1(0)|p,\alpha>$ will not have the 
vacuum quantum numbers). But among states with quantum numbers of
the first particle just this particle should have the smallest 
invariant mass, unless it will be not stable. So $f(p;P) \neq 0$
only then $p_0 \ge \sqrt{m_1^2+{\vec {p}}\;^2}$. Therefore
\begin{eqnarray} &&
f(x;P)= \int dp e^{-i(p-\eta_1 P)x} f(p;P) \Theta (p_0-\sqrt{m_1^2+
{\vec {p}}\;^2})= \nonumber \\ &&               
\int dq e^{-iqx} f(\eta_1 P+q;P) \Theta \left( q_0+\eta_1 P_0-\sqrt{m_1^2+
(\vec {q}+\eta_1 \vec{P})^2}\right) \; .
\nonumber \end{eqnarray}     
\noindent This last equation, if we designate
$$\tilde{f}(q;P)= f(\eta_1 P+q;P) \; , \; \omega_+=\sqrt{m_1^2+
(\vec {q}+\eta_1 \vec{P})^2} -\eta_1P_0 \; , $$
\noindent can be rewritten as
$$f(x;P)=\int d\vec{q} \int_{\omega_+}^\infty dq_0 e^{-iqx} 
\tilde{f}(q;P) .$$
\noindent Let us note
$$\omega_+ \ge m_1-\eta_1P_0=\frac{m_1}{m_1+m_2}(m_1+m_2-P_0)$$
\noindent In the rest frame $P_0$ equals to the bound state mass,
therefore $m_1+m_2>P_0$. That is $\omega_+>0$ and only positive
frequencies contribute in $f(x;P)$.

Analogously, $<0|\phi_2(0)|p,\alpha> \neq 0$ only then $|p,\alpha>$
has quantum numbers of the second particle and stability of this
particle means $g(p;P) \equiv g(p;P) \Theta (p_0-\sqrt{m_2^2+
{\vec {p}}\;^2})$. So
\begin{eqnarray} &&
g(x;P)= \int dp e^{-i(\eta_2 P-p)x} g(p;P) \Theta (p_0-\sqrt{m_2^2+
{\vec {p}}\;^2})= \nonumber \\ &&               
\int dq e^{-iqx} g(\eta_2 P-q;P) \Theta \left(\eta_2 P_0-q_0-\sqrt{m_2^2+
(\eta_2 \vec{P}-\vec {q})^2}\right) \; .
\nonumber \end{eqnarray}     
\noindent If we designate $\tilde{g}(q;P)= g(\eta_2 P-q;P)  , 
\; \omega_-=\eta_2P_0- \sqrt{m_2^2+
(\eta_2 \vec{P}-\vec {q})^2}  , $, then
$$g(x;P)=\int d\vec{q} \int^{\omega_-}_{-\infty} dq_0 e^{-iqx} 
\tilde{g}(q;P) .$$
$\omega_- \le \eta_2P_0-m_2=\frac{m_2}{m_1+m_2}(P_0-m_1-m_2)<0$,
so only negative frequencies contribute in $g(x;P)$ (in the bound
system rest frame). 

As a result we get
\begin{eqnarray} &&
\Phi(x;P)=(2\pi)^{3/2}\Theta(x_0)\int d\vec{q} \int_{\omega_+(\vec{q};P)}
^\infty dq_0 e^{-iqx} \tilde{f}(q;P)+ \nonumber \\ &&
(2\pi)^{3/2}\Theta(-x_0)\int d\vec{q} \int^{\omega_-(\vec{q};P)}_{-\infty} 
dq_0 e^{-iqx} \tilde{g}(q;P). \label{eq17} \end{eqnarray}
In the $\vec{P}=0$ rest frame $\omega_+>0, \;\omega_-<0$, Therefore Eq.17
shows that from the $x_0>0$ positive half-line $\Phi(x;P)$ can be
analytically continued in the bottom half-plane (just then we will have
falling exponent), and from the $x_0<0$ negative half-line -- in the upper
half-plane.

\section*{Analytical properties of the B-S amplitude \\
in the momentum space}
To deal with Eq.16, we need analytical properties of the $\Phi(p,P_0)$
amplitude. Let us consider therefore the momentum space amplitude
$\Phi(p;P)=\int dx e^{ipx}\Phi(x;P)$. Because of Eq.17 we have
$$ \Phi(p;P)=$$ $$(2\pi)^{3/2}\int d\vec{q} \int_{\omega_+(\vec{q})}
^\infty dq_0 \int d\vec{x} e^{-i(\vec{p}-\vec{q})\cdot \vec{x}}
\int_0^\infty dx_0 e^{i(p_0-q_0)x_0}\tilde{f}(q;P)+$$
$$(2\pi)^{3/2}\int d\vec{q} \int_{-\infty}^{\omega_-(\vec{q})}
dq_0 \int d\vec{x} e^{-i(\vec{p}-\vec{q})\cdot \vec{x}}
\int_{-\infty}^0 dx_0 e^{i(p_0-q_0)x_0}\tilde{g}(q;P) .$$
Integrals over $d\vec{x},\; d\vec{q} \; {\rm and} \; dx_0$ can be calculated,
which gives
\begin{eqnarray}
\Phi (p;P) &=& \nonumber \\
i(2\pi)^{9/2}\int_{\omega_+(\vec{p})}^\infty dq_0
\frac{\tilde{f}(q_0,\vec{p};P)}{p_0-q_0+i\epsilon} &-&
i(2\pi)^{9/2}\int^{\omega_-(\vec{p})}_{-\infty} dq_0
\frac{\tilde{g}(q_0,\vec{p};P)}{p_0-q_0-i\epsilon} .
\label{eq18} \end{eqnarray}
While integrating over $dx_0$ the following definition of the integrals
was applied
$$\int_0^\infty dx_0 e^{i(p_0-q_0)x_0} \equiv 
\int_0^\infty dx_0 e^{i(p_0-q_0+i\epsilon)x_0} \; , $$
$$\int^0_{-\infty} dx_0 e^{i(p_0-q_0)x_0} \equiv 
\int^0_{-\infty} dx_0 e^{i(p_0-q_0-i\epsilon)x_0} \; . $$
If now we try to continue $\Phi(p;P)$ analytically in the complex
$p_0$-plane, everything will be O.K. except $p_0-q_0 \pm i\epsilon$
case. To avoid these points, cuts should be assumed in the $p_0$-plane.
In the rest frame we will have the following picture: 
\begin{figure}[htb]
\begin{center}
\epsfig{figure=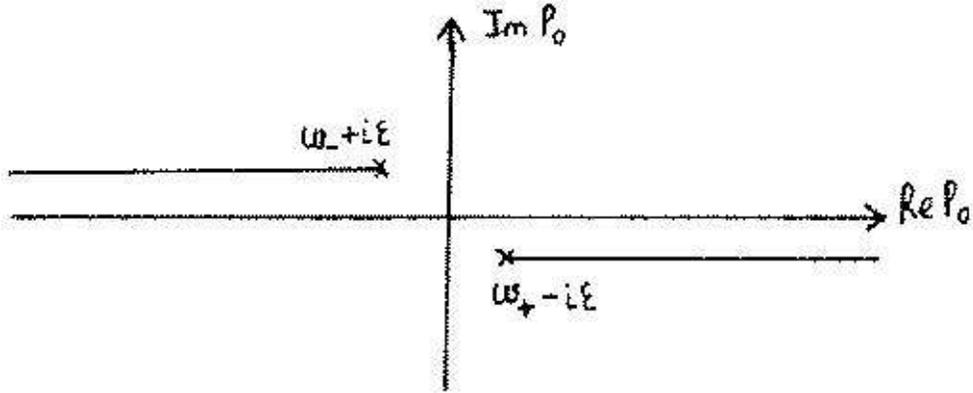,width=13cm}
\caption{Cuts in the $p_0$-plane.}
\end{center}
\label{Fig5}
\end{figure}

In the remaining $p_0$-plane  $\Phi(p;P)$ will be an analytical function.

\section*{Wick rotation}
In the r.h.s. of Eq.16 we have the following integral over $dq_0$
$$\int\limits_{-\infty}^\infty dq_0 \frac{\Phi (q;P_0)}
{\mu^2-(p_0-q_0)^2+(\vec{p}-\vec{q})^2-i\epsilon} \; . $$
\noindent Its integrand has singularities shown in Fig.6.
\begin{figure}[htb]
\begin{center}
\epsfig{figure=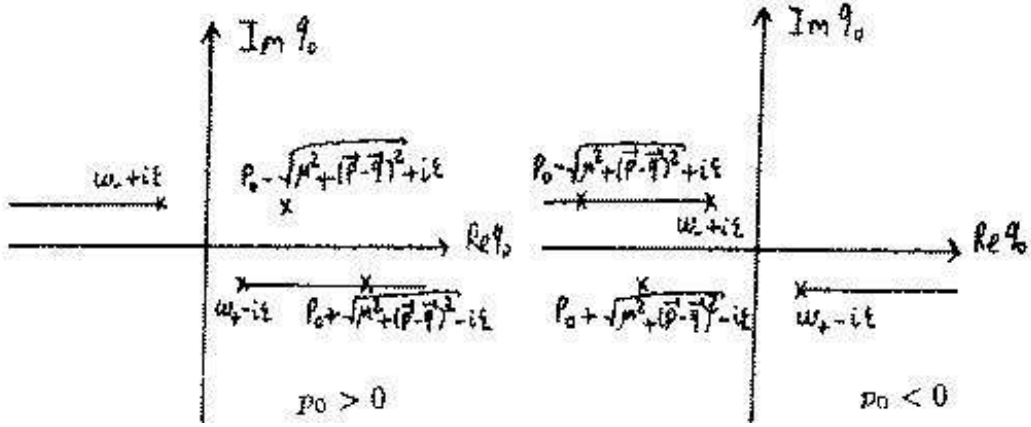,width=14cm}
\caption{Integrand singularities.}
\end{center}
\label{Fig6}
\end{figure}

Therefore 
\begin{eqnarray}
\int \limits_C dq_0 \frac{\Phi (q;P_0)}
{\mu^2-(p_0-q_0)^2+(\vec{p}-\vec{q})^2-i\epsilon} = 0 \; ,
\label{eq19} \end{eqnarray} \noindent
where the integration contour is shown in Fig.7.
\begin{figure}[htb]
\begin{center}
\epsfig{figure=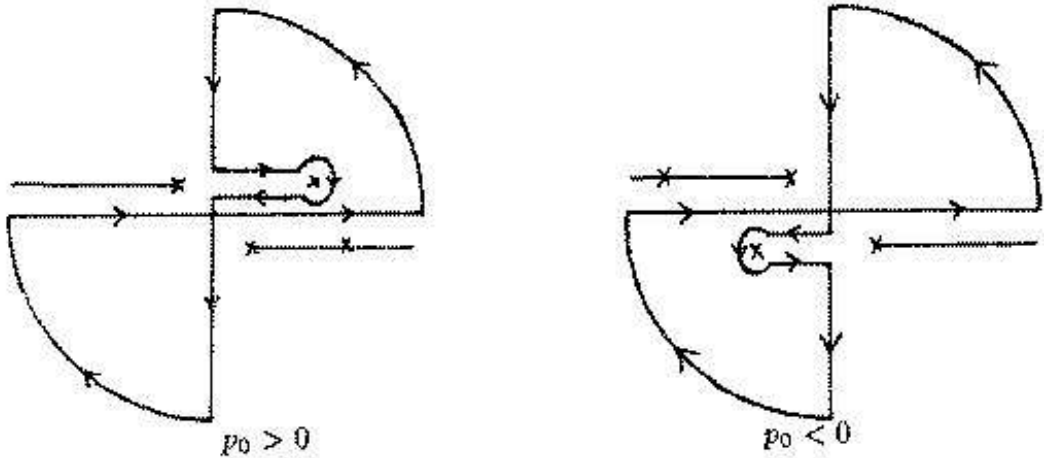,width=14cm}
\caption{Integration contour.}
\end{center}
\label{Fig7}
\end{figure}

The contributions from the two infinite quarter circles tends to zero.
So Eq.19 means that the $dq_0$-integral over the real axis can be
replaced by that over the $C_1$ contour, which follows the imaginary
axis and goes around a $p_0-\sqrt{\mu^2+(\vec{p}-\vec{q})^2}
+i\epsilon$ pole (if $p_0>0$), or  $p_0+\sqrt{\mu^2+(\vec{p}-\vec{q})^2}
-i\epsilon$ pole (if $p_0<0$).

\noindent Therefore
\begin{eqnarray} &&
[m_1^2+{\vec{p}}\,^2-(\eta_1P_0+p_0)^2]
[m_2^2+{\vec{p}}\,^2-(\eta_2P_0-p_0)^2] \Phi (p;P_0)= \nonumber \\ &&
\frac{\lambda}{i\pi^2} \int d\vec{q} \int \limits_{C_1} dq_0
\frac{\Phi (q;P_0)}{\mu^2-(p_0-q_0)^2+(\vec{p}-\vec{q})^2-i\epsilon} 
\; . \label{eq20} \end{eqnarray} \noindent
If now we begin to rotate $p_0$ counterclockwise, both sides of Eq.20
remain well defined. When we reach the imaginary axis $p_0=ip_4$, the
disposition of the integrand singularities with regard to the $C_1$
contour will be such:
\begin{figure}[htb]
\begin{center}
\epsfig{figure=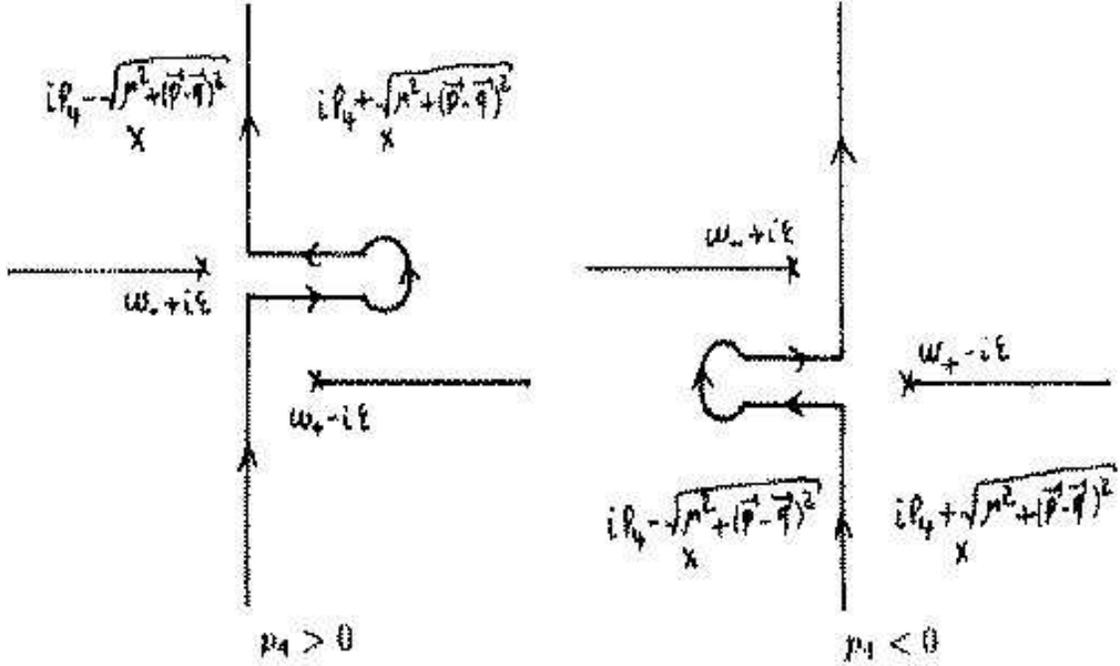,width=15cm}
\caption{Disposition of the integrand singularities.}
\end{center}
\label{Fig8}
\end{figure}

As we see, the dangerous pole has left the $C_1$ contour loop, so this 
contour can be straightened and it will coincide the imaginary axis. 
Therefore Eq.20 becomes:
\begin{eqnarray} &&
[m_1^2+{\vec{p}}\,^2-(\eta_1P_0+ip_4)^2]
[m_2^2+{\vec{p}}\,^2-(\eta_2P_0-ip_4)^2] \Phi (ip_4,\vec{p};P_0)= \nonumber 
\\ &&
\frac{\lambda}{i\pi^2} \int d\vec{q} \int \limits_{-i\infty}^{i\infty} dq_0
 \frac{\Phi (q;P_0)}
{\mu^2-(ip_4-q_0)^2+(\vec{p}-\vec{q})^2-i\epsilon} 
\; . \nonumber \end{eqnarray} \noindent
Let us make $q_0 \to iq_4$ variable change in the integral over $dq_0$:
\begin{eqnarray} &&
[m_1^2+{\vec{p}}\,^2+(p_4-i\eta_1P_0)^2]
[m_2^2+{\vec{p}}\,^2+(p_4+i\eta_2P_0)^2] \Phi (ip_4,\vec{p};P_0)= \nonumber
\\ &&
\frac{\lambda}{\pi^2} \int d\vec{q} \int \limits_{-\infty}^{\infty} dq_4
 \frac{\Phi (iq_4,\vec{q};P_0)}
{\mu^2+(p_4-q_4)^2+(\vec{p}-\vec{q})^2}
\; , \nonumber \end{eqnarray} \noindent
If now introduce Euclidean 4-vectors $\tilde{p}=(\vec{p},p_4), \; \tilde{q}=
(\vec{q},q_4)$ and an amplitude $\tilde \Phi(\tilde{p};P_0)=\Phi
(ip_4,\vec{p};P_0)$ (which is defined by the analytical continuation of 
the B-S amplitude), then the above equation can be rewritten as
\begin{eqnarray} &&
[m_1^2+{\vec{p}}\,^2+(p_4-i\eta_1P_0)^2]     
[m_2^2+{\vec{p}}\,^2+(p_4+i\eta_2P_0)^2] \tilde \Phi (\tilde{p};P_0)= \nonumber     
\\ &&
\frac{\lambda}{\pi^2} \int d\tilde{q}   
 \frac{\tilde \Phi (\tilde{q};P_0)}
{\mu^2+(\tilde{p}-\tilde{q})^2}
\; . \label{eq21} \end{eqnarray} \noindent
This is the Wick rotated, rest frame, ladder Bethe-Salpeter equation. 

\section*{Salpeter equation}
One of the main peculiarities, which distinguishes the B-S equation
from the Schr\"{o}dinger equation, is the presence of the relative
time $x_0$ (in the momentum space the corresponding quantity is 
the relative energy $p_0$). It is clear that one of the physical
sources for its appearance is the retardation of the interaction. If
there are no other physical grounds behind the relative time, its
effect should disappear in the instantaneous approximation, when
the retardation of the interaction is neglected. Let us check this
for an example of two interacting spinorial particles, for which
the B-S equation looks like
\begin{eqnarray} &&
- \left [ m_1-\gamma_{(1)}^\mu (\eta_1 P_\mu+p_\mu) \right ]
\left [ m_2-\gamma_{(2)}^\mu (\eta_2 P_\mu-p_\mu) \right ] 
\Psi(p;P) = \nonumber \\ &&
\int dq I(p,q;P) \Psi(q;P) \; . \label{eq22} \end{eqnarray}
\noindent in the l.h.s. instead of full propagators we have taken
free ones, that is $\Delta (p) =\frac{1}{i(m-\hat p)}$. Besides
$\Psi(p;P)=\int dx e^{ipx} \Psi (x;P)$, and $\Psi (x;P)$ is 
a 16-component reduced B-S amplitude
$$\Psi_{\alpha \beta}(x;P)=(2\pi)^{3/2}
<0|T\psi_{1\alpha}(\eta_2x)\psi_{2\beta}(-\eta_1x)|B> \; . $$
\noindent $\gamma_{(1)}$ matrices act on the first spinor index and
$\gamma_{(2)}$ matrices on the second one.

The instantaneous approximation means that $I(p,q;P) \equiv I(\vec{p},
\vec{q};P)$ does not depend on relative energy. Indeed, if this is the
case, then
$$I(x,y;X)=(2\pi)^{-8}\int dp~dq~dP e^{-ipx} e^{iqy} e^{-iPX} 
I(\vec{p},\vec{q};P) = $$
$$(2\pi)^{-6}\delta(x_0)\delta(y_0)\int d\vec{p}~d\vec{q}~dP
e^{i\vec{p}\cdot\vec{x}} e^{-i\vec{q}\cdot\vec{y}}
e^{-iPX}I(\vec{p},\vec{q};P) \; . $$ 
\noindent Let us multiply both sides of Eq.22 over $\gamma^0_{(1)}
\gamma^0_{(2)}$ and designate $\tilde I=-\gamma^0_{(1)} \gamma^0_{(2)}
I$, then the equation in the rest frame $P_\mu=(E,\vec{0})$ becomes:
\begin{eqnarray} &&
[H_1(\vec{p})-\eta_1E-p_0][H_2(-\vec{p})-\eta_2E+p_0]\Psi(p;P)=
\nonumber \\ &&
\int dq \tilde I(\vec{p},\vec{q};P) \Psi(q;P) \; , \label{eq23}
\end{eqnarray}
\noindent where $H(\vec{p})=\vec {\alpha} \cdot \vec{p}+\beta m$ is
a conventional Dirac Hamiltonian. The right hand side of this equation
can be rewritten as
$$\int dq \tilde I(\vec{p},\vec{q};P) \Psi(q;P)=
\int d\vec{q} \tilde I(\vec{p},\vec{q};P) \Phi(\vec{q};P) \; ,$$
\noindent where
$$\Phi(\vec{q};P)=\int \limits_{-\infty}^{\infty} dq_0 \Psi(q;P)=
\int dq_0~dx e^{iqx}\Psi(x;P)=$$
$$2\pi\int dx \delta(x_0)e^{-i\vec{q}\cdot\vec{x}}\Psi(x_0,\vec{x};P)=
2\pi\int d\vec{x} e^{-i\vec{q}\cdot\vec{x}} \Psi(0,\vec{x};P) .$$
\noindent So $\Phi(\vec{q};P)$ is determined by a simultaneous B-S 
amplitude. 

To express the l.h.s. of the Eq.23 also in terms of $\Phi(\vec{q};P)$, 
the following trick can be used. The projection operators
$$\Lambda_{\pm}(\vec{p})=\frac{1}{2}\left ( 1 \pm \frac{H(\vec{p})}
{\sqrt{m^2+\vec{p}\;^2 }} \right )  $$
have a property $\Lambda_\pm (\vec{p})H(\vec{p})=\pm \sqrt{m^2+\vec{p}
\; ^2 }\Lambda_\pm (\vec{p})$. using this, Eq.23 can be replaced by
a system
\begin{eqnarray} && \hspace*{-2mm}
\left [ \sqrt{m^2_1+\vec{p}\;^2 }-\eta_1E-p_0-i\epsilon \right ]
\left [ \sqrt{m^2_2+\vec{p}\;^2 }-\eta_2E+p_0-i\epsilon \right ]
\Psi_{++}(p;P)= \nonumber \\ && \hspace*{-2mm}
\Lambda_+^{(1)}(\vec{p})\Lambda_+^{(2)}(-\vec{p})
\int d\vec{q} \tilde I(\vec{p},\vec{q};P) \Phi(\vec{q};P) \; ,
\nonumber \\ && \hspace*{-2mm}
\left [ \sqrt{m^2_1+\vec{p}\;^2 }-\eta_1E-p_0-i\epsilon \right ]
\left [ -\sqrt{m^2_2+\vec{p}\;^2 }-\eta_2E+p_0+i\epsilon \right ]
\Psi_{+-}(p;P)= \nonumber \\ && \hspace*{-2mm}
\Lambda_+^{(1)}(\vec{p})\Lambda_-^{(2)}(-\vec{p})
\int d\vec{q} \tilde I(\vec{p},\vec{q};P) \Phi(\vec{q};P) \; ,
\nonumber \\ && \hspace*{-2mm}
\left [- \sqrt{m^2_1+\vec{p}\;^2 }-\eta_1E-p_0+i\epsilon \right ]
\left [ \sqrt{m^2_2+\vec{p}\;^2 }-\eta_2E+p_0-i\epsilon \right ]
\Psi_{-+}(p;P)= \nonumber \\ && \hspace*{-2mm}
\Lambda_-^{(1)}(\vec{p})\Lambda_+^{(2)}(-\vec{p})
\int d\vec{q} \tilde I(\vec{p},\vec{q};P) \Phi(\vec{q};P) \; ,
\nonumber \\ &&  \hspace*{-2mm}
\left [- \sqrt{m^2_1+\vec{p}\;^2 }-\eta_1E-p_0+i\epsilon \right ]
\left [- \sqrt{m^2_2+\vec{p}\;^2 }-\eta_2E+p_0+i\epsilon \right ] 
\Psi_{--}(p;P)= \nonumber \\ && \hspace*{-2mm}
\Lambda_-^{(1)}(\vec{p})\Lambda_-^{(2)}(-\vec{p})
\int d\vec{q} \tilde I(\vec{p},\vec{q};P) \Phi(\vec{q};P) \; .
\nonumber \end{eqnarray} \vspace*{-21mm} 
\begin{flushright} ($23^\prime$) \end{flushright}

\noindent Here
$\Psi_{++}(p;P)=\Lambda_+^{(1)}(\vec{p})\Lambda_+^{(2)}(-\vec{p})\Psi(p;P)$
 and so on. Note the  substitution $ m \to m-i\epsilon$
in the Feynman propagators (positive frequencies propagate forward in
time, while negative frequencies -- backward).

By means of residue theory we get
$$\int \limits_{-\infty}^{\infty} dp_0
\left [ \sqrt{m^2_1+\vec{p}\;^2 }-\eta_1E-p_0-i\epsilon \right ]^{-1}
\left [ \sqrt{m^2_2+\vec{p}\;^2 }-\eta_2E+p_0-i\epsilon \right ]^{-1}=$$
$$-2\pi i\left [ E- \sqrt{m^2_1+\vec{p}\;^2 }-\sqrt{m^2_2+\vec{p}\;^2 }
\right ]^{-1} \; , $$
$$\int \limits_{-\infty}^{\infty} dp_0
\left [-\sqrt{m^2_1+\vec{p}\;^2 }-\eta_1E-p_0+i\epsilon \right ]^{-1}
\left [-\sqrt{m^2_2+\vec{p}\;^2 }-\eta_2E+p_0+i\epsilon \right ]^{-1}=$$
$$2\pi i\left [ E+ \sqrt{m^2_1+\vec{p}\;^2 }+\sqrt{m^2_2+\vec{p}\;^2 }
\right ]^{-1} \; , $$
$$\int \limits_{-\infty}^{\infty} dp_0
\left [\sqrt{m^2_1+\vec{p}\;^2 }-\eta_1E-p_0-i\epsilon \right ]^{-1} 
\left [-\sqrt{m^2_2+\vec{p}\;^2 }-\eta_2E+p_0+i\epsilon \right ]^{-1}=$$ 
$$\int \limits_{-\infty}^{\infty} dp_0
\left [-\sqrt{m^2_1+\vec{p}\;^2 }-\eta_1E-p_0+i\epsilon \right ]^{-1} 
\left [\sqrt{m^2_2+\vec{p}\;^2 }-\eta_2E+p_0-i\epsilon \right ]^{-1}=
0.$$ 
Therefore for the $\Phi(\vec{q};P)$ amplitude the following system holds
$$\left [ E- \sqrt{m^2_1+\vec{p}\;^2 }-\sqrt{m^2_2+\vec{p}\;^2 }\right ]
\Phi_{++}(\vec{p};P) \equiv$$
$$\Lambda_+^{(1)}(\vec{p})\Lambda_+^{(2)}(-\vec{p})
[E-H_1(\vec{p})-H_2(-\vec{p})]\Phi(\vec{p};P)=$$
$$\Lambda_+^{(1)}(\vec{p})\Lambda_+^{(2)}(-\vec{p})
\frac{2\pi}{i}\int d\vec{q} \tilde I(\vec{p},\vec{q};P) \Phi(\vec{q};P),$$
$$\left [ E+\sqrt{m^2_1+\vec{p}\;^2 }+\sqrt{m^2_2+\vec{p}\;^2 }\right ]
\Phi_{--}(\vec{p};P) \equiv$$
$$\Lambda_-^{(1)}(\vec{p})\Lambda_-^{(2)}(-\vec{p})
[E-H_1(\vec{p})-H_2(-\vec{p})]\Phi(\vec{p};P)=$$
$$-\Lambda_-^{(1)}(\vec{p})\Lambda_-^{(2)}(-\vec{p})
\frac{2\pi}{i}\int d\vec{q} \tilde I(\vec{p},\vec{q};P) \Phi(\vec{q};P),$$
$$\Phi_{+-}(\vec{p};P)=\Phi_{-+}(\vec{p};P)=0 \; . $$
\noindent To rewrite the last two equations in the same form, as the first
ones, note that
$$ E- \sqrt{m^2_1+\vec{p}\;^2 }+\sqrt{m^2_2+\vec{p}\;^2 }>0 \; , \;
 E+\sqrt{m^2_1+\vec{p}\;^2 }-\sqrt{m^2_2+\vec{p}\;^2 }>0 \; . $$ 
\noindent Indeed, if $m_1=m_2$, these inequalities are obvious. Let 
$m_1>m_2$, then only
$$ E- \sqrt{m^2_1+\vec{p}\;^2 }+\sqrt{m^2_2+\vec{p}\;^2 }>0$$ \noindent
inequality needs a proof. But
$$ \sqrt{m^2_1+\vec{p}\;^2 }-\sqrt{m^2_2+\vec{p}\;^2} =
\frac{m_1^2-m_2^2}{\sqrt{m^2_1+\vec{p}\;^2 }+\sqrt{m^2_2+\vec{p}\;^2}} \le
\frac{m_1^2-m_2^2}{m_1+m_2}=m_1-m_2,$$ 
\noindent and $E>m_1-m_2$ is a stability condition for the first particle.
Otherwise the first particle decay into the second antiparticle and $|B>$
bound state will be energetically permitted.

Because of the above mentioned inequalities, $\Phi_{+-}(\vec{p};P)=
\Phi_{-+}(\vec{p};P)=0$ equations are equivalent to 
$$0= \left [ E- \sqrt{m^2_1+\vec{p}\;^2 }+\sqrt{m^2_2+\vec{p}\;^2 }\right ]
\Phi_{+-}(\vec{p};P) \equiv$$
$$\Lambda_+^{(1)}(\vec{p})\Lambda_-^{(2)}(-\vec{p})
[E-H_1(\vec{p})-H_2(-\vec{p})]\Phi(\vec{p};P) \; , $$
$$0= \left [ E+ \sqrt{m^2_1+\vec{p}\;^2 }-\sqrt{m^2_2+\vec{p}\;^2 }\right ]
\Phi_{-+}(\vec{p};P) \equiv$$
$$\Lambda_-^{(1)}(\vec{p})\Lambda_+^{(2)}(-\vec{p})
[E-H_1(\vec{p})-H_2(-\vec{p})]\Phi(\vec{p};P) \; . $$
\noindent If we sum all four equations for  $\Phi(\vec{p};P)$ and use
$$\Lambda_+^{(1)}(\vec{p})\Lambda_+^{(2)}(-\vec{p})+
\Lambda_+^{(1)}(\vec{p})\Lambda_-^{(2)}(-\vec{p})+
\Lambda_-^{(1)}(\vec{p})\Lambda_+^{(2)}(-\vec{p})+
\Lambda_-^{(1)}(\vec{p})\Lambda_-^{(2)}(-\vec{p})=1,$$
\noindent then we get the Salpeter equation
\begin{eqnarray} &&
[E-H_1(\vec{p})-H_2(-\vec{p})]\Phi(\vec{p};E)= \nonumber \\ &&
[\Lambda_+^{(1)}(\vec{p})\Lambda_+^{(2)}(-\vec{p})-
\Lambda_-^{(1)}(\vec{p})\Lambda_-^{(2)}(-\vec{p})]
\frac{2\pi}{i}\int d\vec{q} \tilde I(\vec{p},\vec{q};P) \Phi(\vec{q};E) .
\label{eq24} \end{eqnarray}
\noindent So the relative energy is indeed excluded from the equation,
but with the price of $\Lambda_{++}-\Lambda_{--}$ operator introduction.
To understand why relative energy has left such a trace, it is useful 
to compare with the nonrelativistic case.

\section*{Bethe-Salpeter equation in the nonrelativistic theory}
To derive the nonrelativistic Bethe-Salpeter equation, one can use the fact
that quantum field theory in many respects is similar to a second
quantized many particle theory. In graphical representation, to the second
quantized Hamiltonian
$$H=\int d\vec{x} \Psi^+(\vec{x},t)\left (-\frac{\Delta}{2m}\right )
\Psi(\vec{x},t)+$$
$$\frac{1}{2}\int d\vec{x}~d\vec{y}~\Psi^+(\vec{x},t)\Psi^+(\vec{y},t)
V(|\vec{x}-\vec{y}|)\Psi(\vec{x},t)\Psi(\vec{y},t)$$
\noindent there corresponds the free propagator
$$\Delta(x-y)=\frac{i}{(2\pi)^4}\int dp \frac{e^{-ip(x-y)}}{p_0-
(\vec{p}\,^2/2m)+i\epsilon}$$
\noindent and the pair interaction (instantaneous) with the potential $V$:
$$-i\delta(x_0-y_0)V(|\vec{x}-\vec{y}|) \; .$$
Everywhere in the derivation of the B-S equation , $\phi(x)$ field operator
can be replaced with the second quantized operator $\Psi(x)$. As a result,
we end with an equation for the following nonrelativistic Bethe-Salpeter
amplitude
$$\Phi(x;P)=(2\pi)^{3/2}<0|T\Psi(\eta_2x)\Psi(-\eta_1x)|B> \; . $$
Namely, in the ladder approximation, the interaction operator is
$$I(x_1,x_2;y_1,y_2)=-i\delta(x_0)V(|\vec{x}|)\delta(x_1-y_1)
\delta(x_2-y_2)=$$
$$-i\delta(x_0)V(|\vec{x}|)\delta(x-y)\delta(X-Y) \; , $$
\noindent with the corresponding expression in the momentum space
$$I(p,q;P)=(2\pi)^{-4}\int dx~dy~dX e^{ipx}e^{-iqy}e^{iPX}I(x,y;X)=$$
$$-i(2\pi)^{-4}\int e^{-i(\vec{p}-\vec{q})\cdot
\vec{x}}V(|\vec{x}|)~d\vec{x}=\frac{-i}{(2\pi)^4}V(\vec{p}-\vec{q})
\; . $$
Besides, the momentum space free propagator looks like $\Delta(p)=
i/(p_0-\vec{p}\,^2/2m+i\epsilon) $, therefore
the B-S equation, in the rest frame and ladder approximation, will take
the form
$$\left [ (\eta_1E+p_0)-\frac{\vec{p}\,^2}{2m}+i\epsilon \right ]
\left [ (\eta_2E-p_0)-\frac{\vec{p}\,^2}{2m}+i\epsilon \right ]
\Phi(p;E)=$$ $$i(2\pi)^{-4}\int dq~V(\vec{p}-\vec{q})\Phi(q;E)$$
\noindent (note that $m_1=m_2\equiv m $ and so $\eta_1=\eta_2=
\frac{1}{2}$).

Let us introduce integrated over $dp_0$ amplitude $\tilde \Phi (\vec{q};E)=
\int \limits_{-\infty}^\infty dq_0~\Phi(q;E)$. For it the 
Salpeter equation holds
$$\tilde \Phi (\vec{p};E)=\left [ \int \limits_{-\infty}^\infty
\frac{dp_0}{[\eta_1E+p_0-\vec{p}\,^2/2m+i\epsilon]
[\eta_2E-p_0-\vec{p}\,^2/2m+i\epsilon]}\right ] \times $$
$$\frac{i}{(2\pi)^4}\int d\vec{q}~V(\vec{p}-\vec{q})\tilde \Phi (\vec{q};E)
\; . $$
The square bracket integral can be evaluated via residue theory and it
equals $-2\pi i/(E-\vec{p}\,^2/2m)$. So we obtain the following equation
\begin{eqnarray}
\left ( E-\frac{\vec{p}\,^2}{2m} \right ) \tilde \Phi (\vec{p};E)=
(2\pi)^{-3}\int d\vec{q}~V(\vec{p}-\vec{q})\tilde \Phi (\vec{q};E) \; .
\label{eq25} \end{eqnarray}

But this is just momentum space Schr\"{o}dinger equation! Indeed, in
configuration space 
$$\tilde \Phi(\vec{p})=\int e^{-i\vec{p}\cdot\vec{x}}\tilde \Phi(\vec{x})~
d\vec{x}, \; \; \; \; V(\vec{p})=\int e^{-i\vec{p}\cdot\vec{x}}V(|\vec{x}|)
~d\vec{x} \; , $$
\noindent we will have
$$\left [-\frac{\Delta}{m}+V(|\vec{x}|)\right ]\tilde \Phi(\vec{x})=
E\tilde \Phi(\vec{x})\; .$$
Thus, in the nonrelativistic theory, the relative time can be excluded
without any trace. So its introduction is purely formal. 

\section*{Physical meaning of the relative time}
What is the crucial peculiarity, which distinguishes the above given
nonrelativistic model from the instantaneous approximation of the
relativistic one? It is the propagator! There are no antiparticles
in the nonrelativistic case. Therefore the propagator describes only 
forward propagation in time, that is we have retarded Green's function:
$\Delta(x)=0, \; {\rm if} \; x_0<0$. This boundary condition demands
$p_0\to p_0+i\epsilon$ prescription for the propagator poles. Let us see
what will be changed in the Salpeter equation derivation if we replace
the Feynman propagator $i(\hat{p}+m)/(p^2-m^2+i\epsilon)$ by the
retarded Green's function $i(\hat{p}+m)/[(p_0+i\epsilon)^2-\vec{p}\,^2-m^2]$.
Instead of ${\rm Eq}.23^\prime$ system, we will have
\begin{eqnarray} && \hspace*{-2mm}
\left [ \sqrt{m^2_1+\vec{p}\;^2 }-\eta_1E-p_0-i\epsilon \right ]
\left [ \sqrt{m^2_2+\vec{p}\;^2 }-\eta_2E+p_0-i\epsilon \right ]
\Psi_{++}(p;P)= \nonumber \\ && \hspace*{-2mm}
\Lambda_+^{(1)}(\vec{p})\Lambda_+^{(2)}(-\vec{p})
\int d\vec{q} \tilde I(\vec{p},\vec{q};P) \Phi(\vec{q};P) \; ,
\nonumber \\ && \hspace*{-2mm}
\left [ \sqrt{m^2_1+\vec{p}\;^2 }-\eta_1E-p_0-i\epsilon \right ]
\left [ -\sqrt{m^2_2+\vec{p}\;^2 }-\eta_2E+p_0-i\epsilon \right ]
\Psi_{+-}(p;P)= \nonumber \\ && \hspace*{-2mm}
\Lambda_+^{(1)}(\vec{p})\Lambda_-^{(2)}(-\vec{p})
\int d\vec{q} \tilde I(\vec{p},\vec{q};P) \Phi(\vec{q};P) \; ,
\nonumber \\ && \hspace*{-2mm}
\left [- \sqrt{m^2_1+\vec{p}\;^2 }-\eta_1E-p_0-i\epsilon \right ]
\left [ \sqrt{m^2_2+\vec{p}\;^2 }-\eta_2E+p_0-i\epsilon \right ]
\Psi_{-+}(p;P)= \nonumber \\ && \hspace*{-2mm}
\Lambda_-^{(1)}(\vec{p})\Lambda_+^{(2)}(-\vec{p})
\int d\vec{q} \tilde I(\vec{p},\vec{q};P) \Phi(\vec{q};P) \; ,
\nonumber \\ &&  \hspace*{-2mm}
\left [- \sqrt{m^2_1+\vec{p}\;^2 }-\eta_1E-p_0-i\epsilon \right ]
\left [- \sqrt{m^2_2+\vec{p}\;^2 }-\eta_2E+p_0-i\epsilon \right ]
\Psi_{--}(p;P)= \nonumber \\ && \hspace*{-2mm}                
\Lambda_-^{(1)}(\vec{p})\Lambda_-^{(2)}(-\vec{p})
\int d\vec{q} \tilde I(\vec{p},\vec{q};P) \Phi(\vec{q};P) \; .
\nonumber \end{eqnarray}
\noindent (the propagators, which appear in the B-S equation, have
$\eta_1P+p$ and $\eta_2P-p$ as their 4-momenta. Therefore the transition
to the retarded Green's function means the following replacements
$$(\eta_1P_0+p_0)\to (\eta_1P_0+p_0)+i\epsilon \; \; , \; \;
(\eta_2P_0-p_0)\to (\eta_2P_0-p_0)+i\epsilon $$
\noindent that is $i\epsilon$ has the same sign, as $E$). Integration over
$dp_0$ can be performed using
$$\int\limits_{-\infty}^\infty \frac{dp_0}{(a-p_0-i\epsilon)
(b+p_0-i\epsilon)}=2\pi i\frac{1}{a+b} \; , $$
\noindent and we get the system
$$\left [ E- \sqrt{m^2_1+\vec{p}\;^2 }-\sqrt{m^2_2+\vec{p}\;^2 }\right ]
\Phi_{++}(\vec{p};P) \equiv
\Lambda_{++}
[E-H_1(\vec{p})-H_2(-\vec{p})]\Phi(\vec{p};P)$$
$$=\Lambda_{++}
\frac{2\pi}{i}\int d\vec{q} \tilde I(\vec{p},\vec{q};P) \Phi(\vec{q};P),$$
$$\left [ E- \sqrt{m^2_1+\vec{p}\;^2 }+\sqrt{m^2_2+\vec{p}\;^2 }\right ]
\Phi_{+-}(\vec{p};P) \equiv  
\Lambda_{+-}                                        
[E-H_1(\vec{p})-H_2(-\vec{p})]\Phi(\vec{p};P)$$
$$=\Lambda_{+-}                                        
\frac{2\pi}{i}\int d\vec{q} \tilde I(\vec{p},\vec{q};P) \Phi(\vec{q};P),$$
$$\left [ E+ \sqrt{m^2_1+\vec{p}\;^2 }-\sqrt{m^2_2+\vec{p}\;^2 }\right ]
\Phi_{-+}(\vec{p};P) \equiv  
\Lambda_{-+}                                        
[E-H_1(\vec{p})-H_2(-\vec{p})]\Phi(\vec{p};P)$$
$$=\Lambda_{-+}                                        
\frac{2\pi}{i}\int d\vec{q} \tilde I(\vec{p},\vec{q};P) \Phi(\vec{q};P),$$
$$\left [ E+\sqrt{m^2_1+\vec{p}\;^2 }+\sqrt{m^2_2+\vec{p}\;^2 }\right ]
\Phi_{--}(\vec{p};P) \equiv
\Lambda_{--}
[E-H_1(\vec{p})-H_2(-\vec{p})]\Phi(\vec{p};P)$$
$$=\Lambda_{--}
\frac{2\pi}{i}\int d\vec{q} \tilde I(\vec{p},\vec{q};P) \Phi(\vec{q};P).$$
Summing these four equations, we get the Breit equation
\begin{eqnarray} 
[E-H_1(\vec{p})-H_2(-\vec{p})]\Phi(\vec{p};E)= 
\frac{2\pi}{i}\int d\vec{q} \tilde I(\vec{p},\vec{q};P) \Phi(\vec{q};E) .
\label{eq26} \end{eqnarray}

\noindent which is a direct generalization of the two particle
Schr\"{o}dinger equation Eq.25. (The nonrelativistic Hamiltonian
$H(\vec{p})=\vec{p}\,^2/2m$ is replaced by the Dirac Hamiltonian
$H(\vec{p})=\vec{\alpha}\cdot\vec{p}+\beta m $ and $-i(2\pi)^4\tilde
I(\vec{p},\vec{q}) $ plays the role of
potential). Once again, the relative time disappears without any trace left.

Therefore, the second (and more important) source for the essential
relative time dependence of the B-S amplitude is the existence of
antiparticles, that is the possibility for particles to turn back in time
and propagate backward. Because of retardation of the interaction, relative
times of the order of bound system size will be significant, while the
forward-backward motion in time makes essential configurations, for which
individual routes in time are very different for bound state constituting 
particles, and so the relative time is large. $\Lambda_{++}-\Lambda_{--}$
operator in the Salpeter equation just corresponds to the contribution
of these configurations to the bound state amplitude. 

\section*{Wick-Cutkosky model}
This model corresponds to the B-S equation in the ladder approximation
for two scalar particles interacting via massless quanta exchange. After
the Wick rotation, the corresponding rest frame B-S equation takes the 
form
\begin{eqnarray} 
[m_1^2+{\vec{p}}\,^2+(p_4-i\eta_1E)^2]     
[m_2^2+{\vec{p}}\,^2+(p_4+i\eta_2E)^2] \Phi (p)=    
\frac{\lambda}{\pi^2} \int dq   
\frac{\Phi (q)}{(p-q)^2}
\label{eq27} \end{eqnarray} \noindent
where $E$ is the bound state mass and $p,q$ are Euclidean
4-vectors.

To investigate the mathematical structure of Eq.27, let us consider
at first the simplest case $m_1=m_2=m$ and $E=0$ (although, in the rest
frame, $E=0$ is, of course, unphysical: massless bound state doesn't have
rest frame). Eq.27 then becomes
\begin{eqnarray}
(p^2+m^2)^2\Phi(p)=\frac{\lambda}{\pi^2}\int dq
\frac{\Phi (q)}{(p-q)^2} \; .
\label{eq28} \end{eqnarray} \noindent 
Let us show, that one of its solutions is $\phi(p)=(p^2+m^2)^{-3}$. We have
$$A^{-1}B^{-1}=\frac{1}{A-B}\left ( -\frac{1}{A}+\frac{1}{B} \right ) = $$
$$\int\limits_0^1\frac{dx}{[B+(A-B)x]^2}=\int\limits_0^1\frac{dx}
{[xA+(1-x)B]^2} \; , $$
\noindent besides
$$A^{-n}B^{-m}=\frac{(-1)^{n+m}}{(n-1)!(m-1)!}\frac{\partial^{n-1}}{\partial
A^{n-1}}\frac{\partial^{m-1}}{\partial B^{m-1}}(A^{-1}B^{-1})\; , $$
\noindent and we get the following Feynman parameterization
\begin{eqnarray}
A^{-n}B^{-m}=\frac{(n+m-1)!}{(n-1)!(m-1)!}\int\limits_0^1
\frac{x^{n-1}(1-x)^{m-1}}{[xA+(1-x)B]^{n+m}} dx \; .
\label{eq29} \end{eqnarray}
\noindent In particular
$$[(p-q)^2]^{-1}[q^2+m^2]^{-3}=$$
$$3\int\limits_0^1\frac{(1-x)^2 dx}{[x(p-q)^2+(1-x)(q^2+m^2)]^4}=
\int\limits_0^1\frac{3(1-x)^2 dx}{[(q-xp)^2+(1-x)(m^2+xp^2)]^4},$$
\noindent So
$$\int dq \frac{\Phi (q)}{(p-q)^2}=\int dq \int\limits_0^1\frac{3(1-x)^2
~dx}{[(q-xp)^2+(1-x)(m^2+xp^2)]^4}\; . $$
\noindent If now we use
$$A^{-4}=-\frac{1}{3!}\frac{\partial^3}{\partial A^3}\left (\frac{1}{A} \right )
=-\frac{1}{3!}\frac{\partial^3}{\partial A^3}\int\limits_0^\infty e^{-\alpha A}
d\alpha=\frac{1}{3!}\int\limits_0^\infty \alpha^3 e^{-\alpha A} d\alpha
\; , $$
\noindent then we get
$$\int dq \int\limits_0^1\frac{3(1-x)^2
~dx}{[(q-xp)^2+(1-x)(m^2+xp^2)]^4}=$$
$$\frac{1}{2}\int\limits_0^1(1-x)^2dx\int\limits_0^\infty \alpha^3 d\alpha
\int dq \exp {\{ -\alpha [(q-xp)^2+(1-x)(m^2+xp^2)]\} }. $$
\noindent The Gaussian integral over $dq$ equals
$$\int dq \exp {\{ -\alpha [(q-xp)^2+(1-x)(m^2+xp^2)]\} }=$$
$$\frac{\pi^2}{\alpha^2} \exp{\{-\alpha (1-x)(m^2+xp^2)\} }  , $$
\noindent and we are left with an integral over $d\alpha$ of the type
$$\int\limits_0^\infty \alpha e^{-\alpha A} d\alpha=-\frac{\partial}
{\partial A} \int \limits_0^\infty  e^{-\alpha A} d\alpha=A^{-2} \; .$$
\noindent As a final result, we get
$$\int dq \frac{\Phi (q)}{(p-q)^2}=\frac{\pi^2}{2}\int\limits_0^1
\frac{dx}{[m^2+xp^2]^2}=\frac{\pi^2}{2m^2}(m^2+p^2)^{-1} \; . $$
Inserting this into Eq.28, we will see that $\Phi (p)=(p^2+m^2)^{-3} $ is 
indeed a solution and the corresponding eigenvalue is $\lambda=2m^2$.

The most interesting thing about this solution is that we can indicate
its analog in the nonrelativistic hydrogen atom problem. 

The Schr\"{o}dinger equation
$$\left [-\frac{\Delta}{2m}+V(\vec{r})\right ]\Psi(\vec{r})
=E\Psi(\vec{r}) $$
\noindent in the momentum space becomes an integral equation
$$(\vec{p}\,^2-2mE)\Psi(\vec{p})=-\frac{2m}{(2\pi)^{3/2}}\int d\vec{q}~
V(\vec{p}-\vec{q})\Psi(\vec{q}) \; , $$
\noindent where $\Psi(\vec{q})=(2\pi)^{-3/2}\int e^{-i\vec{p}\cdot\vec{r}}
\Psi(\vec{r})d\vec{r} \,$
and $\, V(\vec{p})=(2\pi)^{-3/2}\int e^{-i\vec{p}\cdot\vec{r}}
V(\vec{r})d\vec{r} $. 
$\Delta(r^{-1})=-4\pi\delta(\vec{r})$ identity indicates 
that $V(\vec{r})=-e^2/r $ Coulomb potential has
$V(\vec{p})=-(4\pi/(2\pi)^{3/2})(e^2/\vec{p}\,^2)$ as its momentum space 
image. Therefore a nonrelativistic hydrogen atom is described by the equation:
\begin{eqnarray}
(\vec{p}\,^2+p_0^2)\Psi(\vec{p})=\frac{me^2}{\pi^2}\int 
\frac{\Psi(\vec{q})}{(\vec{p}-\vec{q})^2} d\vec{q} \; ,
\label{eq30} \end{eqnarray}
\noindent where $p_0^2=-2mE$ (we consider a discrete spectrum and therefore 
$E<0$).

One of the solutions of this equation is $\Psi(\vec{p})=(\vec{p}\,^2+p_0^2)
^{-2}$. Indeed, in the similar way
as above we get, after integrating over $d\vec{q}$
$$\int d\vec{q}~ [(\vec{p}-\vec{q})^2]^{-1}[\vec{p}\,^2+p_0^2]^{-2}=$$
$$\pi^{3/2}\int\limits_0^1 (1-x)~dx\int\limits_0^\infty \sqrt{\alpha}
\exp {\{ -\alpha (1-x)(p_0^2+x\vec{p}\,^2)\} }~d\alpha \; . $$
Integral over $d\alpha$ can be evaluated in such a way
$$\int\limits_0^\infty \sqrt{\alpha}e^{-\alpha A}~d\alpha=
2 \int\limits_0^\infty t^2 e^{-t^2 A}~dt=-2 \frac{\partial}{\partial A}
\int\limits_0^\infty
e^{-t^2 A}~dt=$$ 
$$ -\frac{\partial}{\partial A}\sqrt{\frac{\pi}{A}}=
\frac{1}{2A}\sqrt{\frac{\pi}{A}} \; , $$
\noindent therefore
$$\int \frac{\Psi(\vec{q})}{(\vec{p}-\vec{q})^2} d\vec{q}=
\frac{\pi^2}{2}\int\limits_0^1\frac{dx}{(p_0^2+x\vec{p}\,^2)\sqrt{
(1-x)(p_0^2+x\vec{p}\,^2)}}=$$
$$-2\pi^2\frac{\partial^2}{\partial (p_0^2)^2}\int\limits_0^1\sqrt{
\frac{p_0^2+x\vec{p}\,^2}{1-x}}~dx=-2\pi^2\frac{\partial^2}{\partial
(p_0^2)^2}\int\limits_0^1\sqrt { \frac{p_0^2+\vec{p}\,^2}{1-x}-
\vec{p}\,^2 }~dx=$$
$$-2\pi^2\frac{\partial^2}{\partial (p_0^2)^2}\int\limits_1^\infty
\frac{dt}{t^2}\sqrt{(p_0^2+\vec{p}\,^2)t-\vec{p}\,^2 }=
\frac{\pi^2}{2}\int\limits_1^\infty [(p_0^2+\vec{p}\,^2)t-\vec{p}\,^2]^
{-3/2}~dt=$$
$$\frac{\pi^2}{2(p_0^2+\vec{p}\,^2)}\int\limits_{p_0^2}^\infty x^{-3/2}~
dx=\frac{\pi^2}{p_0(p_0^2+\vec{p}\,^2)} \; . $$
Inserting in Eq.30, we see that if $me^2/p_0=1$, then the equation is
fulfilled. Thus $\psi(\vec{p})=(\vec{p}\,^2+p_0^2)^{-2}$ corresponds to
the $E=\frac{-me^4}{2}$ ground state of hydrogen atom.

If the analogy between the found solutions of Eq.28 and Eq.30 is not
accidental, one can expect that the same methods, which are used in
dealing with hydrogen atom, will be useful also for Eq.28 and maybe
even for Eq.27. In particular, it is well known that the nonrelativistic
hydrogen atom possesses a hidden symmetry and its study is more easily
performed in the Fock space, where this symmetry becomes explicit.
For the sake of simplicity, let us illustrate the Fock's method on 
an example of a 2-dimensional hydrogen atom. 
\section*{Fock's method for a 2-dimensional hydrogen atom}
The Scr\"{o}dinger equation in the 2-dimensional momentum space
looks like
$$(\vec{p}\,^2-2mE)\Psi(\vec{p})=-\frac{m}{\pi}\int d\vec{q}~
V(\vec{p}-\vec{q})\Psi(\vec{q}) \; . $$
We have the following connection with the configuration space
$$\Psi(\vec{p})=\frac{1}{2\pi} \int e^{-i\vec{p}\cdot\vec{r}}
\Psi(\vec{r})d\vec{r} \; , \; V(\vec{p})=\frac{1}{2\pi} \int
e^{-i\vec{p}\cdot\vec{r}} V(\vec{r})d\vec{r} \; . $$
For the Coulomb potential $ V(\vec{r})=-e^2/r $ the above given 
integral diverges. But let us note that as bound state wave function 
is concentrated in a finite domain of space it should not feel a difference 
between Coulomb potential and a $(-e^2/r)e^{-\alpha r}$ potential for 
sufficiently small $\alpha$. Therefore, at least for bound states, the 
momentum space image of the Coulomb potential can be defined as
$$V(\vec{p})=\lim_{\alpha \to 0} \frac{1}{2\pi} 
\int e^{-i\vec{p} \cdot\vec{r}}
\left ( -\frac{e^2}{r} \right ) e^{-\alpha r} d\vec{r}=$$
$$-\frac{e^2}{2\pi} \lim_{\alpha \to 0}  \int \limits_0^\infty 
e^{-\alpha r} dr
\int \limits_0^{2\pi} e^{-ipr\cos {\Theta}} d\Theta \; .$$
Using
$$\int \limits_0^{2\pi} e^{-ipr\cos {\Theta}} d\Theta =
2\pi J_0(pr) \; , \; \; \int \limits_0^\infty e^{-\alpha r} J_0(pr)dr=
\frac{1}{\sqrt{p^2+\alpha^2}} , $$
\noindent we get
$$V(\vec{p})=-e^2 \lim_{\alpha \to 0} \frac{1}{\sqrt{p^2+
\alpha^2}}=-\frac{e^2}{|\vec{p}|} \; . $$
So the Schr\"{o}dinger equation for the 2-dimensional Coulomb problem,
in the case of discrete spectrum $E<0$, will be ($p_0^2=-2mE$):
\begin{eqnarray}
(\vec{p}\,^2+p_0^2)\Psi(\vec{p})=\frac{me^2}{\pi}\int 
\frac{\Psi(\vec{q})}{| \vec{p}-\vec{q}|} ~d\vec{q} \; . 
\label{eq31} \end{eqnarray}
The 2-dimensional momentum space can be mapped onto the surface of the 
3-dimensional sphere (which we call the Fock space) by means of the
stereographic projection. The stereographic projection transforms a
2-dimensional momentum $\vec{p}=p_x\vec{i}+p_y\vec{j}$ into a point on the
surface of the 3-dimensional sphere where this surface is crossed by the
line which connects the south pole of the sphere with the $(p_x,p_y)$ point
in the equatorial plane. This is shown schematically in Fig. 9 below.
\begin{figure}[htb]
\begin{center}
\epsfig{figure=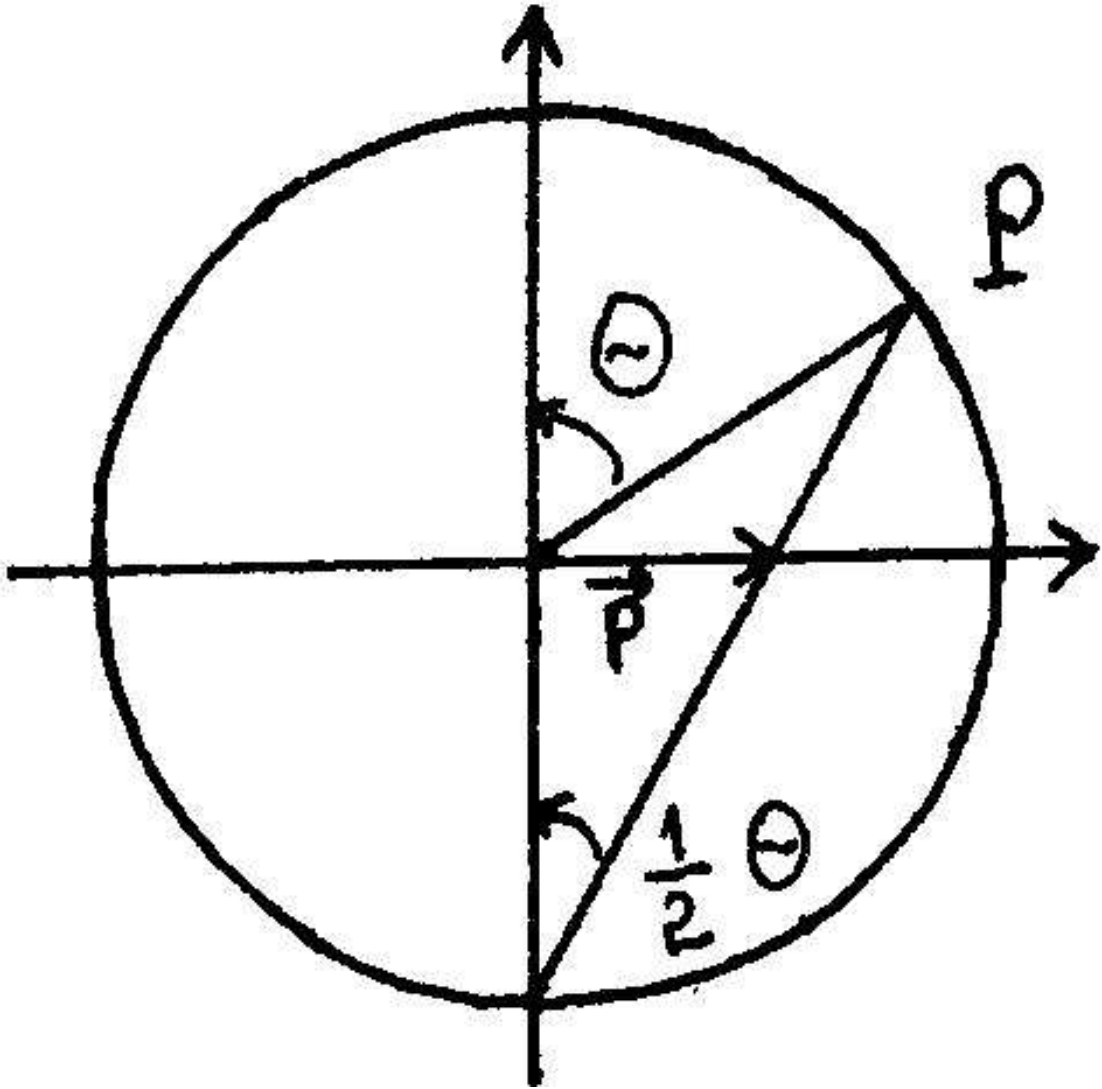,width=15cm}
\caption{The stereographic projection.}
\end{center}
\label{Fig9}
\end{figure}

Let the radius of the sphere be $p_0$. If the polar coordinates of the
vector $\vec{p}$ are $(p,\phi)$ when the point $P$ will have the spherical
coordinates $(p_0,\Theta,\phi)$, where the angle $\Theta$, as it is clear
from the Fig. 9, is determined by equation $|\vec{p}|=p_0~tg(\Theta /2)$.
Therefore
$$ \cos{\Theta}=2\cos^2{\frac{\Theta}{2}} -1=\frac{2}{1+tg^2(\Theta /2)}
-1=\frac{p_0^2-p^2}{p_0^2+p^2} $$
\noindent and
$$\sin{\Theta}=\sqrt{1-\cos^2{\Theta}}=\frac{2p_0p}{p_0^2+p^2},$$
\noindent As for the Cartesian coordinates $P_x=p_0\sin{\Theta}\cos{\phi} 
, \; P_y=p_0\sin{\Theta}\sin{\phi}  , \; P_z=p_0\cos{\Theta}$ of the point
$P$, we will have
\begin{eqnarray}
P_x=\frac{2p_0^2p_x}{p_0^2+p^2} \; , \; P_y=\frac{2p_0^2p_y}{p_0^2+p^2} \; ,
\; P_z=\frac{p_0^2-p^2}{p_0^2+p^2}p_0 \; . 
\label{eq32} \end{eqnarray}
Let $P$ and $Q$ be stereographic projections of the vectors $\vec{p}$
and $\vec{q}$, and let $\alpha$ be an angle between 3-dimensional vectors
$\vec{P}$ and $\vec{Q}$. When, as $|\vec{P}|=|\vec{Q}|=p_0$, the distance
between $P$ and $Q$ points in the 3-dimensional space will be $\sqrt{p_0^2+
p_0^2-2p_0^2\cos{\alpha}}=2p_0\sin{(\alpha/2)}$. On the other hand, the
square of this distance equals $(P_x-Q_x)^2+(P_y-Q_y)^2+(P_z-Q_z)^2$,
therefore
$$\left ( 2p_0\sin{\frac{\alpha}{2}} \right )^2=
\left (\frac{2p_0^2p_x}{p_0^2+p^2}-\frac{2p_0^2q_x}{p_0^2+q^2}\right )^2+
\left (\frac{2p_0^2p_y}{p_0^2+p^2}-\frac{2p_0^2q_y}{p_0^2+q^2}\right )^2+$$
$$\left ( p_0~\frac{p_0^2-p^2}{p_0^2+p^2}-p_0~\frac{p_0^2-q^2}{p_0^2+q^2}
\right )^2=$$
$$\frac{4p_0^4}{(p_0^2+p^2)(p_0^2+q^2)}\left \{ p^2~\frac{p_0^2+q^2}
{p_0^2+p^2}+q^2~\frac{p_0^2+p^2}{p_0^2+q^2}-2\vec{p}\cdot\vec{q}+ \right .$$
$$\left . \frac{p^4}{p_0^2}~\frac{p_0^2+q^2}{p_0^2+p^2}+\frac{q^4}{p_0^2}
~\frac{p_0^2+p^2}{p_0^2+q^2}-2\frac{p^2q^2}{p_0^2} \right \}=$$
$$\frac{4p_0^4}{(p_0^2+p^2)(p_0^2+q^2)}\left \{ \frac{p^2}{p_0^2}
(p_0^2+q^2)+\frac{q^2}{p_0^2}(p_0^2+p^2)-2\vec{p}\cdot\vec{q}-
2\frac{p^2q^2}{p^2} \right \}=$$
$$\frac{4p_0^4}{(p_0^2+p^2)(p_0^2+q^2)}(\vec{p}-\vec{q})^2 \; .$$
\noindent But
$$\frac{4p_0^4}{(p_0^2+p^2)(p_0^2+q^2)}=\frac{1}{(1+tg^2(\Theta/2))
(1+tg^2(\Theta^\prime /2))}=\cos^2{\frac{\Theta}{2}}\cos^2{\frac{
\Theta^\prime}{2}} \; , $$
\noindent where $\Theta^\prime$ spherical coordinate corresponds to the
point $Q$. So we get
\begin{eqnarray}
|\vec{p}-\vec{q}|=p_0\sec{\frac{\Theta}{2}}\sec{\frac{\Theta^\prime}{2}}
\sin{\frac{\alpha}{2}} \; .
\label{eq33} \end{eqnarray}
Now let us express $d\vec{p}=dp_x~dp_y$ through the solid angle element
$d\Omega=\sin{\Theta}~d\Theta~d\phi$.
$$d\Omega=-d(\cos{\Theta})~d\phi=-d\left ( \frac{p_0^2-p^2}{p_0^2+p^2}
\right )~d\phi=\frac{4p_0^2}{(p_0^2+p^2)^2}p~dp~d\phi=$$
$$\left [ \frac{2p_0}{p_0^2+p^2} \right ]~d\vec{p}=\left [ \frac{2}{p_0}
\cos^2{\frac{\Theta}{2}} \right ]^2~d\vec{p} \; . $$
\noindent Thus
\begin{eqnarray}
d\vec{p}=\frac{p_0^2}{4}\sec^4{\frac{\Theta}{2}}~d\Omega \; .
\label{eq34} \end{eqnarray}
With the help of Eq.33 and Eq.34, it is possible to rewrite the 
Schr\"{o}dinger equation (31) in the Fock space
$$\sec^3{\frac{\Theta}{2}}\Psi(\vec{p})=\frac{me^2}{4\pi p_0}
\int \frac{\Psi(\vec{q})\sec^3{\Theta^\prime/2}}{\sin{\alpha/2}}~d\Omega
^\prime \; , $$
\noindent or, if a new unknown function $\Phi(\Omega)=\sec^3{\frac{\Theta}
{2}}\Psi(\vec{p})$ is introduced
\begin{eqnarray}
\Phi(\Omega)=\frac{me^2}{4\pi p_0}\int \frac{\Phi(\Omega^\prime)}
{\sin{\alpha/2}}~d\Omega^\prime \; . 
\label{eq35} \end{eqnarray}
This is the Fock space Schr\"{o}dinger equation. It is invariant with 
respect to a 3-dimensional rotational group $SO(3)$ (the angle $\alpha$
measures an angular distance between two points on the sphere surface and
so it is unchanged under rotations of the sphere). In the momentum space 
this symmetry is hidden: Eq. 31 is explicitly invariant only under a smaller
group $SO(2)$.

A solution of Eq. 35 is proportional to the spherical function $Y_{lm}$.
Indeed, from the generating function of the Legendre polynomials
$$(1-2\epsilon \cos{\alpha}+\epsilon^2)^{-1/2}=\sum \limits_{l=0}
^\infty \epsilon^l P_l(\cos{\alpha}) \; ,$$ \noindent
we will have for $\epsilon=1$
$$\frac{1}{2\sin{(\alpha/2)}}=\sum \limits_{l=0}^\infty P_l(\cos{\alpha})=
\sum \limits_{l=0}^\infty \sum \limits_{m=-l}^l \frac{4\pi}{2l+1}Y_{lm}
(\Omega)Y_{lm}^*(\Omega^\prime) \; , $$
and therefore
\begin{eqnarray}
Y_{lm}(\Omega)=\frac{2l+1}{8\pi}\int \frac{Y_{lm}(\Omega^\prime)}
{\sin{(\alpha/2)}}~d\Omega^\prime \; .
\label{eq36} \end{eqnarray}
So the general solution of Eq.35 has a form $\Phi(\Omega)=\sum_{m=-l}^l
a_m Y_{lm}(\Omega)$ , if the eigenvalue condition $(me^2/p_0)=l+(1/2)$
is fulfilled. The corresponding energy eigenvalues are
$E_l=-\frac{2me^4}{(2l+1)^2}$. Each of the energy levels is $(2l+1)$-fold
degenerate, and this degeneration is of course caused by the above mentioned
hidden symmetry, which in the Fock space becomes explicit.  

\section*{Stereographic projection (general case)}
To generalize the Fock's method for equations (28) and (30), one needs
first of all a definition of the stereographic projection when the 
momentum space dimension is greater than two. But the above given 
definition of the stereographic projection admits in fact a trivial
generalization for the $n$-dimensional momentum space.

Let a momentum space point $p=(p_1,\dots,p_n)$ is transformed into
a $P=(P_1,\dots,P_{n+1})$ point on the sphere surface by the stereographic
projection. $\vec{sp}$ and $\vec{pP}$ vectors are parallel because $P$ lies
on the line connecting the point $p$ to the "south pole"
$s=(0,\dots,0,-p_0)$ of the sphere. Therefore
$(P_1-p_1,\dots,P_n-p_n,P_{n+1})=k(p_1,\dots,p_n,p_0)$, where $k$ is some
constant. So $P_i=(k+1)p_i, \; i=1\div n$ and $P_{n+1}=kp_0$.
But the point $P$ is on the sphere surface and therefore 
$P_1^2+\cdots +P_n^2+P_{n+1}^2=p_0^2$. This enables us to determine $k$
from the equation $(k+1)^2{\vec{p}}\, ^2+k^2p_0^2=p_0^2$ as $k=\frac{p_0^2-p^2}
{p_0^2+p^2}$.

Therefore a generalization of the Eq.32 is 
\begin{eqnarray}
P_1=\frac{2p_0^2p_1}{p_0^2+p^2},\; \dots \;,
P_n=\frac{2p_0^2p_n}{p_0^2+p^2}, \; P_{n+1}=\frac{p_0^2-p^2}{p_0^2+p^2}p_0
\; . \label{eq37} \end{eqnarray} \noindent
It is also straightforward to generalize Eq.33
\begin{eqnarray}
|\vec{p}-\vec{q}|=p_0\sec{\frac{\Theta_n}{2}}\sec{\frac{\Theta_n^\prime}{2}}
\sin{\frac{\alpha}{2}} \; .
\label{eq38} \end{eqnarray} \noindent
($\vec{p}$ and $\vec{q}$ are $n$-dimensional vectors, $\alpha$ is the angle
between $\vec{P}$ and $\vec{Q} \; \;$ $(n+1)$-dimensional vectors). A proof 
of Eq.38 looks like this:
$$2p_0^2\sin^2{\frac{\alpha}{2}}=p_0^2(1-\cos{\alpha})=p_0^2-\vec{P}
\cdot \vec{Q}=$$
$$p_0^2-\frac{4p_0^4\vec{p}\cdot\vec{q}}{(p_0^2+p^2)(p_0^2+q^2)}-
\frac{(p_0^2-p^2)(p_0^2-q^2)}{(p_0^2+p^2)(p_0^2+q^2)}p_0^2=
\frac{2p_0^4(p^2+q^2-2\vec{p}\cdot\vec{q})}{(p_0^2+p^2)(p_0^2+q^2)}=$$
$$2\frac{p_0^2}{p_0^2+p^2}\frac{p_0^2}{p_0^2+q^2}(\vec{p}-\vec{q})^2=
2\cos^2{\frac{\Theta_n}{2}}\cos^2{\frac{\Theta_n^\prime}{2}}
(\vec{p}-\vec{q})^2 \; .$$
Now it is necessary to generalize Eq.34. A connection between spherical and
Cartesian coordinates in the $n$-dimensional space reads
\begin{eqnarray} 
r_1 &=& r\sin{\Theta_{n-1}}\sin{\Theta_{n-2}}\dots\sin{\Theta_3}
\sin{\Theta_2}\cos{\Theta_1} \; , \nonumber \\
r_2 &=& r\sin{\Theta_{n-1}}\sin{\Theta_{n-2}}\dots\sin{\Theta_3}  
\sin{\Theta_2}\sin{\Theta_1} \; , \nonumber \\
r_3 &=& r\sin{\Theta_{n-1}}\sin{\Theta_{n-2}}\dots\sin{\Theta_3}  
\cos{\Theta_2} \; , \nonumber \\
r_4 &=& r\sin{\Theta_{n-1}}\sin{\Theta_{n-2}}\dots\sin{\Theta_4}  
\cos{\Theta_3} \; , \nonumber \\
r_n &=& r\cos{\Theta_{n-1}} \; . \nonumber \end{eqnarray}
At that for the $n$-dimensional volume element $d^{(n)}\vec{r}=dr_1\dots
dr_n$ we will have
\begin{eqnarray}
 & d^{(n)}\vec{r}=r^{n-1}~dr~d^{(n)}\Omega= & \nonumber \\
& r^{n-1}\sin^{n-2}{\Theta_{n-1}}\sin^{n-3}{\Theta_{n-2}} \dots
\sin{\Theta_2}~dr~d\Theta_1\dots d\Theta_{n-1} \; . &
\label{eq39} \end{eqnarray}
This equation can be proved by induction. As for $n=3$ it is correct, it is
sufficient to show that from its correctness there follows an analogous
relation for the $(n+1)$-dimensional case. Let us designate
$r=\sqrt{r_1^2+\dots+r_{n+1}^2}$ and $\rho=\sqrt{r_1^2+\dots+r_n^2}=
r\sin{\Theta_n}$, when  $d^{(n+1)}\vec{r}=d^{(n)}\vec{r}~dr_{n+1}=
\rho^{n-1}d\rho~dr_{n+1}~d^{(n)}\Omega$, but 
$$dr_{n+1}~d\rho=\frac{\partial(r_{n+1},\rho)}{\partial(r,\Theta_n)}dr~
d\Theta_n=\left | \begin{array}{cc} \cos{\Theta_n} & \sin{\Theta_n} \\
-r\sin{\Theta_n} & r\cos{\Theta_n} \\ \end{array}\right | ~dr~d\Theta_n=
r~dr~d\Theta_n \; ,$$ 
\noindent therefore
$$d^{(n+1)}\vec{r}=(r\sin{\Theta_n})^{n-1}rdrd\Theta_nd^{(n)}\Omega=
r^ndr\sin^{n-1}{\Theta_n}d\Theta_nd^{(n)}\Omega=r^ndrd^{(n+1)}\Omega$$
\noindent At a stereographic projection
$$\sin{\Theta_n}=\frac{2p_0p}{p_0^2+p^2} \; \; {\rm and} \; \;
\cos{\Theta_n}=\frac{p_0^2-p^2}{p_0^2+p^2}\; , $$
\noindent therefore
$$d^{(n+1)}\Omega=\sin^{n-1}\Theta_n~d\Theta_nd^{(n)}\Omega=
-\sin^{n-2}\Theta_n~d(\cos{\Theta_n})d^{(n)}\Omega=$$
$$-\left (\frac{2p_0p}{p_0^2+p^2}\right )^{n-2}d\left(
\frac{p_0^2-p^2}{p_0^2+p^2}\right ) d^{(n)}\Omega=$$
$$\frac{2^n}{p_0^n}\left (\frac{p_0^2}{p_0^2+p^2}\right)^n~p^{n-1}dp
d^{(n)}\Omega=\left(\frac{2}{p_0}\right)^n\cos^{2n}{\frac{\Theta_n}{2}}d^{(n)}
\vec{p} \; . $$
\noindent So, Eq.34 is generalized in such a way
\begin {eqnarray}
d^{(n)}\vec{p}=\left ( \frac{p_0}{2} \right )^n \sec^{2n}{\frac{\Theta_n}{2}}
d^{(n+1)}\Omega \; .
\label{eq40} \end{eqnarray}
By using Eq.38 and Eq.40 it is possible to rewrite equations (28) and
(30) in the Fock representation. In particular, Eq.30 for a hydrogen atom,
after an introduction of the new $\Phi(\Omega)=\sec^4{\frac{\Theta_3}{2}}
\Psi (\vec{p})$ unknown function, takes the form
\begin{eqnarray}
\Phi(\Omega)=\frac{me^2}{8\pi^2 p_0}\int \frac{\Phi(\Omega^\prime)}
{\sin^2{(\alpha/2)}}~d\Omega^\prime
\label{eq41} \end{eqnarray}
\noindent where $d\Omega\equiv d^{(4)}\Omega=\sin^2{\Theta_3}\sin{\Theta_2}
d\Theta_1d\Theta_2d\Theta_3$.

As for the Eq.28, it can be rewritten, for the new $\tilde{\Phi}(p)=
\sec^6{\frac{\Theta_4}{2}}\Phi(p)$ function, as
\begin{eqnarray}
\tilde \Phi(\Omega)=\frac{\lambda}{16\pi^2 p^2_0}\int 
\frac{\tilde \Phi(\Omega^\prime)}
{\sin^2{(\alpha/2)}}~d\Omega^\prime
\label{eq42} \end{eqnarray}

\noindent here $p_0=m$ and $d\Omega \equiv d^{(5)}\Omega=
\sin^3{\Theta_4}\sin^2{\Theta_3}\sin{\Theta_2}d\Theta_1d\Theta_2
d\Theta_3d\Theta_4$.
Note that, as it turned out, Eq.28 possesses a hidden $SO(5)$ symmetry,
which became explicit in the Fock space.

So, as we see,there is indeed a close analogy with the nonrelativistic
hydrogen atom problem.

To solve equations (41) and (42), one needs a generalization of the 
$Y_{lm}$ spherical functions for the $n>3$ dimensions, as the example 
of Eq.35 suggests. 

\section*{Spherical functions in a general case}
It is well known that in a 3-dimensional space spherical functions
are connected to solutions of the Laplace equation. Namely, ${\cal Y}
_{lm}(\vec{r})=r^lY_{lm}(\theta,\phi)$ harmonic polynomials (solid
harmonics) are homogeneous polynomials of rank $l$ with regard to
the variables $r_x,r_y,r_z$ and obey the Laplace equation
$\Delta{\cal Y}_{lm}(\vec{r})=0$. For the multidimensional generalization
we will use just this property of the spherical functions.

Let us define $n$-dimensional spherical function 
$Y^{(n)}_{l_{n-1},\dots,l_1}(\Theta_{n-1},\dots,\Theta_1)$ 
by requirement that 
${\cal Y}^{(n)}_{l_{n-1},\dots,l_1}(\vec{r})=r^{l_{n-1}}
Y^{(n)}_{l_{n-1},\dots,l_1}(\Theta_{n-1},\dots,\Theta_1)$
turns up to be a harmonic polynomial, that is obeys the equation
\begin{eqnarray}
\Delta^{(n)}{\cal Y}^{(n)}_{l_{n-1}\dots l_1}(\vec{r})=0 \; ,
\label{eq43} \end{eqnarray}
\noindent where $\Delta^{(n)}=\frac{\partial^2}{\partial r_1^2}+
\dots +\frac{\partial^2}{\partial r_n^2}$ is n-dimensional Laplacian.
For it we have the following decomposition into a radial and angular
parts:
\begin{eqnarray}
\Delta^{(n)}=\frac{\partial^2}{\partial r^2}+\frac{n-1}{r}
\frac{\partial}{\partial r} -\frac{1}{r^2} \delta^{(n)}  \; ,
\label{eq44} \end{eqnarray}
\noindent $\delta^{(n)}$ being the angular part of the Laplacian.

Let us prove Eq.44 by induction. For $n=3$ it is correct. So let Eq.44
is fulfilled and consider
$$\Delta^{(n+1)}=\Delta^{(n)}+\frac{\partial^2}{\partial r^2_{n+1}}=
\frac{\partial^2}{\partial \rho^2}+\frac{n-1}{\rho}
\frac{\partial}{\partial \rho} -\frac{1}{\rho^2} \delta^{(n)}+
\frac{\partial^2}{\partial r^2_{n+1}} \; . $$
\noindent here $\rho=\sqrt{r_1^2+\dots +r_n^2}$. From $r_{n+1}=r
\cos{\Theta_n}$ and $\rho=r\sin{\Theta_n}$ we get
$$\frac{\partial}{\partial \rho}=\frac{\partial}{\partial r}
\frac{\partial r}{\partial \rho}+\frac{\partial}{\partial \Theta_n}
\frac{\partial \Theta_n}{\partial \rho}=\sin{\Theta_n}
\frac{\partial}{\partial r}+\frac{1}{r}\cos{\Theta_n}
\frac{\partial}{\partial \Theta_n} \; , $$
$$\frac{\partial}{\partial r_{n+1}}=\frac{\partial}{\partial r}
\frac{\partial r}{\partial r_{n+1}}+\frac{\partial}{\partial \Theta_n}
\frac{\partial \Theta_n}{\partial r_{n+1}}=\cos{\Theta_n}
\frac{\partial}{\partial r}-\frac{1}{r}\sin{\Theta_n}
\frac{\partial}{\partial \Theta_n} \; , $$
\noindent and for the second derivatives
$$\frac{\partial^2}{\partial \rho^2}=\sin^2{\Theta_n}
\frac{\partial^2}{\partial r^2}+\frac{1}{r^2}\cos^2{\Theta_n}
\frac{\partial^2}{\partial \Theta_n^2}-\frac{2}{r^2}\cos{\Theta_n}
\sin{\Theta_n}\frac{\partial}{\partial \Theta_n}+ $$
$$ \frac{2}{r}\cos{\Theta_n}\sin{\Theta_n}\frac{\partial^2}
{\partial r \partial \Theta_n}+\frac{1}{r}\cos^2{\Theta_n}
\frac{\partial}{\partial r} \, $$

$$\frac{\partial^2}{\partial r_{n+1}^2}=\cos^2{\Theta_n}
\frac{\partial^2}{\partial r^2}+\frac{1}{r^2}\sin^2{\Theta_n}
\frac{\partial^2}{\partial \Theta_n^2}+\frac{2}{r^2}\cos{\Theta_n}
\sin{\Theta_n}\frac{\partial}{\partial \Theta_n}- $$
$$ \frac{2}{r}\cos{\Theta_n}\sin{\Theta_n}\frac{\partial^2}
{\partial r \partial \Theta_n}+\frac{1}{r}\sin^2{\Theta_n}
\frac{\partial}{\partial r} \, $$
Substitution of these expressions into $\Delta^{n+1}$ gives
$$\Delta^{(n+1)}=\frac{\partial^2}{\partial r^2}+\frac{n}{r}
\frac{\partial}{\partial r}-\frac{1}{r^2}\left [
\frac{\delta^{(n)}}{\sin^2{\Theta_n}}-(n-1)ctg~{\Theta_n}
\frac{\partial}{\partial \Theta_n}-\frac{\partial^2}{\partial 
\Theta_n^2} \right ] \; . $$
So the following recurrent relation holds
\begin{eqnarray}
\delta^{(n+1)}=
\frac{\delta^{(n)}}{\sin^2{\Theta_n}}-(n-1)ctg~{\Theta_n}
\frac{\partial}{\partial \Theta_n}-\frac{\partial^2}{\partial 
\Theta_n^2}  \; . 
\label{eq45} \end{eqnarray}
From Eq.43 and the decomposition (44) it follows that a $n$-dimensional
spherical function obeys an eigenvalue equation
\begin{eqnarray}
\delta^{(n)}Y^{(n)}_{l_{n-1},\dots,l_1}(\Theta_{n-1},\dots,\Theta_1)=
l_{n-1}(l_{n-1}+n-2)Y^{(n)}_{l_{n-1},\dots,l_1}
(\Theta_{n-1},\dots,\Theta_1).
\label{eq46} \end{eqnarray}
Eq.45 suggests that the spherical function has a factorized appearance
$$Y^{(n+1)}_{l_n,\dots,l_1}(\Theta_n,\dots,\Theta_1)=
f(\Theta_n)Y^{(n)}_{l_{n-1},\dots,l_1}(\Theta_{n-1},\dots,\Theta_1)
\; , $$
Insertion of this into Eq.46 gives the following equation for $f$:
$$\left [ \frac{l_{n-1}(l_{n-1}+n-2)}{\sin^2{\Theta_n}}-(n-1)ctg~{\Theta_n}
\frac{\partial}{\partial \Theta_n}-\frac{\partial^2}{\partial \Theta_n^2}
\right ] f(\Theta_n)=$$
$$l_n(l_n+n-1) f(\Theta_n) \; . $$
Since
$$ctg~{\Theta_n}\frac{\partial}{\partial \Theta_n}=-\cos{\Theta_n}
\frac{\partial}{\partial \cos{ \Theta_n}} \; \; {\rm and} \; \;
\frac{\partial^2}{\partial \Theta_n^2}=\frac{\partial}{\partial \Theta_n}
(-\sin{\Theta_n})\frac{\partial}{\partial \cos{ \Theta_n}}=$$
$$-\cos{\Theta_n}\frac{\partial}{\partial \cos{ \Theta_n}}-\sin{\Theta_n}
\frac{\partial}{\partial \Theta_n}\frac{\partial}{\partial \cos{ \Theta_n}}
=$$ $$-\cos{\Theta_n}\frac{\partial}{\partial \cos{ \Theta_n}}+
\sin^2{\Theta_n}\frac{\partial^2}{\partial( \cos{ \Theta_n})^2} $$
\noindent this equation can be rewritten as
$$(1-x^2)\frac{d^2 f}{dx^2}-nx\frac{d f}{dx}+\left [l_n(l_n+n-1)-
\frac{l_{n-1}(l_{n-1}+n-2)}{1-x^2} \right ] ~ f=0 \; , $$
\noindent where $x=\cos{\Theta_n}$. Let us take $f(x)=(1-x^2)^{l_{n-1}/2}
g(x)$, when the equation for $g$ will be
$$(1-x^2)\frac{d^2 g}{dx^2}-[n+2l_{n-1}]x\frac{d g}{dx}+
(l_n-l_{n-1})(l_n+l_{n-1}+n-1)g=0 \; . $$
It should be compared with equation defining the Gegenbauer polynomials:
$$(1-x^2)\frac{d^2}{dx^2}C_N^{(\alpha)}(x)-(2\alpha+1)x\frac{d }{dx}
C_N^{(\alpha)}(x)+N(N+2\alpha)C_N^{(\alpha)}(x)=0 \; . $$
As we see, $g(x)$ is proportional to $C_{l_n-l_{n-1}}^{(l_{n-1}+
(n-1)/2)}(x)$. Therefore a multidimensional spherical function can be 
defined through a recurrent relation
\begin{eqnarray} 
Y^{(n+1)}_{l_n\dots l_1}(\Theta_n,\dots,\Theta_1) =&& \nonumber \\
A_{l_n,l_{n-1}}\sin^{l_{n-1}}{\Theta_n}C_{l_n-l_{n-1}}^{(l_{n-1}+(n-1)/2)}
(\cos{\Theta_n}) && \hspace*{-6mm}
Y^{(n)}_{l_{n-1}\dots l_1}(\Theta_{n-1},\dots,
\Theta_1) \; , \label{eq47} \end{eqnarray}
\noindent where $A_{l_n,l_{n-1}}$ constant is determined by the following 
normalization condition \linebreak $\int |Y_{l_n\dots l_1}^{(n+1)}|^2d^{(n+1)}
\Omega=1$, which, if we take into account
$$d^{(n+1)}\Omega=\sin^{n-1}{\Theta_n}d\Theta_nd^{(n)}\Omega=-
\sin^{n-2}{\Theta_n}d(\cos{\Theta_n})d^{(n)}\Omega $$
\noindent can be rewritten as 
$$A^2_{l_n,l_{n-1}}\int\limits_{-1}^1 \left [
C_{l_n-l_{n-1}}^{(l_{n-1}+(n-1)/2)}(x) \right ]^2(1-x^2)^{(1/2)
(n+2l_{n-1}-2)}dx=1 \; . $$
But for the Gegenbauer polynomials we have
$$\int\limits_{-1}^1 (1-x^2)^{\alpha-1/2} [C_N^{(\alpha)}(x)]^2 dx=
\frac{\pi 2^{1-2\alpha}\Gamma (N+2\alpha)}{N!(N+\alpha)[\Gamma(
\alpha)]^2} \; , $$
\noindent therefore:
\begin{eqnarray}
A^2_{l_n,l_{n-1}}=\frac{(l_n-l_{n-1})!(2l_n+n-1)
[\Gamma(l_{n-1}+(n-1)/2)]^2}{\pi 2^{3-n-2l_{n-1}}
\Gamma(l_n+l_{n-1}+n-1)} \; .
\label{eq48} \end{eqnarray}
This expression together with Eq.47 completely determines multidimensional
spherical function.

Namely, 4-dimensional spherical function looks like
\begin{eqnarray}
Y_{nlm}(\Theta_3,\Theta_2,\Theta_1)= && \nonumber \\
\sqrt {\frac{2^{2l+1}(n+1)(n-l)!(l!)^2}{\pi (n+l+1)!}}
(\sin {\Theta_3})^l C_{n-l}^{(l+1)}&& \hspace*{-6mm}(\cos {\Theta_3})
Y_{lm}(\Theta_2,\Theta_1) \; ,
\label{eq49} \end{eqnarray}
\noindent and a 5-dimensional one:
\begin{eqnarray} &
Y_{Nnlm}(\Theta_4,\Theta_3,\Theta_2,\Theta_1)= 
\sqrt {\frac{2^{2n+1}(2N+3)(N-n)![\Gamma(n+3/2)]^2}{\pi (N+n+2)!}} \times &
\nonumber \\ &
(\sin {\Theta_4})^n C_{N-n}^{(n+3/2)}(\cos {\Theta_4})
Y_{nlm}(\Theta_3,\Theta_2,\Theta_1) \; , &
\label{eq50} \end{eqnarray}
It is clear from these equations that $l \le n \le N$. In general, as 
a lower index of the Gegenbauer polynomial coincides to the polynomial rank
and so is not negative, from Eq.47 it follows the following condition
on the quantum numbers $l_n \ge l_{n-1}$.

For spherical functions the following addition theorem holds: 
\begin{eqnarray} &
\sum \limits_{l_{n-2}=0}^{l_{n-1}} \cdots \sum \limits_{l_2=0}^{l_3}
\sum \limits_{l_1=-l_2}^{l_2}
Y^{(n)}_{l_{n-1}\dots l_1}(\Theta_{n-1},\dots, \Theta_1) 
Y^{(n)*}_{l_{n-1}\dots l_1}(\Theta_{n-1}^\prime,\dots, \Theta_1
^\prime)=& \nonumber \\ &
\frac{2l_{n-1}+n-2}{4\pi^{n/2}}\Gamma\left (\frac{n-2}{2}\right )
C_{l_{n-1}}^{((n-2)/2)}(\cos {\alpha}) \; , & 
\label{eq51} \end{eqnarray}
\noindent where $\alpha$ is the angle between $n$-dimensional unit vectors,
which are defined by spherical coordinates $(\Theta_1,\dots,\Theta_{n-1})$
and $(\Theta_1^\prime,\dots,\Theta_{n-1}^\prime)$.

If $n=3$, Eq.51 gives the addition theorem for the 3-dimensional spherical
functions $Y_{lm}$, since $C_l^{(1/2)}(\cos{\alpha})=P_l(\cos{\alpha})$.
So let us try to prove Eq.51 by induction, that is suppose Eq.51 is correct
and consider $(n+1)$-dimensional case:
$$\sum \limits_{l_{n-1},\dots,l_1}
Y^{(n+1)}_{l_n\dots l_1}(\Theta_n,\dots,\Theta_1) 
Y^{(n+1)*}_{l_n\dots l_1}(\Theta_n^\prime,\dots, \Theta_1
^\prime)=\sum \limits_{l_{n-1}=0}^{l_n}A^2_{l_n,l_{n-1}}\times $$
$$\sin^{l_{n-1}}{\Theta_n}\sin^{l_{n-1}}{\Theta_n^\prime}
C_{l_n-l_{n-1}}^{(l_{n-1}+(n-1)/2)}(\cos{\Theta_n})
C_{l_n-l_{n-1}}^{(l_{n-1}+(n-1)/2)}(\cos{\Theta_n^\prime}) \times $$
$$\sum \limits_{l_{n-2},\dots,l_1}
Y^{(n)}_{l_{n-1}\dots l_1}(\Theta_{n-1},\dots,\Theta_1)
Y^{(n)*}_{l_{n-1}\dots l_1}(\Theta_{n-1}^\prime,\dots, \Theta_1
^\prime)=$$
$$\frac{2^{n-3}(2l_n+n-1)\Gamma((n-2)/2)}{4\pi^{(n/2+1)}}\times $$
$$\sum \limits_{l_{n-1}=0}^{l_n} 2^{l_{n-1}}\frac{(l_n-l_{n-1})!}
{\Gamma(l_n+l_{n-1}+n-1)}(2l_{n-1}+n-2) \times $$
$$[\Gamma(l_{n-1}+(n-1)/2)]^2\sin^{l_{n-1}}{\Theta_n}
C_{l_n-l_{n-1}}^{(l_{n-1}+(n-1)/2)}(\cos{\Theta_n}) \times $$
$$\sin^{l_{n-1}}{\Theta_n^\prime}
C_{l_n-l_{n-1}}^{(l_{n-1}+(n-1)/2)}(\cos{\Theta_n^\prime})
C_{l_{n-1}}^{((n-2)/2)}(\cos{\omega}) \; . $$
\noindent $\omega$ is the angle between n-dimensional unit vectors and it 
is connected with the angle $\alpha$ between $(n+1)$-dimensional unit vectors
as $\cos{\alpha}=\cos{\Theta_n}\cos{\Theta_n^\prime}+
\sin{\Theta_n}\sin{\Theta_n^\prime}\cos{\omega}$ (if $\vec{e}$ is 
$(n+1)$-dimensional unit vector, determined by $(\Theta_1,\dots,\Theta_n)$
angles and $\vec{f}$ -- $n$-dimensional unit vector, determined by \linebreak
$(\Theta_1,\dots,\Theta_{n-1})$ angles, when $\cos{\alpha}=\vec{e}\cdot
\vec{e}^{\, \prime} \; , \; \; \cos{\omega}=\vec{f}\cdot \vec{f}^{\,
\prime} ,$ and $\vec{e}=\cos{\Theta_n}\vec{e}_{n+1}+\sin{\Theta_n}\vec{f}$).
 
Using the addition theorem for the Gegenbauer polynomials
$$C_n^{(\alpha)}(\cos{\Theta}\cos{\Theta^\prime}+
\sin{\Theta}\sin{\Theta^\prime}\cos{\phi})=$$
$$\sum \limits_{m=0}^n 2^{2m}(2\alpha+2m-1)
\frac{(n-m)![\Gamma(\alpha +m)]^2 \Gamma(2\alpha-1)}
{[\Gamma(\alpha)]^2 \Gamma(2\alpha+n+m)}\sin^m{\Theta}
C_{n-m}^{(\alpha+m)}(\cos{\Theta}) \times $$
$$\sin^m{\Theta^\prime}C_{n-m}^{(\alpha+m)}(\cos{\Theta^\prime})
C_m^{(\alpha-1/2)}(\cos {\phi}) \; , $$
\noindent we get:
$$\sum \limits_{l_{n-1},\dots,l_1}
Y^{(n+1)}_{l_n\dots l_1}(\Theta_n,\dots,\Theta_1) 
Y^{(n+1)*}_{l_n\dots l_1}(\Theta_n^\prime,\dots, \Theta_1
^\prime)=$$
$$2^{n-3}\frac{2l_n+n-1}{4\pi^{(n/2+1)}}\frac{[\Gamma((n-1)/2)]^2
\Gamma((n-2)/2)}{\Gamma(n-2)}C_{l_n}^{((n-1)/2)}(\cos{\alpha}) \; , $$
\noindent but $\Gamma(z)\Gamma(z+1/2)=(\pi^{1/2}/2^{2z-1})\Gamma(2z)$
formula gives $\Gamma((n-1)/2)\Gamma((n-2)/2)=  
(\pi^{1/2}/2^{n-3})\Gamma(n-2)$
and we end with the desired result:
$$\sum \limits_{l_{n-1},\dots,l_1}
Y^{(n+1)}_{l_n\dots l_1}(\Theta_n,\dots,\Theta_1) 
Y^{(n+1)*}_{l_n\dots l_1}(\Theta_n^\prime,\dots, \Theta_1
^\prime)=$$
$$\frac{2l_n+n-1}{4\pi^{(n+1)/2}}\Gamma((n-1)/2)
C_{l_n}^{((n-1)/2)}(\cos{\alpha}) \; . $$

In a 4-dimensional space the addition theorem for the spherical functions 
looks like
$$\sum\limits_{l=0}^n\sum\limits_{m=-l}^l Y_{nlm}
(\Theta_3,\Theta_2,\Theta_1)Y_{nlm}^*
(\Theta_3^\prime,\Theta_2^\prime,\Theta_1^\prime)=$$
$$\frac{n+1}{2\pi^2}C_n^{(1)}(\cos{\alpha})=\frac{n+1}
{2\pi^2}\,\frac{\sin {(n+1)\alpha}}{\sin{\alpha}} \; , $$
Using this, we can prove, that a solution of Eq.41 is proportional to 
a spherical function $Y_{nlm}(\Omega)$. Indeed, from the generating function
of the Gegenbauer polynomials $(1-2\epsilon \cos{\alpha}+\epsilon^2)^{-\nu}=
\sum_{n=0}^\infty C_n^{(\nu)}(\cos{\alpha})\epsilon^n$, we get in the 
$\epsilon=\nu=1$ case $1/4\sin^2{(\alpha/2)}=\sum_{n=0}^\infty C_n^{(1)}
(\cos{\alpha})$
Note that it is not rigorously correct to take $\epsilon =1$
because in this case the series stops to be convergent. But
$$\sum\limits_{n=0}^\infty C_n^{(1)}(\cos{\alpha})=
\sum\limits_{n=0}^\infty\frac{\sin {(n+1)\alpha}}{\sin{\alpha}}=
\lim_{N \to \infty}\frac{\sin{((N+1)/2)\alpha}\sin{(N\alpha/2)}}
{\sin{\alpha}\sin{(\alpha/2)}} \; . $$
\noindent of course a limit doesn't exist, but
$$\frac{\sin{\frac{N+1}{2}\alpha}\sin{\frac{N\alpha}{2}}}
{\sin{\alpha}\sin{(\alpha/2)}}=\frac{\cos{\frac{\alpha}{2}}-
\cos{(N+\frac{1}{2})\alpha}}{2\sin{\alpha}\sin(\alpha/2)}=
\frac{1}{4\sin^2{(\alpha/2)}}-\frac{\cos{(N+\frac{1}{2})\alpha}}
{2\sin{\alpha}\sin(\alpha/2)} $$
\noindent and the second term oscillates more and more quickly as $N$
increases. So its multiplication on any normal function and integration
will give a result which tends to zero as $N \to \infty$. Therefore
$1/4\sin^2{(\alpha/2)}=\sum_{n=0}^\infty C_n^{(1)}(\cos{\alpha})$
relation can be used under integration without any loss of rigor.
Now $C_n^{(1)}(\cos{\alpha})$ can be replaced by using the addition 
theorem and we get because of the orthonormality property of the spherical
functions:
$$\int \frac{Y_{nlm}(\Omega^\prime)}{\sin^2{\alpha/2}}
d\Omega^\prime=\frac{8\pi^2}{n+1}Y_{nlm}(\Omega) \; . $$
So a 4-dimensional spherical function is indeed a solution of the Eq.41
and the corresponding eigenvalue is determined from the
$p_0=me^2/(n+1)$ condition ($m$ being electron mass), which gives a well
known expression for the hydrogen atom levels $E=-me^4/2(n+1)^2$,
$ \; n=0,1,\dots$.

Now it should be no surprise that a solution of the Eq.42 is proportional
to a 5-dimensional spherical function. To prove this, as the previous 
experience suggests, it is necessary to decompose $(1-\cos{\alpha})^{-1}$
into a series of Gegenbauer polynomials and afterwards, by using the addition
theorem (51), replace them by the spherical functions.

Suppose $1/(1-x)=\sum_{N=0}^\infty A_N C_N^{(3/2)}(x)$, when because of 
the orthogonality of the Gegenbauer polynomials and of the normalization
$$\int\limits_{-1}^1 (1-x^2)[C_N^{(3/2)}(x)]^2 dx=
\frac{2(N+1)(N+2)}{2N+3} $$
\noindent we get
$$A_N=\frac{2N+3}{2(N+1)(N+2)} \int\limits_{-1}^1 (1+x)C_N^{(3/2)}
(x)dx \; . $$
\noindent The integral can be evaluated by using relations
$$C_N^{(3/2)}(x)=\frac{d}{dx}P_{N+1}(x), \; P_n(1)=1, \; \; {\rm and}
\; \int\limits_{-1}^1 P_n(x)dx=0 \; {\rm if}\; n \neq 0.$$
\noindent So
$$A_N=\frac{2N+3}{(N+1)(N+2)} \; , \; \; \frac{1}{1-\cos{\alpha}}=
\sum\limits_{N=0}^\infty \frac{2N+3}{(N+1)(N+2)}C_N^{(3/2)}(\cos{\alpha})
=$$
$$\sum\limits_{Nnlm}\frac{8\pi^2}{(N+1)(N+2)}
Y_{Nnlm}(\Omega)Y_{Nnlm}^*(\Omega^\prime) \; , $$
\noindent Therefore
$$\int \frac{Y_{Nnlm}(\Omega^\prime)}{\sin^2{\alpha/2}}
d\Omega^\prime=\frac{16\pi^2}{(N+1)(N+2)}Y_{Nnlm}(\Omega) \; . $$
\noindent and eigenvalues of the Eq.42 will be $\lambda_N=(N+1)(N+2)m^2$
(here $m$ stands for the particle mass, not an index of the spherical 
function), with the corresponding solution
$$\Phi_N(\Omega)=\sum\limits_{nlm}A_{nlm}Y_{Nnlm}(\Omega) \; . $$

\section*{Stereographic projection in the Wick-Cutkosky model}
Let us return to Eq.27 and consider again the equal mass case
\begin{eqnarray} 
\left [ m^2+\left (p+\frac{i}{2}P \right )^2 \right ]     
\left [ m^2+\left (p-\frac{i}{2}P \right )^2 \right ] \Phi (p)=    
\frac{\lambda}{\pi^2} \int dq   
\frac{\Phi (q)}{(p-q)^2} \; ,
\label{eq52} \end{eqnarray} \noindent
here $P=(\vec{0},M)$ - Euclidean 4-vector and $M$ is the bound 
state mass.

Equation (52) exhibits explicitly only $SO(3)$ symmetry (because it
contains a fixed 4-vector $P$), but now we know that a higher 
symmetry can be hidden. To find out this, let us transform the 
4-dimensional momentum space onto the 5-dimensional sphere surface
by the stereographic projection. R.h.s. of the equation will be
changed at that according to formulas (38) and (40)
$$\frac{\lambda}{\pi^2} \int dq \frac{\Phi (q)}{(p-q)^2}=
\frac{\lambda p_0^2}{16\pi^2}\sec^{-2}{\frac{\Theta_4}{2}}
\int \frac{\sec^6{(\Theta_4^\prime/2)}\Phi (q)}{\sin^2{\alpha/2}}
d\Omega^\prime \; . $$
As for the l.h.s. , we will have
$$\left [ m^2+\left (p+\frac{i}{2}P \right )^2 \right ]     
\left [ m^2+\left (p-\frac{i}{2}P \right )^2 \right ] = $$
$$ m^4-\frac{1}{2}m^2M^2+\frac{1}{16}M^4+2m^2p^2-\frac{1}{2}M^2p^2+
p^4+(p\cdot P)^2= $$
$$\left (m^2-\frac{1}{4}M^2\right )^2+2p_0^2~{\rm tg}^2{\frac{\Theta_4}{2}}
\left (m^2-\frac{1}{4}M^2\right)+ $$
$$p_0^2~{\rm tg}^2{\frac{\Theta_4}{2}}\left ( p_0^2~{\rm tg}^2
{\frac{\Theta_4}{2}}
+M^2\cos^2{\Theta_3}\right ) \; . $$   
It is convenient to choose $p_0$ - 5-sphere radius as:
$p_0^2=m^2-(1/4)M^2$, when
$$\left [ m^2+\left (p+\frac{i}{2}P \right )^2 \right ]     
\left [ m^2+\left (p-\frac{i}{2}P \right )^2 \right ] =
p_0^2 \left [p_0^2\sec^4{\frac{\Theta_4}{2}}+ \right . $$
$$\left . M^2~{\rm tg}^2\frac{\Theta_4}{2}\cos^2{\Theta_3}\right ]=
p_0^2\sec^4{\frac{\Theta_4}{2}}\left [p_0^2+\frac{M^2}{4}\sin^2{\Theta_4}
\cos^2{\Theta_3}\right ] \; . $$
\noindent and after introduction of the new  $F(\Omega)\equiv
F(\Theta_1,\Theta_2,\Theta_3,\Theta_4)=\sec^6{\frac{\Theta_4}{2}}
\Phi(p)$  function , Eq.52 takes the form
\begin{eqnarray}
\left [p_0^2+\frac{M^2}{4}\sin^2{\Theta_4}\cos^2{\Theta_3}\right ]
F(\Omega)=\frac{\lambda}{16\pi^2}\int \frac{F(\Omega^\prime)}
{\sin^2{\alpha/2}} d\Omega^\prime \; . 
\label{eq53} \end{eqnarray}
Note that $\sin{\Theta_4}\cos{\Theta_3}=\tilde P_4/p_0$ ($\tilde P$ is
that point, which corresponds to the $p$ 4-vector in the stereographic 
projection ), therefore Eq.53 is invariant under such rotations of the
5-dimensional coordinate system, which don't change the 4-th axis.
So Eq.52 possesses a hidden $SO(4)$ symmetry. This symmetry was not 
explicit because $p$-momentum space is not orthogonal to the 4-th axis 
$\vec{e}_4$. But with the inverse stereographic projection, if we 
choose its pole on the axis $\vec{e}_4$, we get $k$-momentum space,
which is orthogonal to $\vec{e}_4$. This is schematicly shown in Fig.10:
\begin{figure}[htb]
\begin{center}
\epsfig{figure=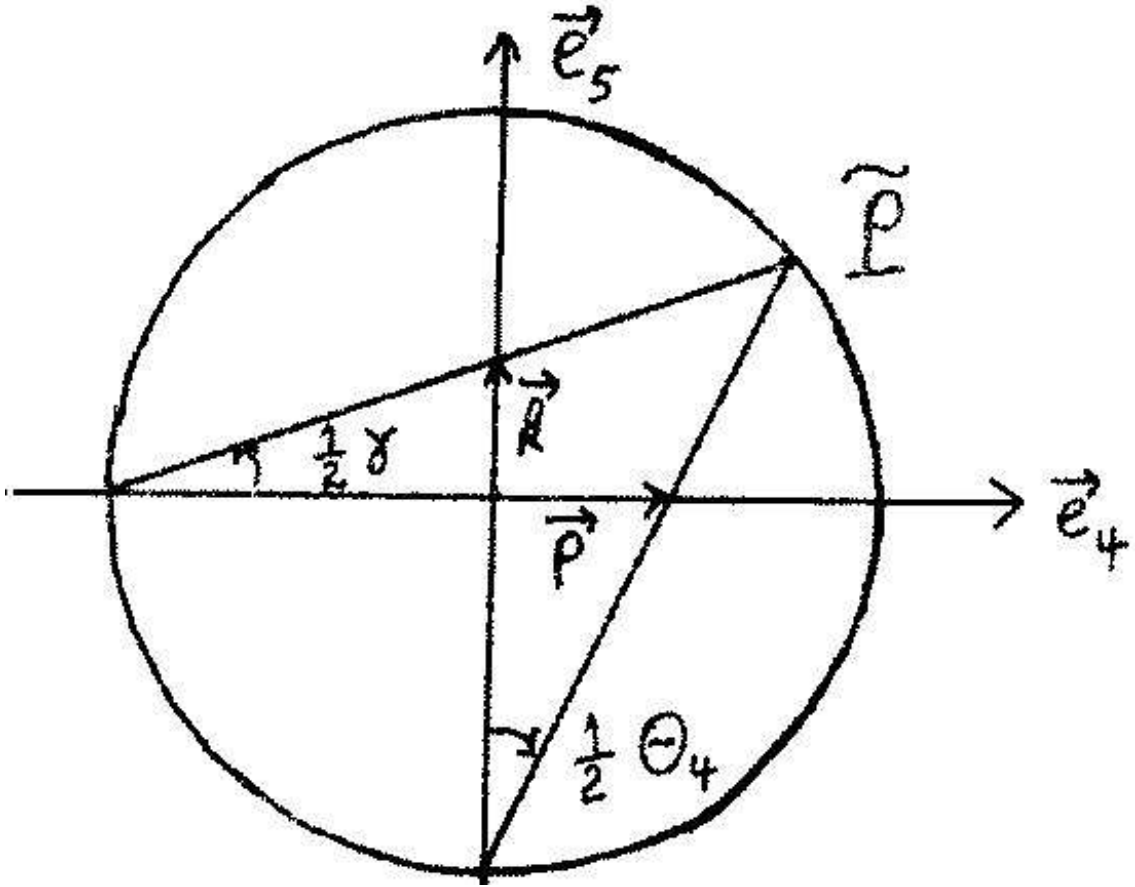,width=15cm}
\caption{$k$-momentum space.}
\end{center}
\label{Fig10}
\end{figure}

\noindent Therefore Eq.52, rewritten in terms of the $k$-variables,
should become explicitly $SO(4)$ invariant. Let us get this new equation.
According to the last equality in (37)
$$\sin{\Theta_4}\cos{\Theta_3}=\cos{\gamma}=\frac{\tilde{P_4}}
{p_0}=\frac{p_0^2-k^2}{p_0^2+k^2} \; . $$
\noindent Furthermore, because of Eq.38
$$\frac{1}{\sin^2{\alpha/2}}=\frac{p_0^2}{(k-k^\prime)^2}~\sec^2
{\frac{\gamma}{2}}~\sec^2{\frac{\gamma^\prime}{2}} \; , $$
\noindent and according to Eq.40
$$d\Omega^\prime=\frac{16}{p_0^4}~\sec^{-8}\left(\frac{\gamma^\prime}{2}
\right) dk^\prime \; . $$

The substitution all of these into the Eq.53, after introduction of the
new $Q(k)=\cos{(\gamma/2)}F(\Omega)$ unknown function, will give
$$\left [p_0^2+\frac{M^2}{4}~\frac{(p_0^2-k^2)^2}{(p_0^2+k^2)^2}\right ]
\sec^4{\frac{\gamma}{2}} \; Q(k)=\frac{\lambda}{\pi^2p_0^2}\int
\frac{Q(k^\prime)}{(k-k^\prime)^2} ~dk^\prime \; . $$
\noindent But
$$\sec^4{\frac{\gamma}{2}}=\left(1+{\rm tg}^2{\frac{\gamma}{2}}\right)^2
=\left(1+\frac{k^2}{p_0^2}\right)^2=\frac{1}{p_0^4}(p_0^2+k^2)^2 \; . $$
\noindent and we finally get
$$\left [(p_0^2+k^2)^2+\frac{M^2}{4p_0^2}(p_0^2-k^2)^2 \right ] \; Q(k)
\equiv \vspace*{-4mm} $$
\begin{eqnarray}
\left [ m^2p_0^2+\frac{m^2}{p_0^2}(k^2)^2+2p_0^2k^2
\left(2-\frac{m^2}{p_0^2}\right)\right ] \; Q(k)=\frac{\lambda}{\pi^2}\int
\frac{Q(k^\prime)}{(k-k^\prime)^2} ~dk^\prime \; . 
\label{eq54} \end{eqnarray}
Since the equation just obtained is $SO(4)$ invariant, its solution
should have the form $Q(k)=f(k^2)Y_{nlm}(\Omega)$, which enables to
perform the angular integration in Eq.54 and, as a result, to get 
a one-dimensional integral equation for $f(k^2)$. Let us denote $x=
\sqrt{k^2}$ and let $\alpha$ be the angle between $k$ and $k^\prime$
4-vectors, then
$$\frac{1}{(k-k^\prime)^2}=$$
$$ \left \{  \matrix {
\frac{1}{x^2}~\frac{1}{1-2\frac{x^\prime}{x}\cos{\alpha}+
\frac{x^{\prime 2}}{x^2}}=\frac{1}{x^2}\sum_{n^\prime=0}^\infty
\left(\frac{x^\prime}{x}\right)^{n^\prime}C_{n^\prime}^{(1)}
(\cos{\alpha}),~{\rm if}~ x > x^\prime \cr
\frac{1}{x^{\prime 2}}~\frac{1}{1-2\frac{x}{x^\prime}\cos{\alpha}+
\frac{x^2}{x^{\prime 2}}}=\frac{1}{x^{\prime 2}}\sum_{n^\prime=0}^\infty
\left(\frac{x}{x^\prime}\right)^{n^\prime}C_{n^\prime}^{(1)}
(\cos{\alpha}),~{\rm if}~ x < x^\prime 
} \right \} = $$
$$\sum\limits_{n^\prime=0}^\infty C_{n^\prime}^{(1)}(\cos{\alpha})
\left[ \frac{(x^\prime)^{n^\prime}}{x^{n^\prime+2}}\Theta(x-x^\prime)+
\frac{x^{n^\prime}}{(x^\prime)^{n^\prime+2}}\Theta(x^\prime-x)
\right] \; . $$
\noindent According to the addition theorem 
$$C_{n^\prime}^{(1)}(\cos{\alpha})=\frac{2\pi^2}{n^\prime+1}
\sum\limits_{l^\prime m^\prime}Y_{n^\prime l^\prime m^\prime }
(\Omega)Y^*_{n^\prime l^\prime m^\prime }(\Omega^\prime) \; , $$
\noindent therefore
$$\frac{1}{(k-k^\prime)^2}=\sum\limits_{n^\prime l^\prime m^\prime}
\frac{2\pi^2}{n^\prime+1}
\left[ \frac{(x^\prime)^{n^\prime}}{x^{n^\prime+2}}\Theta(x-x^\prime)+
\right . $$ $$ \left .
\frac{x^{n^\prime}}{(x^\prime)^{n^\prime+2}}\Theta(x^\prime-x)
\right]Y_{n^\prime l^\prime m^\prime }(\Omega)
Y^*_{n^\prime l^\prime m^\prime }(\Omega^\prime) \; . $$
\noindent The substitution of this and $Q(k)=f(x)Y_{nlm}(\Omega)$
into Eq.54 will give (because of orthonormality of the spherical
functions and $dk^\prime=|k^\prime|^3d|k^\prime|d^{(4)}\Omega 
^\prime)$)
$$\left [(p_0^2+x^2)^2+\frac{M^2}{4p_0^2}(p_0^2-x^2)^2 \right ] \; 
f(x)=$$ $$\frac{2\lambda}{n+1}\int\limits_0^\infty
\left[ \frac{y^{n+3}}{x^{n+2}}\Theta(x-y)+
\frac{x^n}{y^{n-1}}\Theta(y-x) \right]~ f(y)~dy \; . $$
\noindent Since
$$(p_0^2+x^2)^2+\frac{M^2}{4p_0^2}(p_0^2-x^2)^2 =(p_0^2+x^2)^2
\left(1+\frac{M^2}{4p_0^2}\right)-4\frac{M^2}{4p_0^2}p_0^2x^2=$$
$$(p_0^2+x^2)^2\frac{m^2}{p_0^2}-M^2x^2=\frac{m^2}{p_0^2}
\left(p_0^2+x^2+\frac{p_0M}{m}x\right)
\left(p_0^2+x^2-\frac{p_0M}{m}x\right) $$
\noindent then this integral equation can be rewritten also as
$$\left(p_0^2+x^2+\frac{p_0M}{m}x\right)
\left(p_0^2+x^2-\frac{p_0M}{m}x\right)~f(x)=$$
$$\frac{2\lambda p_0^2}{m^2(n+1)}\int\limits_0^\infty
\left[ \frac{y^{n+3}}{x^{n+2}}\Theta(x-y)+
\frac{x^n}{y^{n-1}}\Theta(y-x) \right]~ f(y)~dy \; . $$
\noindent or introducing $s=x/p_0$ undimensional variable:
\begin{eqnarray} &
\left(1+s^2+\frac{M}{m}s\right)
\left(1+s^2-\frac{M}{m}s\right)~f(s)= & \nonumber \\
& \frac{2\lambda }{m^2(n+1)}\int\limits_0^\infty
\left[ \frac{r^{n+3}}{s^{n+2}}\Theta(s-r)+
\frac{s^n}{r^{n-1}}\Theta(r-s) \right]~ f(r)~dr \; . &
\label{eq55} \end{eqnarray}
\noindent It is clear from this equation that for the $f_n(s)\equiv
f(s)$ function, if $n\neq 0$, we have $f_n(0)=0$ boundary condition
at the point $s=0$.

Let us show that Eq.55 is equivalent to a second order differential
equation. Let
$$R(s,r)=\left[ \frac{r^{n+3}}{s^{n+2}}\Theta(s-r)+
\frac{s^n}{r^{n-1}}\Theta(r-s) \right] \; .$$
\noindent Since $(d/ds)\Theta(s-r)=\delta(s-r)$, for the derivatives
over $s$ we will have:
$$\frac{d}{ds}R(s,r)=-(n+2)\frac{r^{n+3}}{s^{n+3}}\Theta(s-r)
+n\frac{s^{n-1}}{r^{n-1}}\Theta(r-s) \; , $$
$$\frac{d^2}{ds^2}R(s,r)=(n+2)(n+3)\frac{r^{n+3}}{s^{n+4}}
\Theta(s-r)+$$ $$n(n-1)\frac{s^{n-2}}{r^{n-1}}\Theta(r-s)-
2(n+1)\delta(s-r) \; . $$
\noindent Let us try to choose $A$ and $B$ coefficients such that
$$\left [\frac{d^2}{ds^2}+\frac{A}{s}~\frac{d}{ds}+
\frac{B}{s^2} \right]~R(s,r)=-2(n+1)\delta(s-r) $$
\noindent which means the following system
$$(n+2)A-B=(n+2)(n+3) \; , \; \; nA+B=-n(n-1) $$
\noindent with $A=3$ and $B=-n(n+2)$ as the solution. Therefore
$$\left [\frac{d^2}{ds^2}+\frac{3}{s}~\frac{d}{ds}-
\frac{n(n+2)}{s^2} \right]
\left [(1+s^2)^2-\frac{M^2}{m^2}s^2 \right ] f(s)= $$
$$-\frac{4\lambda}{m^2}\int\limits_0^\infty \delta(s-r)f(r)dr=
-\frac{4\lambda}{m^2}f(s) \; .$$
\noindent Let us introduce the new variable $t=s^2$ and the new 
unknown function $\phi(t)=t[(1+t)^2-(M^2/m^2)t]f(t)$. Because
$$\frac{1}{s}~\frac{d}{ds}=2\frac{d}{dt} \; \; {\rm and} \; \;
\frac{d^2}{ds^2}=\frac{d}{ds}2s\frac{d}{dt}=2\frac{d}{dt}+
2s\frac{d}{ds}~\frac{d}{dt}=2\frac{d}{dt}+4t\frac{d^2}{dt^2} 
\;. $$
\noindent We finally get
\begin{eqnarray}
\frac{d^2}{dt^2}\phi_n(t)+\left [ \frac{\lambda}{t
[m^2(1+t)^2-M^2t]}-\frac{n(n+2)}{4t^2} \right ] \phi_n(t)=0
\; . \label{eq56} \end{eqnarray}
The asymptotic form of this equation (when $n\neq 0$) in the
$t\to 0$ or $t\to \infty$ limit is
$$\frac{d^2\phi_n}{dt^2}-\frac{n(n+2)}{4t^2}\phi_n=0 \; , $$
\noindent with solutions $t^{1+n/2}$ and $t^{-n/2}$. At the origin
$\phi_n\sim t^{-n/2}$ behavior is not adequate, because then
$f_n \sim t^{-(n/2)-1}$, which contradicts to the $f_n(0)=0$ 
condition. Let us show that at the infinity, on the contrary,
$\phi_n\sim t^{1+n/2}$ behavior is not good, because then
$f_n \sim t^{(n/2)-2}=s^{n-4}$, which contradicts to the integral
equation (55). Indeed, when $s\to \infty$, left hand size of the
equation will be of the order of $s^n$, and the r.h.s. of the order of 
$$\frac{1}{s^{n+2}}\int\limits_0^s r^{n+3}f_n(r)dr+
s^n\int\limits_s^\infty \frac{r^{n-4}}{r^{n-1}}dr=\frac{1}{2}s^{n-2}
+\frac{1}{s^{n+2}}\int\limits_0^s r^{n+3}f_n(r)dr \; . $$
\noindent Let $s_0$ be some big enough number, so that when $r>s_0$
we can use the asymptotic form of the solution. Then
$$\int\limits_0^s r^{n+3}f_n(r)dr \approx \int\limits_0^{s_0}
r^{n+3}f_n(r)dr+\int\limits_{s_0}^s r^{2n-1}dr=$$
$$\frac{1}{2n}(s^{2n}-s_0^{2n})+\int\limits_0^{s_0}
r^{n+3}f_n(r)dr \; . $$
\noindent Therefore
$$\frac{1}{s^{n+2}}\int\limits_0^s r^{n+3}f_n(r)dr \sim s^{n-2} \; . $$
\noindent So the r.h.s. of the Eq.55 turns out to be of the order of
$s^{n-2}$ and it can't be equal to the l.h.s.

As we see, Eq.56 should be accompanied by the following boundary 
conditions
\begin{eqnarray} &
\phi_n(t) \sim t^{1+n/2} \; , \; {\rm if} \; \; t \to 0 \; , & \nonumber 
\\ &\phi_n(t) \sim t^{-n/2} \; , \; {\rm if} \; \; t \to \infty \; . &
\label{eq57} \end{eqnarray}

In Eq.56 $t$-variable changes from $0$ to $\infty$. For numeric 
calculations it is more convenient to have a finite interval. So let us
introduce the new $z=(1-t)/(1+t)$ variable, which changes from $-1$ to 
$1$. Using
$$ \frac{d}{dt}=-\frac{2}{(1+t)^2}~\frac{d}{dz} \; , \;
\frac{d^2}{dt^2}=\frac{d}{dt}\frac{-2}{(1+t)^2}\frac{d}{dz}= $$
$$\frac{4}{(1+t)^3}~\frac{d}{dz}+\frac{4}{(1+t)^4}~\frac{d^2}{dz^2}
=\frac{1}{2}(1+z)^3\frac{d}{dz}+\frac{1}{4}(1+z)^4\frac{d^2}{dz^2}$$
\noindent we get
$$(1-z^2)\frac{d^2\phi_n}{dz^2}+2(1-z)\frac{d\phi_n}{dz}+
\left [ \frac{\lambda}{m^2-(M^2/4)(1-z^2)}-\frac{n(n+2)}{1-z^2}
\right ] \phi_n=0 \; , $$
\noindent which can be somewhat simplified if we introduce the new
$g_n(z)=(1+z)(1-z^2)^{n/2}\phi_n(z)$ function, for which the equation
looks like
\begin{eqnarray}
(1-z^2)\frac{d^2g_n}{dz^2}+2nz\frac{dg_n}{dz}+
\left [ \frac{\lambda}{m^2-(M^2/4)(1-z^2)}-n(n+1)
\right ] g_n=0 \; .
\label{eq58} \end{eqnarray}
 
Eq.57 indicates that at $z\to 1$ we should have $g_n\sim(1-z)^{n/2}
(1-z)^{1+n/2}=(1-z)^{n+1}$, and at $z\to -1$ -- $g_n\sim(1+z)^{1+n/2}
(1+z)^{n/2}=(1+z)^{n+1}$. Therefore the boundary conditions for the
Eq.58 are $g_n(-1)=g_n(1)=0$.

\section*{Integral representation method}
Transition from Eq.52 to an one-dimensional integral equation can
be performed also in another way, which is not directly connected
with the hidden symmetry of the equation. This time let us consider
a general case of unequal masses and let $m_1=m+\Delta , \; \;
m_2=m-\Delta$. Eq.27 takes the form
$$ \Phi(p)= \vspace*{-4mm} $$
\begin{eqnarray} 
\frac{\lambda}{\pi^2} \left [ \left (
(m+\Delta)^2+(p-i\eta_1P)^2 \right ) \left (
(m-\Delta)^2+(p+i\eta_2P)^2 \right ) \right]^{-1}
\int dq \frac{\Phi(q)}{(p-q)^2} \; . 
\label{eq59} \end{eqnarray}
\noindent Note, that the solution of the equation
$$\Phi(p)=\frac{\lambda}{\pi^2}(p^2+m^2)^{-2}\int dq \frac{\Phi(q)}
{(p-q)^2} \; , $$
\noindent is, as we have seen earlier
$$\Phi(p) \sim \cos^6{\frac{\Theta_4}{2}}Y_{Nnlm}(\Omega)\sim
(1+\cos{\Theta_4})^3\sin^n{\Theta_4}\sin^l{\Theta_3}Y_{lm}
(\Theta_2,\Theta_1)\sim $$
$$(p^2+m^2)^{-n-3}|\vec{p}|^lY_{lm}(\Theta_2,\Theta_1)=
\frac{{\cal{Y}}_{lm}(\vec{p})}{(p^2+m^2)^{n+3}} \; ,$$
\noindent because
$$\cos{\Theta_4}=\frac{m^2-p^2}{m^2+p^2} \; , \; \;
\sin{\Theta_4}=\frac{2m|p|}{m^2+p^2} \; \; {\rm and} \; \;
\sin{\Theta_3}=\frac{|\vec{p}|}{|p|} \; . $$
\noindent (Here $|p|=\sqrt{p^2}$ is the length of the 4-vector
$p=(\vec{p},p_4)$ ).

Using the following parametric representation
$$\frac{1}{AB}=\frac{1}{2}\int\limits_{-1}^1\frac{dz}
{[(1/2)(A+B)+(1/2)(A-B)z]^2} \; , $$
\noindent we will have
$$[(m-\Delta)^2+(p+i\eta_2P)^2]^{-1}
[(m+\Delta)^2+(p-i\eta_1P)^2]^{-1}= \vspace*{-4mm} $$
\begin{eqnarray}
\frac{1}{2} \int \limits_{-1}^1 dz
\left [(1+\frac{\Delta^2}{m^2}-2\frac{\Delta}{m}z)p_0^2+p^2+
i(z-\frac{\Delta}{m})p\cdot P\right ]^{-2} \; . 
\label{eq60} \end{eqnarray}
\noindent Here $p_0^2=m^2-(1/4)M^2$, as earlier.

All of this gives a hint to check, how changes, when inserted in the 
r.h.s. of the Eq.59, the following function 
$$\Phi_{nlm}(p,z)=\frac{{\cal{Y}}_{lm}(\vec{p})}
{[(1+\frac{\Delta^2}{m^2}-2\frac{\Delta}{m}z)p_0^2+p^2+
i(z-\frac{\Delta}{m})p\cdot P]^{n+3}} \; . $$

First of all let us evaluate $\int dq \frac{\Phi_{nlm}(q,z)}
{(p-q)^2}$. According to Eq.29
$$\int dq \frac{\Phi_{nlm}(q,z)}{(p-q)^2}= $$
$$\int dq {\cal{Y}}_{lm}(\vec{q})
\left [(1+\frac{\Delta^2}{m^2}-2\frac{\Delta}{m}z)p_0^2+q^2+
i(z-\frac{\Delta}{m})q\cdot P\right ]^{-(n+3)} \times $$
$$[(p-q)^2]^{-1}=(n+3)\int\limits_0^1 u^{n+2}du\int dq
{\cal{Y}}_{lm}(\vec{q}) \times $$
$$\left[u\{(1+\frac{\Delta^2}{m^2}-2\frac{\Delta}{m}z)p_0^2+q^2+
i(z-\frac{\Delta}{m})q\cdot P \} +(1-u)(p-q)^2 \right ]^{-(n+4)}
\; , $$
\noindent which can be transformed further as
$$u\left[ (1+\frac{\Delta^2}{m^2}-2\frac{\Delta}{m}z)p_0^2+q^2+
i(z-\frac{\Delta}{m})q\cdot P \right ] +(1-u)(p-q)^2 = $$
$$\left [ q+\frac{i}{2}u(z-\frac{\Delta}{m})P-(1-u)p \right ]^2+
u(1-u)p^2+\frac{1}{4}u^2(z-\frac{\Delta}{m})^2P^2+$$
$$up_0^2(1+\frac{\Delta^2}{m^2}-2\frac{\Delta}{m}z)+iu(1-u)
(z-\frac{\Delta}{m})p\cdot P \; , $$
\noindent therefore let us introduce the new $k=q+\frac{i}{2}
u(z-\frac{\Delta}{m})P-(1-u)p$ integration variable:
$$\int dq \frac{\Phi_{nlm}(q,z)}{(p-q)^2}=(n+3)\int\limits_0^1
u^{n+2}du\int dk~{\cal{Y}}_{lm}(\vec{k}+(1-u)\vec{p})
[k^2+u(1-u)p^2+$$ $$ \frac{1}{4}u^2(z-
\frac{\Delta}{m})^2P^2+
up_0^2(1+\frac{\Delta^2}{m^2}-2\frac{\Delta}{m}z)+iu(1-u)
(z-\frac{\Delta}{m})p\cdot P ]^{-(n+4)} \; . $$
\noindent For the solid harmonics the following equation holds
$${\cal{Y}}_{lm}(\vec{a}+\vec{b})= \sum\limits_{k=0}^l
\sum\limits_{\mu=-k}^k  
{\cal{Y}}_{l-k,m-\mu}(\vec{a}){\cal{Y}}_{k\mu}(\vec{b}) \times $$
$$\left [\frac{4\pi(2l+1)(l+m)!(l-m)!}{(2k+1)(2l-2k+1)(k+\mu)!
(k-\mu)!(l+m-k-\mu)!(l-m-k+\mu)!} \right ]^{\frac{1}{2}} \; . $$
\noindent Let us decompose ${\cal{Y}}_{lm}((1-u)\vec{p}+\vec{k})$
according to this relation and take into account that if $l\neq 0$,
then $\int Y_{lm}(\vec{k})d^{(3)}\Omega=0$. Therefore only the first
term
$$\sqrt{4\pi}{\cal{Y}}_{lm}[(1-u)\vec{p}]{\cal{Y}}_{00}(\vec{k})=
{\cal{Y}}_{lm}[(1-u)\vec{p}]=(1-u)^l{\cal{Y}}_{lm}(\vec{p})$$
\noindent of this decomposition will give a nonzero contribution. Using
also $dk=(1/2)tdtd^{(4)}\Omega$, where $t=k^2$, and $\int d^{(4)}\Omega
=2\pi^2$, we will get 
$$\int dq \frac{\Phi_{nlm}(q,z)}{(p-q)^2}=\pi^2 (n+3)~{\cal{Y}}_{lm}
(\vec{p})\int\limits_0^1 u^{n+2}(1-u)^ldu \times $$ $$\int\limits_0^\infty
dt~t\left [ t+u(1-u)p^2+\frac{1}{4}u^2(z-\frac{\Delta}{m})^2P^2 +
\right . $$ $$ \left . up_0^2(1+\frac{\Delta^2}{m^2}-2\frac{\Delta}{m}z)+
iu(1-u)(z-\frac{\Delta}{m})p\cdot P \right ]^{-(n+4)} \; . $$
\noindent Over $t$ the integral is of the type
$$\int\limits_0^\infty \frac{tdt}{[t+a]^{n+4}}=
\int\limits_0^\infty \frac{dt}{[t+a]^{n+3}}-
a\int\limits_0^\infty \frac{dt}{[t+a]^{n+4}}=
\frac{1}{(n+2)(n+3)}a^{-(n+2)} \; . $$
\noindent That is
$$\int dq \frac{\Phi_{nlm}(q,z)}{(p-q)^2}=
\frac{\pi^2}{n+2}~{\cal{Y}}_{lm}(\vec{p})\int\limits_0^1(1-u)^l
\left [ (1-u)p^2+\frac{1}{4}u(z-\frac{\Delta}{m})^2P^2 +
\right . $$ $$ \left . p_0^2(1+\frac{\Delta^2}{m^2}-2\frac{\Delta}{m}z)+
i(1-u)(z-\frac{\Delta}{m})p\cdot P \right ]^{-(n+2)}du \; . $$
\noindent or, after introduction of the new $t=1-u$ integration variable:
$$\int dq \frac{\Phi_{nlm}(q,z)}{(p-q)^2}=
\frac{\pi^2}{n+2}~{\cal{Y}}_{lm}(\vec{p})\int\limits_0^1t^l
\left [ t\left \{ p^2+i(z-\frac{\Delta}{m})p\cdot P- 
\right .\right . $$
$$\left . \left . \frac{1}{4}(z-\frac{\Delta}{m})^2P^2 \right \}+
p_0^2(1+\frac{\Delta^2}{m^2}-2\frac{\Delta}{m}z)+
\frac{1}{4}(z-\frac{\Delta}{m})^2P^2 \right ]^{-(n+2)}dt \; . $$
\noindent Let us denote
$$a=(1+\frac{\Delta^2}{m^2}-2\frac{\Delta}{m}z)p_0^2+p^2+
i(z-\frac{\Delta}{m})p\cdot P $$
\noindent and
$$b=(1+\frac{\Delta^2}{m^2}-2\frac{\Delta}{m}z)p_0^2+
\frac{1}{4}(z-\frac{\Delta}{m})^2P^2 \; . $$
\noindent We have
$$\int\limits_0^1 \frac{t^ldt}{[t(a-b)+b]^{n+2}}=(-1)^l\frac{(n+1-l)!}
{(n+1)!}~\frac{\partial^l}{\partial a^l}\int\limits_0^1 \frac{dt}
{[t(a-b)+b]^{n-l+2}}=$$
$$(-1)^l\frac{(n+1-l)!}{(n+1)!}~\frac{\partial^l}{\partial a^l}
\frac{1}{(a-b)^{n-l+2}}\int\limits_0^1 \frac{dt}
{t+b/(a-b)]^{n-l+2}}=$$
$$(-1)^l\frac{(n-l)!}{(n+1)!}~\frac{\partial^l}{\partial a^l}
~\frac{1}{a-b}~\left [\frac{1}{b^{n-l+1}}-\frac{1}{a^{n-l+1}} 
\right]= $$
$$(-1)^l\frac{(n-l)!}{(n+1)!}~\frac{\partial^l}{\partial a^l}~
\frac{a^{n-l}+a^{n-l-1}b+a^{n-l-2}b^2+\dots +b^{n-l}}
{a^{n-l+1}b^{n-l+1}}=$$
$$(-1)^l\frac{(n-l)!}{(n+1)!}~\frac{\partial^l}{\partial a^l}~
\sum\limits_{k=0}^{n-l}a^{-(n-l+1-k)}b^{-(k+1)}=$$
$$\frac{(n-l)!}{(n+1)!}~\sum\limits_{k=0}^{n-l}
\frac{(n-k)!}{(n-l-k)!}a^{-(n-k+1)}b^{-(k+1)} \; . $$
\noindent where $n \ge l$ supposition was done. Thus
$$\int dq \frac{\Phi_{nlm}(q,z)}{(p-q)^2}=\pi^2\frac{(n-l)!}
{(n+2)!}\sum\limits_{k=0}^{n-l}\frac{(n-k)!}{(n-l-k)!}\times $$
$$\left[ p_0^2(1+\frac{\Delta^2}{m^2}-2\frac{\Delta}{m}z)+
\frac{1}{4}(z-\frac{\Delta}{m})^2P^2 \right]^{-(k+1)}\times $$
$$\frac{{\cal Y}_{lm}(\vec{p})}
{[(1+\frac{\Delta^2}{m^2}-2\frac{\Delta}{m}z)p_0^2+p^2+
i(z-\frac{\Delta}{m})p\cdot P]^{n-k+1}} \; , $$
\noindent but
$${\cal Y}_{lm}(\vec{p})\left[
(1+\frac{\Delta^2}{m^2}-2\frac{\Delta}{m}z)p_0^2+p^2+
i(z-\frac{\Delta}{m})p\cdot P \right]^{-(n-k+1)}=$$
$$\Phi_{n-k-2,lm}(p,z) \; , $$
\noindent therefore
$$\int dq \frac{\Phi_{nlm}(q,z)}{(p-q)^2}=\pi^2\frac{(n-l)!}
{(n+2)!}\sum\limits_{k=0}^{n-l}\frac{(n-k)!}{(n-l-k)!}\times $$
$$\frac{\Phi_{n-k-2,lm}(p,z)}{\left[
p_0^2(1+\frac{\Delta^2}{m^2}-2\frac{\Delta}{m}z)+
\frac{1}{4}(z-\frac{\Delta}{m})^2P^2 \right]^{k+1} } \; . $$

The result is encouraging, because we have got a linear combination of
$\Phi_{nlm}$ functions, and therefore it can be expected that the
solution of the Eq.59 is expressible through these functions. But, before
this conclusion is done, we should check that nothing wrong happens
after multiplication over the 
$$[(m+\Delta)^2+(p-i\eta_1P)^2 ]^{-1}
[(m-\Delta)^2+(p+i\eta_2P)^2 ]^{-1} \; . $$
According to formulas (60) and (29):
$$[(m+\Delta)^2+(p-i\eta_1P)^2 ]^{-1}
[(m-\Delta)^2+(p+i\eta_2P)^2 ]^{-1} \times $$
$$\left[
(1+\frac{\Delta^2}{m^2}-2\frac{\Delta}{m}z)p_0^2+p^2+
i(z-\frac{\Delta}{m})p\cdot P \right]^{-(n-k+1)}= $$
$$\frac{1}{2}(n-k+1)(n-k+2)\int\limits_{-1}^1dt\int\limits_0^1 dx 
\times $$ $$x(1-x)^{n-k}\left [ ix(t-z)p\cdot P+
(1+\frac{\Delta^2}{m^2}-2\frac{\Delta}{m}z)p_0^2+p^2+ \right . $$
$$\left . i(z-\frac{\Delta}{m})p\cdot P-2\frac{\Delta}{m}x(t-z)p_0^2 
\right ]^{-(n-k+3)}= $$
$$\frac{1}{2}(n-k+1)(n-k+2)\int\limits_{-1}^1dt\int\limits_0^1 dx \times $$
$$\frac{x(1-x)^{n-k}}{\{ [1+\frac{\Delta^2}{m^2}-2\frac{\Delta}{m}
(xt+z-xz)]p_0^2+p^2+i(xt+z-xz-\frac{\Delta}{m})p\cdot P\}^{n-k+3}}\; ,$$
\noindent therefore
$$[ (m+\Delta)^2+(p-i\eta_1P)^2 ]^{-1}
[(m-\Delta)^2+(p+i\eta_2P)^2 ]^{-1}
\int dq \frac{\Phi_{nlm}(q,z)}{(p-q)^2}= $$
$$\frac{\pi^2}{2}\sum\limits_{k=0}^{n-l}\frac{(n-l)!(n-k+2)!}
{(n+2)!(n-l-k)!}\int\limits_{-1}^1dt \times $$
$$\int\limits_0^1\frac{x(1-x)^{n-k}\Phi_{n-k,lm}
(p,xt+z-xz)dx}{\left[
p_0^2(1+\frac{\Delta^2}{m^2}-2\frac{\Delta}{m}z)+
\frac{1}{4}(z-\frac{\Delta}{m})^2P^2 \right]^{k+1} } \; . $$
\noindent Note, that when $0\le x \le 1$, then ${\rm min}(z,t) \le
xt+z-xz \le {\rm max}(z,t)$, therefore if the $z$-parameter of the 
$\Phi_{nlm}(p,z)$ function changes in the $-1\le z \le 1$ range, then 
$-1\le xt+z-xz \le 1$ and the above given equation can be rewritten as
$$[ (m+\Delta)^2+(p-i\eta_1P)^2 ]^{-1}
[(m-\Delta)^2+(p+i\eta_2P)^2 ]^{-1}
\int dq \frac{\Phi_{nlm}(q,z)}{(p-q)^2}= $$
$$\frac{\pi^2}{2}\sum\limits_{k=0}^{n-l}\frac{(n-l)!(n-k+2)!}
{(n+2)!(n-l-k)!}\int\limits_{-1}^1dt\int\limits_0^1dx~x(1-x)^{n-k}
\int\limits_{-1}^1d\zeta \times $$
$$\frac{\delta (\zeta-xt-z+xz)}{\left[
p_0^2(1+\frac{\Delta^2}{m^2}-2\frac{\Delta}{m}z)+
\frac{1}{4}(z-\frac{\Delta}{m})^2P^2 \right]^{k+1} }\Phi_{n-k,lm}
(p,\zeta) \; . $$
\noindent So we see that if $\Phi_{nlm}(q,z)$ is substituted in the 
r.h.s. of the Eq.59, the result will be a superposition of the same 
kind functions. Therefore the solution of Eq.59 can be expressed in the 
form
$$\Phi_{nlm}(p)=\sum\limits_{k=0}^{n-l}\int\limits_{-1}^1
g_{nl}^k(z)\Phi_{n-k,lm}(p,z)dz \; . $$
\noindent Then
$$[ (m+\Delta)^2+(p-i\eta_1P)^2 ]^{-1}
[(m-\Delta)^2+(p+i\eta_2P)^2 ]^{-1}
\int dq \frac{\Phi_{nlm}(q,z)}{(p-q)^2}= $$
$$\sum\limits_{\nu=0}^{n-l}\int\limits_{-1}^1 dz g_{nl}^\nu(z)
[ (m+\Delta)^2+(p-i\eta_1P)^2 ]^{-1} \times $$
$$[(m-\Delta)^2+(p+i\eta_2P)^2 ]^{-1}
\int dq \frac{\Phi_{n-\nu,lm}(q,z)}{(p-q)^2}= $$
$$\sum\limits_{\nu=0}^{n-l}\sum\limits_{\tau=0}^{n-l-\nu}
\frac{\pi^2}{2}\frac{(n-l-\nu)!(n-\nu-\tau+2)!}
{(n-\nu+2)!(n-\nu-\tau-l)!} \times $$
$$\int\limits_{-1}^1 dz g_{nl}^\nu(z)\int\limits_{-1}^1dt
\int\limits_0^1dx~x(1-x)^{n-\nu-\tau}\times $$
$$\int\limits_{-1}^1d\zeta
\frac{\delta (\zeta-xt-z+xz)\Phi_{n-\nu-\tau,lm}(p,\zeta)}{\left[
p_0^2(1+\frac{\Delta^2}{m^2}-2\frac{\Delta}{m}z)+
\frac{1}{4}(z-\frac{\Delta}{m})^2P^2 \right]^{\tau+1} } \; . $$
\noindent Let us denote $k=\nu+\tau$ and take into account that
$$\sum\limits_{\nu=0}^N\sum\limits_{\tau=0}^{N-\nu}A(\nu,\tau)=
\sum\limits_{k=0}^N\sum\limits_{\nu=0}^{k}A(\nu,k-\nu)$$
\noindent therefore Eq.59 can be rewritten as
$$\sum\limits_{k=0}^{n-l}\int\limits_{-1}^1
g_{nl}^k(\zeta)\Phi_{n-k,lm}(p,\zeta)d\zeta=$$
$$\frac{\lambda}{2}\sum\limits_{k=0}^{n-l}\sum\limits_{\nu=0}^{k}
\frac{(n-l-\nu)!(n-k+2)!}{(n-\nu+2)!(n-l-k)!}
\int\limits_{-1}^1 dz g_{nl}^\nu(z)\int\limits_{-1}^1dt
\int\limits_0^1dx~x(1-x)^{n-k}\times $$
$$\int\limits_{-1}^1d\zeta
\frac{\delta (\zeta-xt-z+xz)\Phi_{n-k,lm}(p,\zeta)}{\left[
p_0^2(1+\frac{\Delta^2}{m^2}-2\frac{\Delta}{m}z)+
\frac{1}{4}(z-\frac{\Delta}{m})^2P^2 \right]^{k-\nu+1} } \; , $$
\noindent from this we get the following system of non-homogeneous
integral equations for the $g_{nl}^k(z)$ coefficient functions:
$$g_{nl}^k(\zeta)=\frac{\lambda}{2}\sum\limits_{\nu=0}^{k}         
\frac{(n-l-\nu)!(n-k+2)!}{(n-\nu+2)!(n-l-k)!}           
\int\limits_{-1}^1dt  \int\limits_0^1dx~x(1-x)^{n-k}\times \vspace*{-4mm}$$
\begin{equation}
\int\limits_{-1}^1
\frac{\delta (\zeta-xt-z+xz)g_{nl}^\nu(z)dz}{\left[       
p_0^2(1+\frac{\Delta^2}{m^2}-2\frac{\Delta}{m}z)+
\frac{1}{4}(z-\frac{\Delta}{m})^2P^2 \right]^{k-\nu+1} } \; , 
\label{eq61} \end{equation} 
\noindent Note that we have interchanged integrations over $dz$ and
$d\zeta$. If now integration over $dz$ is further interchanged with
integrations over $dx$ and $dt$, Eq.61 can be rewritten as (after
$z \leftrightarrow \zeta$ replacement)
$$g_{nl}^k(z)=\frac{\lambda}{2}\sum\limits_{\nu=0}^{k}
\frac{(n-l-\nu)!(n-k+2)!}{(n-\nu+2)!(n-l-k)!}\times $$
$$\int\limits_{-1}^1      
\frac{I(z,\zeta)g_{nl}^\nu(\zeta)d\zeta}{\left[       
p_0^2(1+\frac{\Delta^2}{m^2}-2\frac{\Delta}{m}\zeta)+
\frac{1}{4}(\zeta-\frac{\Delta}{m})^2P^2 \right]^{k-\nu+1} } \; , $$
\noindent where
$$I(z,\zeta)=\int\limits_{-1}^1dt\int\limits_0^1dx~x(1-x)^{n-k}
\delta (z-xt-\zeta+x\zeta)=$$
$$\int\limits_0^1dx~x(1-x)^{n-k}\int\limits_{-1}^1
\delta (z-xt-\zeta+x\zeta)dt \; . $$
\noindent Let us introduce instead of t the new $y=xt+(1-x)\zeta$
integration variable:
$$I(z,\zeta)=\int\limits_0^1dx~(1-x)^{n-k}
\int\limits_{-x+(1-x)\zeta}^{x+(1-x)\zeta}\delta(z-y)dy=
\int\int_D (1-x)^{n-k}\delta(z-y)dxdy \; . $$
\noindent The integration area $D=D_1 \cup D_2$ is shown in Fig. 11.
\begin{figure}[htbp]
\begin{center}
\epsfig{figure=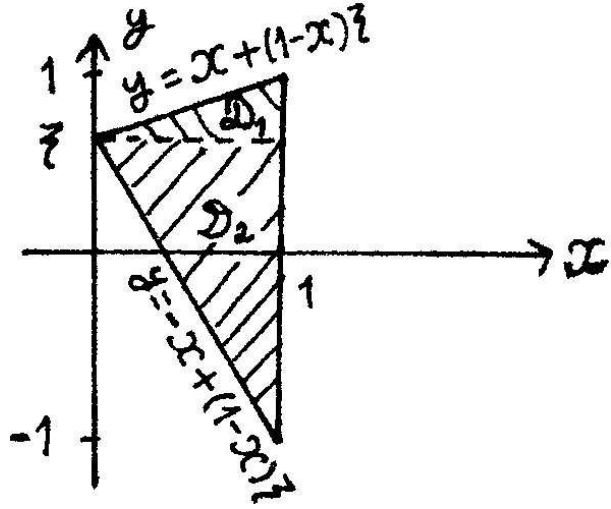,width=8cm}
\caption{The integration domain.}
\end{center}
\label{Fig11}
\end{figure}

It is easy to calculate integrals over $D_1$ and $D_2$:
$$\int\int_{D_1}(1-x)^{n-k}\delta(z-y)dxdy =\int\limits_\zeta^1
dy\delta(z-y)\int\limits_{\frac{y-\zeta}{1-\zeta}}^1(1-x)^{n-k}dx=$$
$$\frac{1}{n-k+1}\int\limits_\zeta^1\delta(z-y)\left (
\frac{1-y}{1-\zeta}\right )^{n-k+1}dy=$$
$$\frac{1}{n-k+1}\left ( \frac{1-z}{1-\zeta}
\right )^{n-k+1}\Theta(z-\zeta) \; , $$
\noindent and
$$\int\int_{D_2}(1-x)^{n-k}\delta(z-y)dxdy =\int\limits^\zeta_{-1}
dy\delta(z-y)\int\limits_{\frac{\zeta-y}{1+\zeta}}^1(1-x)^{n-k}dx=$$
$$\frac{1}{n-k+1}\int\limits^\zeta_{-1}\delta(z-y)\left ( 
\frac{1+y}{1+\zeta}\right )^{n-k+1}dy=$$
$$\frac{1}{n-k+1}\left (\frac{1+z}{1+\zeta}
\right )^{n-k+1}\Theta(\zeta-z) \; , $$
\noindent Therefore
$$I(z,\zeta)=\frac{1}{n-k+1}\left [\left (\frac{1-z}{1-\zeta}
\right )^{n-k+1}\Theta(z-\zeta)+\left ( \frac{1+z}{1+\zeta}
\right )^{n-k+1}\Theta(\zeta-z)\right ]\; . $$
\noindent Let us introduce
$$R(z,\zeta)=\frac{1-z}{1-\zeta}
\Theta(z-\zeta)+\frac{1+z}{1+\zeta}\Theta(\zeta-z) $$
\noindent and define the $\Theta$-function at the origin as $\Theta (0) 
=\frac{1}{2}$, then $I(z,\zeta)=\frac{[R(z,\zeta)]^{n-k+1}}{n-k+1}$ 
and our system of integral 
equations  can be rewritten as  
$$ g_{nl}^k(z)=\frac{\lambda}{2}\sum\limits_{\nu=0}^{k}
\frac{(n-l-\nu)!(n-k)!}{(n-\nu+2)!(n-l-k)!} \vspace*{-4mm} $$
\begin{eqnarray}
\int\limits_{-1}^1      
\frac{(n-k+2)[R(z,\zeta)]^{n-k+1}g_{nl}^\nu(\zeta)d\zeta}{\left[       
p_0^2(1+\frac{\Delta^2}{m^2}-2\frac{\Delta}{m}\zeta)+
\frac{1}{4}(\zeta-\frac{\Delta}{m})^2P^2 \right]^{k-\nu+1} } \; ,   
\label{eq62} \end{eqnarray}
\noindent In particular, $g_{nl}^0(z)\equiv g_n(z)$ satisfies a homogeneous
integral equation:
\begin{eqnarray}
g_n(z)=\frac{\lambda}{2(n+1)}\int\limits_{-1}^1      
\frac{[R(z,\zeta)]^{n+1}g_n(\zeta)d\zeta}{
p_0^2(1+\frac{\Delta^2}{m^2}-2\frac{\Delta}{m}\zeta)+
\frac{1}{4}(\zeta-\frac{\Delta}{m})^2P^2 } \; .
\label{eq63} \end{eqnarray}
\noindent Note that $R(1,\zeta)=R(-1,\zeta)=0$, if $|\zeta|\neq 1$. 
Therefore $g_n(-1)=g_n(1)=0$.

\noindent From Eq.63 $\lambda$ eigenvalues are determined. Inverting
$\lambda = \lambda (M)$ dependence, we get a mass spectrum $M=M(\lambda)$.

Let us see, to which second order differential equation is equivalent our
integral equation (63). We have $\frac{d}{dz}R(z,\zeta)=\Theta(\zeta-z)
\frac{1}{1+\zeta}-\Theta(z-\zeta)\frac{1}{1-\zeta}$ and $\frac{d^2}{dz^2}
R(z,\zeta)=-\frac{2}{1-z^2}\delta(z-\zeta)$. Furthermore
$$(1-z^2)\left[ \frac{dR}{dz} \right]^2+2zR\frac{dR}{dz}-R^2=
\left[ \frac{dR}{dz} \right]^2-\left ( R-z\frac{dR}{dz}\right )^2=$$
$$\left [(1-z)\frac{dR}{dz}+R \right ]\left [(1+z)\frac{dR}{dz}-R 
\right ]=-\frac{4}{1-\zeta^2}\Theta(z-\zeta)\Theta(\zeta-z)\; . $$
\noindent The r.h.s. differs from zero only at the $z=\zeta$ point. So
$$(1-z^2)\left[ \frac{dR(z,\zeta)}{dz} \right]^2+
2zR(z,\zeta)\frac{dR(z,\zeta)}{dz}-R^2(z,\zeta)=$$
$$-\frac{4}{1-\zeta^2}\Theta(z-\zeta)\Theta(\zeta-z)\; . $$
\noindent Since
$$\left \{ (1-z^2)\frac{d^2}{dz^2}+2z(n-1)\frac{d}{dz}-n(n-1) \right \}
R^n(z,\zeta)=$$
$$n(n-1)R^{n-2}\left \{ \left[ \frac{dR}{dz} \right]^2+2zR\frac{dR}{dz}
-R^2\right \}+n(1-z^2)R^{n-1}\frac{d^2R}{dz^2} \; , $$
\noindent and $R(z,z)=1$, we get finally
$$ \left \{ (1-z^2)\frac{d^2}{dz^2}+2z(n-1)\frac{d}{dz}-n(n-1) \right \}
R^n(z,\zeta)= \vspace*{-4mm} $$
\begin{eqnarray}
-2n\delta(z-\zeta)-\frac{4n(n-1)}{1-z^2}\Theta(z-\zeta)
\Theta(\zeta-z)\; .  
\label{eq64} \end{eqnarray}
Using this equality, we can transform Eq.63 into a differential equation
(note that if $f(\zeta)$ is a normal function, without $\delta(z-\zeta)$
type singularities, then $\int_{-1}^1 \Theta(z-\zeta) \Theta(\zeta-z)
f(\zeta)d\zeta=0$)
$$(1-z^2)\frac{d^2g_n}{dz^2}+2nz\frac{dg_n}{dz}-n(n+1)g_n+ \vspace*{-3mm}$$
\begin{eqnarray}
\frac{\lambda}{p_0^2(1+\frac{\Delta^2}{m^2}-2\frac{\Delta}{m}z)+
\frac{1}{4}(z-\frac{\Delta}{m})^2P^2 }g_n=0 \; .  
\label{eq65}\end{eqnarray}
\noindent But
$$p_0^2\left(1+\frac{\Delta^2}{m^2}-2\frac{\Delta}{m}z\right)+
\frac{1}{4}\left(z-\frac{\Delta}{m}\right)^2P^2=$$
$$\left(m^2-\frac{1}{4}M^2\right)\left(1+\frac{\Delta^2}{m^2}-
2\frac{\Delta}{m}z\right)+\frac{1}{4}(z-\frac{\Delta}{m})^2M^2=$$
$$m^2\left(1+\frac{\Delta^2}{m^2}-2\frac{\Delta}{m}z\right)-
\frac{1}{4}M^2(1-z^2) \; , $$
\noindent therefore Eq.65 can be rewritten also as
$$(1-z^2)\frac{d^2g_n}{dz^2}+2nz\frac{dg_n}{dz}+ \vspace*{-3mm}$$ 
\begin{eqnarray}
\left [
 \frac{\lambda}{m^2(1+\frac{\Delta^2}{m^2}-2\frac{\Delta}{m}z)-
\frac{1}{4}M^2(1-z^2) }-n(n+1) \right ] g_n=0 \; . 
\label{eq66}\end{eqnarray} 

From This equation, when $\Delta=0$, we get already known to us Eq.58.
It turns out that even in a general case of unequal masses Eq.66 can be
transformed into an equation of the Eq.58 type by suitable change of
variables. To guess this variable change, it is better to rewrite Eq.66
in the form of Eq.56. Introducing $\phi_n(z)=(1+z)^{-1}(1-z^2)^{-n/2}
g_n(z)$ function, we get  
$$(1-z^2)\frac{d^2\varphi_n}{dz^2}+2(1-z)\frac{d\varphi_n}{dz}+$$ 
$$\left [ \frac{\lambda}{m^2(1+\frac{\Delta^2}{m^2}-2\frac{\Delta}{m}z)-
\frac{1}{4}M^2(1-z^2) }-\frac{n(n+2)}{1-z^2} \right ] \varphi_n=0 \; . $$
\noindent and after the $t=\frac{1-z}{1+z}$ variable change
(let us note that
$\frac{d}{dz}=-\frac{(1+t)^2}{2}~\frac{d}{dt}$  and
$\frac{d^2}{dz^2}=\frac{(1+t)^4}{4}~\frac{d^2}{dt^2}+\frac{(1+t)^3}{2}~
\frac{d}{dt}$ ): 
$$\frac{d^2\varphi_n}{dt^2}+\frac{1}{t^2}
\left [ \frac{\lambda}{m^2\left[\frac{(1+t)^2}{t}\left(1+\frac{\Delta^2}
{m^2}\right)-2\frac{\Delta}{m}~\frac{1-t^2}{t}\right]-M^2}
-\frac{n(n+2)}{4} \right ] \varphi_n=0 \; . $$
\noindent But
$$\frac{1}{t}m^2\left[(1+t)^2\left(1+\frac{\Delta^2}
{m^2}\right)-2\frac{\Delta}{m}(1-t^2)\right]=$$
$$\frac{1}{t}m^2\left[t^2\left(1+\frac{\Delta}{m}\right)^2+
\left(1-\frac{\Delta}{m}\right)^2+2t\left(1+\frac{\Delta^2}{m^2}\right)
\right]=$$
$$\frac{1}{t}m^2\left\{\left[t\left(1+\frac{\Delta}{m}\right)+
\left(1-\frac{\Delta}{m}\right)\right]^2+4t\frac{\Delta^2}{m^2}\right\}=$$
$$4\Delta^2+m^2\left(1-\frac{\Delta^2}{m^2}\right)
\frac{[1+((1+\frac{\Delta}{m})/(1-\frac{\Delta}{m}))t]^2}
{((1+\frac{\Delta}{m})/(1-\frac{\Delta}{m}))t} \; , $$
\noindent therefore, if one more variable change $\tilde t=\frac
{1+\Delta/m}{1-\Delta/m}~t$ is made, we get an equation of the Eq.56
type:
\begin{equation}
\frac{d^2\varphi_n}{d\tilde t^2}+
\left [ \frac{\lambda/\left(1-\frac{\Delta^2}{m^2}\right)}
{\tilde t [m^2(1+\tilde t)^2-\left(
(M^2-4\Delta^2)/\left(1-\frac{\Delta^2}{m^2}\right)\right) \tilde t]}
-\frac{n(n+2)}{4\tilde t^2} \right ] \varphi_n=0 \; .
\label{eq67} \end{equation}

As we see, it is enough to find $\lambda=F(M^2)$ spectrum for equal
masses. Then the spectrum for the unequal masses case can be found
from the relation
\begin{equation}
\frac{\lambda}{1-\frac{\Delta^2}{m^2}}=F\left ( \frac{M^2-4\Delta^2}
{1-\frac{\Delta^2}{m^2}}\right ) \; .
\label{eq68} \end{equation}
\noindent Note, that $t \to \tilde t$ variable change implies the
following transformation for the variable $z: \; \; z \to \tilde z=
\frac{1-\tilde t}{1+\tilde t}=\frac{z-\frac{\Delta}{m}}
{1-\frac{\Delta}{m}z}$.

Using Eq.64, we can transform Eq.62 also into a system of second order
differential equations 
$$\left \{ (1-z^2)\frac{d^2}{dz^2}+2(n-k)z\frac{d}{dz}-(n-k)(n-k+1)
\right \}g_{nl}^k(z)+$$
$$\lambda \sum\limits_{\nu=0}^{k}
\frac{(n-l-\nu)!(n-k+2)!}{(n-\nu+2)!(n-l-k)!} \times $$
$$\frac{g_{nl}^\nu(z)}{[m^2(1+\frac{\Delta^2}{m^2}-2\frac{\Delta}{m}z)-
\frac{1}{4}M^2(1-z^2)]^{k-\nu+1} }=0 \;. $$

\section*{Variable separation in bipolar coordinates}
There exists one more method for Eq.27 investigation. It is based on the 
following idea: transform integral Wick-Cutkosky equation into a partial
differential equation and try to separate variables in some special 
coordinate system.

Transition from the integral to the differential equation can be done 
by using the fact that in a 4-dimensional Euclidean space
\begin{equation}
\Delta \frac{1}{x^2}=-4\pi^2\delta(x) \; . 
\label{eq69} \end{equation}
Let us show that this is indeed correct. Derivatives of $\frac{1}{x^2}$
are singular at the $x=0$ point. Therefore $\Delta \frac{1}{x^2}$ should
be specially defined at this point, for example, as
$$\Delta \frac{1}{x^2}=\lim_{\epsilon \to 0}~\Delta \frac{1}
{x^2+\epsilon} \; . $$
\noindent
But $\frac{\partial}{\partial x_i}\frac{1}{x^2+\epsilon}=
-\frac{2x_i}{(x^2+\epsilon)^2}$ and
$$\Delta \frac{1}{x^2+\epsilon}=\frac{\partial^2}{\partial x_i 
\partial x_i}\frac{1}{x^2+\epsilon}=-\frac{8}{(x^2+\epsilon)^2}+
\frac{8x^2}{(x^2+\epsilon)^3}=-\frac{8\epsilon}{(x^2+\epsilon)^3}
\; . $$
\noindent Therefore
$$\lim_{\epsilon \to 0}~\Delta \frac{1}{x^2+\epsilon}=
\left \{  \matrix {
0 \; , \; \; {\rm if} \; \; x \neq 0 \cr
-\infty \; , \; \; {\rm if} \; \; x=0 } \right . $$
This suggests that $\Delta \frac{1}{x^2}$ is proportional to $-\delta(x)$.
But
$$-8\epsilon\int \frac{dx}{(x^2+\epsilon)^3}=-16\epsilon\pi^2
\int\limits_0^\infty \frac{r^3dr}{(r^2+\epsilon)^3}=
-8\epsilon\pi^2\int\limits_0^\infty \frac{tdt}{(t+\epsilon)^3}=
-4\pi^2 \; , $$
\noindent so Eq.69 holds.

Acting on the both sides of Eq.27 by the operator $\Delta_p=\frac
{\partial^2}{\partial p_1^2}+\frac{\partial^2}{\partial p_2^2}+
\frac{\partial^2}{\partial p_3^2}+\frac{\partial^2}{\partial p_4^2}$ and
using $\Delta_p \frac{1}{(p-q)^2}=-4\pi^2\delta(p-q)$, we get
$$\Delta_p [m_1^2+{\vec{p}}\,^2+(p_4-i\eta_1M)^2]
[m_2^2+{\vec{p}}\,^2+(p_4+i\eta_2M)^2] \Phi (p)=
-4\lambda\Phi (p) \; , $$
\noindent or, after the introduction of a new function 
$$\Psi(p)=[m_1^2+{\vec{p}}\,^2+(p_4-i\eta_1M)^2]
[m_2^2+{\vec{p}}\,^2+(p_4+i\eta_2M)^2] \Phi (p) \; ,  $$
\begin{equation}
\Delta \Psi(p)+\frac{4\lambda}{[m_1^2+{\vec{p}}\,^2+(p_4-i\eta_1M)^2]
[m_2^2+{\vec{p}}\,^2+(p_4+i\eta_2M)^2]}\Psi(p)=0 \; .
\label{eq70} \end{equation}

So the partial differential equation is found. Now we should care about
variable separation. It turns out that so called bipolar coordinates can 
be used for this goal.

On a plane, the bipolar coordinates $\tau,\alpha$ are defined as follows:
$\alpha$ is the angle indicated in Fig.12 and $\exp{\tau}=\frac{r_1}{r_2}$.
\begin{figure}[htb]
\begin{center}
\epsfig{figure=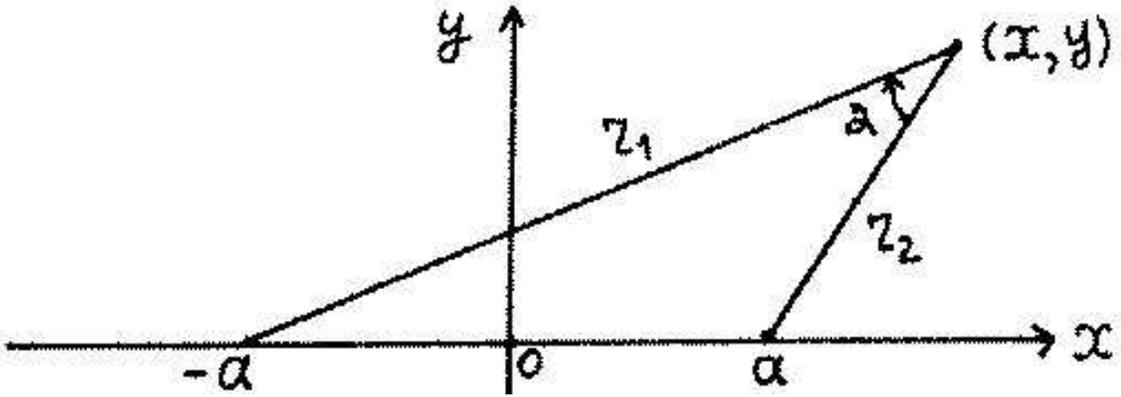,width=15cm}
\caption{Bipolar coordinates.}
\end{center}
\label{Fig12}
\end{figure}

Because $r_1=\sqrt{(x+a)^2+y^2}$ and $r_2=\sqrt{(x-a)^2+y^2}$, this 
definition implies
$$\tau=\frac{1}{2}\ln{\frac{(x+a)^2+y^2}{(x-a)^2+y^2}} \; , \; \; \;
\alpha=\arccos{\frac{x^2+y^2-a^2}{\sqrt{(x+a)^2+y^2}\sqrt{(x-a)^2+y^2}}}
\; . $$

To inverse these relations, let us note that
$$\cosh{\tau}=\frac{1}{2}(e^\tau+e^{-\tau})=\frac{x^2+y^2+a^2}
{r_1r_2} \; , \; \; \sinh{\tau}=\sqrt{\cosh^2{\tau}-1}=
\frac{2ax}{r_1r_2} \; , $$
$$\cos{\alpha}=\frac{x^2+y^2-a^2}{r_1r_2} \; , $$
$$\sin{\alpha}=\sqrt{1-\cos^2{\alpha}}=\frac{2ay}{r_1r_2} \; \;
{\rm and} \; \; \cosh{\tau}-\cos{\alpha}=\frac{2a^2}{r_1r_2}\; . $$
\noindent So
$$x=\frac{a\sinh{\tau}}{\cosh{\tau}-\cos{\alpha}} \; , \; \;
y=\frac{a\sin{\alpha}}{\cosh{\tau}-\cos{\alpha}} \; . $$

In a 4-dimensional Euclidean momentum space the bipolar coordinates are
defined through
$$P_1=\frac{a\sin{\alpha}}{\cosh{\tau}-\cos{\alpha}}\sin{\Theta}
\cos{\varphi} \; , \; \;
P_2=\frac{a\sin{\alpha}}{\cosh{\tau}-\cos{\alpha}}\sin{\Theta}
\sin{\varphi} \; ,   $$
\begin{equation} 
P_3=\frac{a\sin{\alpha}}{\cosh{\tau}-\cos{\alpha}}\cos{\Theta}
\;, \; \;
P_4=\frac{a\sinh{\tau}}{\cosh{\tau}-\cos{\alpha}} \; .
\label{eq71} \end{equation}
\noindent It is convenient to choose the $a$ parameter in such a way to
have
$$a^2=m_1^2-\eta_1^2M^2=m_2^2-\eta_2^2M^2 \; . $$
\noindent If $\eta_i=\frac{m_i}{m_1+m_2}$ (as it was assumed so far), this
is impossible. But let us recall from the BS equation derivation that 
$\eta_1$ and $\eta_2$ are subject to only one $\eta_1+\eta_2$ condition.
In other respects they are arbitrary. So let us demand $m_1^2-\eta_1^2M^2=
m_2^2-\eta_2^2M^2$, which gives
$$\eta_1=\frac{M^2+m_1^2-m_2^2}{2M^2} \; , \; \; 
\eta_2=\frac{M^2-m_1^2+m_2^2}{2M^2} \; , $$
$$ a=\frac{\sqrt{[(m_1+m_2)^2-M^2][M^2-(m_1-m_2)^2]}}{2M} \; . $$

Now Eq.70 should be rewritten in the new coordinates. We have
$$m_1^2+\vec{p}\, ^2+(p_4-i\eta_1M)^2=a^2+\frac{a^2\sin^2{\alpha}}
{(\cosh{\tau}-\cos{\alpha})^2}+\frac{a^2\sinh^2{\tau}}
{(\cosh{\tau}-\cos{\alpha})^2}-$$
$$2i\eta_1M\frac{a\sinh{\tau}}{\cosh{\tau}-\cos{\alpha}}=a^2+
a^2\frac{1-\cos^2{\alpha}}{(\cosh{\tau}-\cos{\alpha})^2}+
a^2\frac{\cosh^2{\tau}-1}{(\cosh{\tau}-\cos{\alpha})^2}-$$
$$2i\eta_1M\frac{a\sinh{\tau}}{\cosh{\tau}-\cos{\alpha}}=
\frac{2a}{\cosh{\tau}-\cos{\alpha}}(a\cosh{\tau}-i\eta_1M\sinh{\tau})
\;  . $$
Analogously $m_2^2+\vec{p}\, ^2+(p_4+i\eta_2M)^2=
\frac{2a}{\cosh{\tau}-\cos{\alpha}}(a\cosh{\tau}+i\eta_2M\sinh{\tau})
$. Therefore
$$[m_1^2+{\vec{p}}\,^2+(p_4-i\eta_1M)^2]
[m_2^2+{\vec{p}}\,^2+(p_4+i\eta_2M)^2]= $$
$$ \frac{4a^2}{b(\cosh{\tau}-\cos{\alpha})^2}
(a\cosh{\tau}-i\eta_1M\sinh{\tau})(a\cosh{\tau}+i\eta_2M\sinh{\tau})
\; . $$
Now we should express the Laplacian in the bipolar coordinates. Denoting
$r=\frac{a\sin{\alpha}}{\cosh{\tau}-\cos{\alpha}}$, we have
$$\Delta_p=\frac{\partial^2}{\partial p_4^2}+\frac{1}{r}
\frac{\partial^2}{\partial r^2}r-\frac{\delta^{(3)}}{r^2}=
\frac{1}{r}\left ( \frac{\partial^2}{\partial p_4^2}+
\frac{\partial^2}{\partial r^2}\right )r-\frac{\delta^{(3)}}{r^2}
\; . $$
\noindent Here $\delta^{(3)}$ is the angular part of the 3-dimensional
Laplacian $\Delta^{(3)}=\frac{\partial^2}{\partial p_1^2}+
\frac{\partial^2}{\partial p_2^2}+\frac{\partial^2}{\partial p_3^2}$.
Let $z=p_4+ir$, then $\frac{\partial}{\partial \bar z}=\frac{1}{2}\left (
\frac{\partial}{\partial p_4}+i\frac{\partial}{\partial r} \right ),\;$
$\frac{\partial}{\partial z}=\frac{1}{2}\left (
\frac{\partial}{\partial p_4}-i\frac{\partial}{\partial r} \right ),$ and
$\frac{\partial^2}{\partial p_4^2}+\frac{\partial^2}{\partial z^2}=
4\frac{\partial}{\partial z}\frac{\partial}{\partial \bar z}$. So $\alpha$
and $\tau$ should be expressed through $\bar z$ and $z$ 
$$\tau=\frac{1}{2}\ln{\frac{(p_4+a)^2+r^2}{(p_4-a)^2+r^2}}=
\frac{1}{2}\ln{\frac{(z+a)(\bar z +a)}{(z-a)(\bar z -a)}} \; . $$
\noindent As for $\alpha$, using $\arccos{A}=i\ln{[A+\sqrt{A^2-1}]}$, we get
$$\alpha=\arccos{\frac{p_4^2+r^2-a^2}{\sqrt{(p_4+a)^2+r^2}
\sqrt{(p_4-a)^2+r^2}}}=$$
$$\arccos{\frac{z\bar z -a^2}{\sqrt{(z^2-a^2)(\bar z^2-a^2)}}}=
\frac{i}{2}\ln{\frac{(z+a)(\bar z -a)}{(z-a)(\bar z +a)}} \; . $$
\noindent It follows from the above given equations that
$$\frac{\partial \alpha}{\partial z}=\frac{-ia}{z^2-a^2} \; , \; \;
\frac{\partial \alpha}{\partial \bar z}=\frac{ia}{\bar z^2-a^2} \; ,
\; \; \frac{\partial \tau}{\partial z}=-\frac{a}{z^2-a^2} \; , \; \; 
\frac{\partial \tau}{\partial \bar z}=-\frac{a}{\bar z^2-a^2} \; . $$
\noindent Thus
$$\frac{\partial }{\partial z}=\frac{\partial \alpha}{\partial z}
\frac{\partial }{\partial \alpha}+\frac{\partial \tau}{\partial z}
\frac{\partial }{\partial \tau}=-\frac{a}{z^2-a^2}\left (
\frac{\partial }{\partial \tau}+i\frac{\partial }{\partial \alpha}
\right ) $$
\noindent and
$$\frac{\partial }{\partial \bar z}=-\frac{a}{\bar z^2-a^2}\left (
\frac{\partial }{\partial \tau}-i\frac{\partial }{\partial \alpha}
\right ) \; . $$
\noindent Therefore
$$\frac{\partial^2}{\partial p_4^2}+\frac{\partial^2}{\partial r^2}=
\frac{4a^2}{(z^2-a^2)(\bar z^2-a^2)}\left (\frac{\partial^2}
{\partial \tau^2}+\frac{\partial^2}{\partial \alpha^2} \right ) \; , $$
\noindent (note that $\frac{\partial }{\partial \tau}+i\frac{\partial }
{\partial \alpha}=2\frac{\partial }{\partial (\tau -i\alpha)}$ commutes
with $\frac{1}{\bar z^2-a^2}$, because $\bar z=p_4-ir=a\frac
{\sinh{\tau}-\sinh{(i\alpha)}}{\cosh{\tau}-\cosh{(i\alpha)}}=a{\rm coth}
\frac{\tau+i\alpha}{2}$). But 
$$(z^2-a^2)(\bar z^2-a^2)=(z+a)(\bar z +a)(z-a)(\bar z -a)=$$
$$[(p_4+a)^2+r^2][(p_4-a)^2+r^2]=\frac{4a^4}{(\cosh{\tau}-
\cos{\alpha})^2} \; , $$
\noindent and for the Laplacian we finally get
$$\Delta=\frac{(\cosh{\tau}-\cos{\alpha})^3}{a^2\sin{\alpha}}
\left (\frac{\partial^2}{\partial \tau^2}+\frac{\partial^2}{\partial
\alpha^2} \right ) \frac{\sin{\alpha}}{\cosh{\tau}-\cos{\alpha}}-
\frac{(\cosh{\tau}-\cos{\alpha})^2}{a^2\sin^2{\alpha}}\delta^{(3)} 
\; . $$
\noindent This could be rewritten in a more convenient form by using
$$\frac{\partial^2}{\partial \alpha^2}\sin{\alpha}=\sin{\alpha}
\left ( \frac{\partial^2}{\partial \alpha^2}+2{\rm ctg}\alpha
\frac{\partial}{\partial \alpha} -1 \right ) $$
\noindent and Eq.45 :
$$\Delta=\frac{(\cosh{\tau}-\cos{\alpha})^3}{a^2}
\left ( \frac{\partial^2}{\partial \tau^2}-\delta^{(4)}-1 \right )
\frac{1}{(\cosh{\tau}-\cos{\alpha})} \; . $$
\noindent So Eq.70 in the bipolar coordinates takes the form
$$\left [(\cosh{\tau}-\cos{\alpha})  
\left ( \frac{\partial^2}{\partial \tau^2}-\delta^{(4)}-1 \right )
\frac{1}{(\cosh{\tau}-\cos{\alpha})}+ \right . $$ $$ \left . \frac{\lambda}
{(a\cosh{\tau}-i\eta_1M\sinh{\tau})(a\cosh{\tau}+i\eta_2M\sinh{\tau})}
\right ] \Psi=0 \; . $$
\noindent It is possible to separate variables in this equation. In 
particular, if we take
\begin{equation}
\Psi(\tau,\alpha,\Theta,\varphi)=(\cosh{\tau}-\cos{\alpha})f(\tau)
Y_{nlm}(\alpha,\Theta,\varphi) \; .
\label{eq72} \end{equation}
\noindent and use $\delta^{(4)}Y_{nlm}=n(n+2)Y_{nlm}$ equation, we get for
the function $f$ the following equation
\begin{equation}
\frac{d^2f}{d\tau^2}+\left [
\frac{\lambda}
{(a\cosh{\tau}-i\eta_1M\sinh{\tau})(a\cosh{\tau}+i\eta_2M\sinh{\tau})}
-(n+1)^2 \right ] f =0 \; .
\label{eq73} \end{equation}
\noindent To simplify, note that because
$$\sinh{2\tau}=2\sinh{\tau}\cosh{\tau} \; , \; \;
1+\cosh{2\tau}=2\cosh^2{\tau} \; \; {\rm and} \; \;
\cosh{2\tau}-1=2\sinh^2{\tau}$$
\noindent the following holds
$$
(a\cosh{\tau}-i\eta_1M\sinh{\tau})(a\cosh{\tau}+i\eta_2M\sinh{\tau})
=$$
$$a^2\cosh^2{\tau}+\eta_1\eta_2M^2\sinh^2{\tau}+iMa(\eta_2-\eta_1)
\cosh{\tau}\sinh{\tau}= $$
$$\frac{1}{2}\{(a^2+\eta_1\eta_2M^2)\cosh{2\tau}+
iaM(\eta_2-\eta_1)\sinh{2\tau}-(\eta_1\eta_2M^2-a^2) \} \; . $$
\noindent But
$$(a^2+\eta_1\eta_2M^2)^2-i^2a^2M^2(\eta_2-\eta_1)^2=$$
$$a^4+\eta_1^2\eta_2^2M^4+a^2M^2\eta_2^2+a^2M^2\eta_1^2=
(a^2+\eta_1^2M^2)(a^2+\eta_2^2M^2)=m_1^2m_2^2 \; , $$
\noindent therefore there exists a complex number $\nu$ such that
$$\cosh{\nu}=\frac{1}{m_1m_2}(a^2+\eta_1\eta_2M^2) \; \; {\rm and}
\; \; \sinh{\nu}=\frac{i}{m_1m_2}aM(\eta_2-\eta_1) \; . $$
\noindent Besides
$$\eta_1\eta_2M^2-a^2=\eta_1\eta_2M^2-m_1^2+\eta_1^2M^2=
\eta_1M^2-m_1^2=$$
$$\frac{1}{2}(M^2+m_1^2-m_2^2)-m_1^2=\frac{1}{2}(M^2-m_1^2-m_2^2)
\; , $$
\noindent and, using $\cosh{(x+y)}=\cosh{x}\cosh{y}+\sinh{x}\sinh{y}$,
we get
$$
(a\cosh{\tau}-i\eta_1M\sinh{\tau})(a\cosh{\tau}+i\eta_2M\sinh{\tau})=
$$
$$\frac{m_1m_2}{2}\left [\cosh{(2\tau+\nu)}-\frac{1}{2m_1m_2}
(M^2-m_1^2-m_2^2) \right ] \; . $$
\noindent Therefore Eq.73, after the introduction of a new variable
$\sigma=\tau+\frac{\nu}{2}$, will look as
\begin{equation}
\frac{d^2f}{d\sigma^2}+\left [
\frac{\lambda}
{m_1m_2\cosh^2{\sigma}-\frac{1}{4}[M^2-(m_1-m_2)^2]} -(n+1)^2 \right ]
f =0 \; .
\label{eq74} \end{equation}
Its form reveals the already known to us fact that the unequal mass case
is equivalent to the equal mass problem with $m^2=m_1m_2$ and $M^{\prime 2}
=M^2-(m_1-m_2)^2$.

Thus, it is sufficient to consider the following equation (for equal masses
$\nu=0$ and $\sigma=\tau$)
\begin{equation}
\frac{d^2f}{d\tau^2}+\left [
\frac{\lambda}
{m^2\cosh^2{\tau}-\frac{1}{4}M^2} -(n+1)^2 \right ] f =0 \; .
\label{eq75} \end{equation}
As is clear from the bipolar coordinates definition $-\infty < \tau <
\infty$, therefore Eq.75 should be supplied with boundary conditions at
$\tau \to \pm \infty$. These conditions follow from the initial integral
equation, but instead we will show that Eq.75 is equivalent to the Eq.56.
Indeed, let us introduce a new variable $t=\sinh{2\tau}+\cosh{2\tau}=
e^{2\tau}$, then $\frac{\partial^2}{\partial \tau^2}=4t\frac{\partial}
{\partial \tau}+4t^2\frac{\partial^2}{\partial \tau^2}$. But $(1+t)^2=
(2\cosh^2{\tau}+2\cosh{2\tau}\sinh{2\tau})^2=4\cosh^2{\tau}(\cosh{2\tau}
+\sinh{2\tau})^2=4(\cosh^2{\tau})t$, so $\cosh^2{\tau}=\frac{(1+t)^2}
{4t}$, and Eq.75 takes the form
$$\frac{d^2f}{dt^2}+\frac{1}{t}\frac{df}{dt}+\left \{
\frac{\lambda}{t[m^2(1+t)^2-M^2t]}-\frac{(n+1)^2}{4t^2} \right \}
f=0 \; , $$
\noindent which after introduction of a new function $\phi=\sqrt{t}f$
coincides to the Eq.56. Therefore Eq.57 implies the following boundary 
conditions for the Eq.75
\begin{equation}
f(\tau) \sim e^{-(n+1)|\tau|} \; , \; \; {\rm when} \; \;
\tau \to \pm \infty \; .
\label{eq76} \end{equation}

Appearance of the spherical function in Eq.72 indicates that the
Wick-Cutkosky model possesses the $SO(4)$ hidden symmetry even for unequal
masses. The variable separation is possible just because of this symmetry.

\vspace*{3mm}
\centerline{{\bf Note about references}}

\vspace*{2mm}
Only the sources of this review are presented here. Exhaustive bibliography
about the Bethe-Salpeter equation and in particular about the Wick-Cutkosky
model can be found in \cite{13}.

\else
\fi

\end{document}